\pdfoutput=1
\RequirePackage{ifpdf}
\ifpdf 
\documentclass[pdftex]{sigma}
\else
\documentclass{sigma}
\fi

\usepackage{mathrsfs}
\input{xy}
\xyoption{all}

\numberwithin{equation}{section}

\newcommand{\mbf}[1]{{\boldsymbol {#1} }}
\newcommand{\im}{\operatorname{Im}}
\newcommand{\rank}{\operatorname{rank}}
\def\ii{{{\rm i}}}
\def\e{{\rm e} }

\newcommand{\DT}{{\tt DT}}

\newcommand{\GV}{{\tt GV}}

\newcommand{\tr}{{\rm tr}}
\newcommand{\torus}{{\mathbb T}}
\newcommand{\id}{{1\!\!1}}

\begin{document}

\allowdisplaybreaks

\newcommand{\arXivNumber}{1511.01421}

\renewcommand{\PaperNumber}{052}

\FirstPageHeading

\ShortArticleName{BPS Spectra, Barcodes and Walls}

\ArticleName{BPS Spectra, Barcodes and Walls}

\Author{Michele CIRAFICI~$^{\dag\ddag\S}$}

\AuthorNameForHeading{M.~Cirafici}

\Address{$^\dag$~Department of Mathematics and Geoscience, Universit\`a di Trieste, and INFN,\\
\hphantom{$^\dag$}~Sezione di Trieste, Via A.~Valerio 12/1, I-34127 Trieste, Italy}
\EmailD{\href{mailto:michelecirafici@gmail.com}{michelecirafici@gmail.com}}

\Address{$^\ddag$~CAMGSD, Instituto Superior T\'ecnico, Universidade de Lisboa,\\
\hphantom{$^\ddag$}~Av. Rovisco Pais, 1049-001 Lisboa, Portugal}

\Address{$^\S$~Institut des Hautes \'Etudes Scientifiques, Le Bois-Marie, 35 route de Chartres,\\
\hphantom{$^\S$}~F-91440 Bures-sur-Yvette, France}

\ArticleDates{Received November 12, 2018, in final form July 04, 2019; Published online July 09, 2019}

\Abstract{BPS spectra give important insights into the non-perturbative regimes of supersymmetric theories. Often from the study of BPS states one can infer properties of the geometrical or algebraic structures underlying such theories. In this paper we approach this problem from the perspective of persistent homology. Persistent homology is at the base of topological data analysis, which aims at extracting topological features out of a set of points. We use these techniques to investigate the topological properties which characterize the spectra of several supersymmetric models in field and string theory. We discuss how such features change upon crossing walls of marginal stability in a few examples. Then we look at the topological properties of the distributions of BPS invariants in string compactifications on compact threefolds, used to engineer black hole microstates. Finally we discuss the interplay between persistent homology and modularity by considering certain number theoretical functions used to count dyons in string compactifications and by studying equivariant elliptic genera in the context of the Mathieu moonshine.}


\Classification{83E30; 81Q60; 55N99}

\section{Introduction}

In supersymmetric theories one often can gain deep insights by studying the properties of protected states. States which preserve a fraction or all of the supersymmetries can be used to get exact results about quantities of physical interest. Such states are usually directly related to geometrical quantities, such as enumerative invariants, or to the mathematical structures underlying the physical models.

For example the BPS spectral problem in quantum field theories is deeply related to the structure of the quantum vacuum and plays an important role in understanding various dualities \cite{Seiberg:1994rs}. In black hole physics, the exact enumeration of microstates is a problem of prime importance as it provides a quantum statistical derivation of gravitational thermodynamics \cite{Katz:1999xq, Strominger:1996sh}. Furthermore in many cases the duality properties of a theory directly imply modular properties of the partition functions. In all these cases the counting problem has roots in various areas of mathematics and has important physical consequences.

In general one is lead to investigate the structures underlying such counting problems. For example if the counting can be organized according to the representation theory of some group or algebra, then one has identified a fundamental principle in the physical theory. The typical situation is however less direct and often a consequence of several structures simultaneously.

Consider for example the case of $\mathcal{N}=2$ ${\rm SU}(N)$ super Yang--Mills theories on $\mathbb{R}^3 \times S^1$. The moduli space of vacua is the Hitchin moduli space and is hyperK\"ahler \cite{Seiberg:1996nz}. The latter condition is guaranteed by the fact that the BPS spectrum of the theory obeys a wall-crossing formula \cite{Gaiotto:2008cd}. The latter originates from the theory of generalized Donaldson--Thomas invariants, which encodes the behavior of stable objects in a (derived) category under a change of stability conditions \cite{Kontsevich:2008fj}. As a consequence the moduli space of vacua is locally a (generalized) cluster variety, the overlap transformations between charts being dictated by a cluster algebra \cite{Gaiotto:2009hg}. The BPS spectrum can also be seen as the set of stable representations of the quiver underlying this cluster algebra \cite{Alim:2011kw} and has a deep connection with integrable systems \cite{Cecotti:2014zga}.

In many physical cases the situation is similar and the study of the structure of the set of supersymmetric states leads to several layers of increasing complexity. In this paper we take a step back and ask the following question: is there any structure at the \textit{topological} level? In particular we can consider a collection of supersymmetric states simply as a set and study its properties using topological methods. The purpose of this note is to investigate the presence (or absence) of any noticeable topological feature in certain samples of supersymmetric states. In particular we will be interested in how these features change as the parameters of the theory are changed, or if one considers a similar problem in different settings.

To be more precise, with topological features we mean the properties of the spectra as seen from the perspective of persistent homology~\cite{ELZ02,ZC05}. Persistence is a relatively new approach to homological features of a space or a set, and is at the core of what is by now known as topological data analysis~\cite{C09,EH08,G14}. This field proposes to handle multidimensional and large sets of data using methods based on topology. This approach has been quite useful in disparate fields, such as biology, neuroscience or complex systems \cite{jholes,CIdSZ08,CCR13,NLC11,MA15}. In this note we will apply such methods to supersymmetric spectra, computed directly or extracted from certain number theoretical functions.

In essence topological data analysis is a multi-scale approach to extracting homological features out of a set of data, focused on identifying those features which persist over a long range of scales. The idea consists in defining a family of simplicial complexes which depend on a continuous proximity parameter $\epsilon$. For each value of $\epsilon$ one can pass to the homology of the complex and study how it varies as a function of $\epsilon$. At each length scale the homology is characterized by its homology classes; as the length scale changes new homology classes can form or already existing classes can disappear, depending on the evolution of the underlying simplicial complex. The set of data is characterized by the lifespans, or persistence, of said homology classes. These lifespans can be more easily visualized as a collection of intervals on the $\epsilon$ line, which begin at the value of $\epsilon$ at which the homology class appears and end when it disappear. Such collections of intervals are called \textit{barcodes}.

We will consider a supersymmetric spectrum as a dataset, where each point is labelled by the charges or the relevant quantum numbers of a state, and by its degeneracy, or BPS enumerative invariant. We will then proceed to apply the methods of topological data analysis to compute the homology of this set as a function of a proximity parameter $\epsilon$ and compute its barcodes for each non-trivial homology group. The set of barcodes gives a complete characterization of the persistent homological features of the supersymmetric spectra. We will then discuss how these topological characteristics vary between different datasets. Roughly speaking we will do so in two ways: or by comparing spectra obtained within the same physical theory but as the parameters are varied; or by studying different models which can be however associated with a~very similar physical problem.

We will discuss at length how the expected physical features show up in the topological analysis. In many cases this will be apparent from the barcodes, in others a bit more care will be required. Overall we will learn how to apply the methods of topological data analysis to BPS counting problems and argue which kind of information we can hope to extract. We will do so with many examples. In this paper we will focus more in detail on supersymmetric spectra, but the formalism is more general and can be applied to other counting problems. Applications of persistent homology to the study of string vacua will appear in~\cite{TDAvacua}.

To keep this note readable we have included very brief reviews of the physical problems at hand. This material is known, but we have chosen to present it in such a way as to highlight the role played by BPS invariants. All the computations in persistent homology have been done with the program \textsc{matlab} using the library \textsc{javaplex}, made available in~\cite{javaplex}. The accompanying software and datasets are available in~\cite{programs}. For the extraction of the BPS invariants and the manipulations of the relevant series we have used \textsc{mathematica}.

This paper is organized as follows. Section~\ref{BPS} will give some background about BPS states, their wall-crossing behavior and their enumerative interpretation. In Section~\ref{persistence} we will give an elementary introduction to the ideas of persistence and discuss the methods of topological data analysis. This section is meant to be readable by non-experts and to quickly convey the main ideas. Section~\ref{classS} contains our first application, to $\mathcal{N}=2$ ${\rm SU}(3)$ super Yang--Mills, where we compare the topological features of the BPS spectrum in two adjacent chambers. The interplay between persistent homology and wall-crossing is also the focus of Section~\ref{conifold}, which discusses the case of the conifold in detail, using some approximation schemes introduced in Section~\ref{witness}. Section~\ref{qgeom} takes a different approach; here we compute the Donaldson--Thomas invariants for a few distinct one parameter compact Calabi--Yaus, compare their distributions and discuss the implications for black hole physics. Section~\ref{dyons} is about a different class of black holes, in $\mathcal{N}=4$ string compactifications. In this case the relevant partition functions have modular properties and we discuss the interplay between modularity and topology. Section~\ref{mathieu} takes a~similar approach, now in the context of elliptic genera in the Mathieu moonshine correspondence. In this case, the technical details of the topological analysis are postponed to the Appendix~\ref{MoonshineBar}. We summarize our finding in Section~\ref{discussion}.

\section{BPS states and wall-crossing} \label{BPS}

In this note we will consider certain field and string theories with extended supersymmetry. In this section we will quickly review some general properties and postpone a more detailed description to later sections on a case by case basis. The theories we shall consider all have moduli spaces of quantum vacua~$\mathcal{M}$. These moduli spaces often have a direct geometrical interpretation, for example parametrizing deformation of a compactification manifold or solutions of certain differential equations. Determining $\mathcal{M}$ captures the vacuum structure of the theory. In theories with extended supersymmetry one can often give a remarkably precise local description of the moduli spaces $\mathcal{M}$, in the form of an answer determined at weak coupling plus a series of quantum corrections.

On top of the geometry of~$\mathcal{M}$, there is other physical information which can be computed exactly. In this note we will be interested in the spectrum of BPS states, which is very closely related to the series of quantum corrections which determine the moduli space of quantum vacua~$\mathcal{M}$. These quantities are particularly important because due to the amount of supersymmetry preserved, they allow for the extrapolation of weakly coupled computations to strong coupling. In other words they are one of the few available sources of non-perturbative information in quantum theories.

Supersymmetric theories have a Hilbert space of states $\mathcal{H}$ upon which the supersymmetry generators act as operators. States in~$\mathcal{H}$ can be organized according to the representation theory of the supersymmetry algebra. BPS states are characterized by the fact that a certain number of supersymmetry generators are represented trivially. The fact that a BPS state is annihilated by certain operators is a rather strong constraint, in many cases strong enough to reduce quantum corrections to a computable form.

To be concrete consider $\mathcal{N}=2$ theories in four dimensions. We denote by~$\Gamma$ the lattice of electric and magnetic conserved charges, as measured at spatial infinity and at a point in the moduli space of vacua. For example in a string theory compactification or engineering, this lattice can be realized as the homology of a certain variety. The lattice of charges is endowed with the antisymmetric Dirac pairing
\begin{gather*}
\langle \, , \, \rangle \colon \ \Gamma \times \Gamma \longrightarrow \mathbb{Z} .
\end{gather*}
This pairing vanishes identically on the charges of particles which are mutually local. The conserved charges of $\Gamma$ divide the Hilbert space of states into superselection sectors. The BPS degeneracies $\Omega (\gamma ; u)$ count with signs the number of BPS states with charge $\gamma \in \Gamma$. They are defined as traces over the single particle BPS Hilbert space $\mathcal{H}^{\rm BPS}_u = \bigoplus_{\gamma \in \Gamma} \mathcal{H}^{\rm BPS}_{\gamma;u}$, filtered by the charge measured at spatial infinity.

The single particle Hilbert spaces $\mathcal{H}_{\gamma;u}$ and the BPS degeneracies $\Omega (\gamma ; u)$ depend explicitly on a point $u \in \mathcal{M}$. The constraints arising from supersymmetry are such that $\Omega (\gamma ; u)$ has a~very specific dependence on $u \in \mathcal{M}$: it is a piecewise constant function, almost independent on the physical parameters except for certain codimension one walls in $\mathcal{M}$, at which it jumps suddenly. This is the wall-crossing phenomenon. At walls of marginal stability the change in the BPS degeneracies $\Omega (\gamma ; u)$ describe physical processes of fusion or fission of BPS particles from or into elementary constituents. The wall-crossing of the BPS degeneracies is a very strong constraint on the consistency of a theory at the quantum level~\cite{Seiberg:1994rs}. The moduli space of vacua~$\mathcal{M}$ is divided by the walls of marginal stability into chambers $\mathscr{C}$. Solving the BPS spectrum of a~theory amounts in finding the $\Omega (\gamma ; u)$ in each chamber.

Walls of marginal stability $\mathsf{MS}$ are defined as the loci in moduli space where the central charges of two or more BPS particles become parallel. In theories with extended supersymmetry the central charge is realized as an holomorphic function over the moduli space $\mathcal{M}$,
\begin{gather*}
\mathcal{Z} \colon \ \mathcal{M} \longrightarrow \mathrm{Hom} (\Gamma ; \mathbb{C}) .
\end{gather*}
For example, a two body decay of a state with charge $\gamma$ into two elementary constituents~$\gamma_1$ and~$\gamma_2$ is kinematically allowed at the locus
\begin{gather*}
\mathsf{MS} (\gamma_1 , \gamma_2) = \left\{ u \in \mathcal{M} \vert \arg \mathcal{Z}_{\gamma_1} (u) = \arg \mathcal{Z}_{\gamma_2} (u) \right\} .
\end{gather*}
In many cases the central charge function has a very explicit description: in four dimensional quantum field theories is given by the integral of the Seiberg--Witten differential $\lambda$ over a cycle of the Seiberg--Witten curve whose homology class correspond to a charge $\gamma \in \Gamma$. Similarly in Calabi--Yau compactifications of the type II string it is given by an integral of the holomorphic $(3,0)$-form over 3-cycles.

More formally we can usually describe BPS states as objects in some abelian category $\mathsf{A}$. In concrete examples $\mathsf{A}$ could be the category of representations of a certain quiver with potential $\mathsf{rep} (Q,\mathcal{W})$, or the category of coherent sheaves on a Calabi--Yau threefold $X$, $\mathsf{coh} (X)$. This is not completely correct as a more precise account would require objects in the bounded derived category $\mathsf{D (A)}$, but for the purpose of this section we will neglect these issues. In these cases there is an isomorphism which identifies the lattice of conserved charges $\Gamma$ with the topological Grothendieck group $K (\mathsf{A})$. In the above cases the isomorphism is given by the Chern character in the case of $\mathsf{coh} (X)$ and by the identification of the simple representations with a basis of BPS states in the case of $\mathsf{rep} (Q,\mathcal{W})$. In any case we can regard the central charge as a stability function on $\mathsf{A}$, at fixed $u \in \mathcal{M}$
\begin{gather*}
\mathcal{Z}_u \colon \ K (\mathsf{A}) \longrightarrow \mathbb{C} ,
\end{gather*}
which to any object $E \in \mathsf{A}$ associates a complex number $\mathcal{Z}_u (E)$. We say that a BPS state described by an object $E$ is $\mathcal{Z}$-stable if $\arg \mathcal{Z}_u (F) < \arg \mathcal{Z}_u (E)$ for any proper sub-object~$F$ of~$E$. Note that the stability condition explicitly depends on $u \in \mathcal{M}$ and therefore on the parameters of the theory. As these parameters are varied, the stability condition changes and a~stable object may become unstable.

Since all that matters is the phase of the complex number $\mathcal{Z}_u (E)$, we will loosely speak of a~BPS state as a~BPS ray $\ell_\gamma$ associated with the BPS state of charge $\gamma$, a vector in the complex plane~$\mathbb{C}$ (which we will refer as the central charge plane), as is by now common use \cite{Alim:2011kw, Gaiotto:2008cd}.

The change in $\Omega (\gamma ; u)$ across a wall of marginal stability is governed by a wall-crossing formula \cite{Joyce:2008pc,Kontsevich:2008fj,Manschot:2010qz}. The Kontsevich-Soibelman wall-crossing formula (KSWCF) states that a certain product of operators, which depends on the stable BPS charges and on the BPS degeneracies, remains invariant across walls of marginal stability as to compensate for the change in the degeneracies $\Omega (\gamma ; u)$. To describe the KSWCF we need a few more ingredients. We introduce the torus $\mathbb{T}_\Gamma = \Gamma \otimes_\mathbb{Z} \mathbb{C}^*$ and formal variables $X_{\gamma}$ for each $\gamma \in \Gamma$, which enjoy the property
\begin{gather*}
X_{\gamma_i} X_{\gamma_j} = (-1)^{\langle \gamma_i , \gamma_j \rangle} X_{\gamma_i + \gamma_j} .
\end{gather*}
The operators $\mathcal{K}_{\gamma}$ are automorphisms of the algebra of functions on~$\mathbb{T}_\Gamma$ which act as
\begin{gather*}
\mathcal{K}_{\gamma} (X_\delta) = \big(1 - (-1)^{\langle \gamma , \delta \rangle} X_{\gamma}\big)^{\langle \gamma , \delta \rangle} X_{\delta} .
\end{gather*}
To state the KSWCF we choose an angular sector $A$ in the central charge plane. Then the KSWCF states that the phase ordered product
\begin{gather} \label{KSWCF}
\prod_{\gamma \colon \arg \mathcal{Z}_{\gamma} (u) \in A} \mathcal{K}_\gamma^{\Omega(\gamma ; u)}
\end{gather}
is invariant across walls of marginal stability, under the assumptions that no BPS state enter or leaves the sector $A$. See \cite{MooreFelix} for a more in depth review.

The situation for $\mathcal{N}=4$ theories is similar. One can still define a central charge function $\mathcal{Z}$ as a moduli dependent function which at a point in the moduli space associates to a state a charge dependent complex vector. The BPS condition now depends on the amount of supersymmetry preserved. It is customary to use the notation $(\mbf P, \mbf Q)$ to indicate the charge of a generic 1/4 BPS dyon, while 1/2 BPS states are necessarily purely electrically or magnetically charged. The degeneracies of BPS states can be defined as certain helicity supertraces over the Hilbert space of states. The main difference in the wall-crossing behavior respect to the $\mathcal{N}=2$ case is that now only two bodies decays are allowed, namely of a 1/4 dyon into two 1/2 BPS states.

In this section we have review very briefly some basic properties of BPS states in supersymmetric theories. The set of stable BPS states has clearly a lot of structure, which has lead to deep physical insights and beautiful mathematics. These structures have deep algebraic and geometrical origin in the theory of generalized Donaldson--Thomas invariants and of wall-crossing structures \cite{Kontsevich:2008fj,Kontsevich:2013rda}. In this note we want to investigate their features from a rather different perspective: we will look at the set of BPS states as a distribution of points and try to understand its topological properties, and how these change upon crossing walls of marginal stability. But first we have to set up the appropriate tools.

\section{Persistent homology} \label{persistence}

In this section we will introduce the concept of persistent homology and explain its uses in the context of topological data analysis. The idea behind persistence is to study topological features of a space as a function of the length scale~\cite{ELZ02,ZC05}. When applied to a set of datas, the topological analysis extract qualitative features, which are independent on any particular metric or coordinate system used, and robust to noise. In our exposition we will mainly follow the reviews~\cite{C09,EH08,G14}.

The techniques of topological data analysis are increasingly common in many fields such as biology, neuroscience, complex systems or the study of language, see \cite{jholes,CIdSZ08,CCR13,NLC11,MA15} for a sample of the literature. An application of these techniques to the study of string vacua appears in \cite{TDAvacua}.

\subsection{Homology of simplicial complexes}

Homology captures intrinsic topological information of a space. To a topological space $X$ we assign a collection of abelian groups~$H_i (X)$ whose independent elements formally correspond to topological features of $X$, such as its number of components or holes. The computation of the homology of a space is a standard procedure to study its topology. There are several ways to do this, as well as several homology theories which can be defined.

A very convenient approach uses \textit{simplicial complexes}. We can think of a simplicial complex as a triangulation of a space, whose elements are vertices, edges and faces and so on, and which can be studied with combinatorial or algebraic techniques. A simplicial complex $S$ is a~pair consisting of a finite set $V$ of vertices and a family $\Sigma$ of non-empty subsets of $V$. The collection~$\Sigma$ is defined by the property that if $\sigma \in \Sigma$ and $\tau \subseteq \sigma$, then $\tau \in \Sigma$, which for example implies that if a certain simplex is part of~$\Sigma$, so are its faces. The $k$-simplexes of $\Sigma$ form the subset $\Sigma_k$ of simplexes with cardinality $k+1$.

For example a standard simplicial complex associated to a metric space $X$ is the \v{C}ech complex. Let $B_\epsilon (x)$ be the standard ball of radius $\epsilon$ centered at $x \in X$. Assume that we can find a~set $V \subset X$ so that $X = \bigcup_{v \in V} B_\epsilon (v)$. Then the \v{C}ech complex is defined as
\begin{gather*}
\textsf{\v{C}ech}_\epsilon (X) = \left\{
\sigma = [ v_0 , \dots , v_k ] \,\vert\, \bigcap_{i=0}^k B_\epsilon (v_i) \neq \varnothing\right\} .
\end{gather*}
This is a particular example of the \textit{nerve} construction.

In this note we will be interested in a version of this construction, applied to a very particular case. We define a \textit{point cloud} $\mathsf{X}$ as a collection of points $\{ \mathsf{x}_i \}_{i \in I}$ in $\mathbb{R}^N$. To a point cloud $\mathsf{X}$ we associate the \textit{Vietoris--Rips complex} $\mathsf{VR}_\epsilon (\mathsf{X})$ as
\begin{gather*}
\mathsf{VR}_\epsilon (\mathsf{X}) = \left\{
\sigma = [ v_0 , \dots , v_k ] \,\vert \, d ( v_i , v_j ) \le \epsilon \ \text{for all} \ i, j \right\} ,
\end{gather*}
where $d ( \, , \, )$ is the standard distance function on $\mathbb{R}^N$. In other words a simplex is identified by the pairwise intersection of radius $\epsilon / 2$ balls. Note that we can generalize immediately the definition of the \v{C}ech complex to point clouds. The fact that the Vietoris--Rips complex $\mathsf{VR}_\epsilon (\mathsf{X})$ is defined only in terms of pairwise intersection makes it much more amenable to algorithmic computations than the \v{C}ech complex $\textsf{\v{C}ech}_\epsilon (\mathsf{X})$. These complexes are related by the inclusions
\begin{gather*}
\mathsf{\check{C}ech}_\epsilon (\mathsf{X}) \subseteq \mathsf{VR}_{2 \epsilon} (\mathsf{X}) \subseteq \mathsf{\check{C}ech}_{2 \epsilon} (\mathsf{X}),
\end{gather*}
which imply that the Vietoris--Rips complex is a good approximation to the \v{C}ech complex. Note also that the two complexes have a natural orientation which follows by declaring that a~$k$-simplex $[v_0 , \dots , v_k]$ changes sign under an odd permutation.

To a simplicial complex $S$ we can associate its homology as follows. We define the group of $k$-chains $C_k$ by taking formal linear combinations of $k$-simplices as $c = \sum_i a_i \sigma_i$ where $a_i \in \mathbb{Z}_p$ for a prime number $p$. On $C_k$ we define the boundary operator $\partial_k \colon C_k \longrightarrow C_{k-1}$
\begin{gather*}
\partial_k ( [v_0 , v_1 , \dots , v_k] ) = \sum_{i=0}^k (-1)^i [v_0 , \dots , \hat{v}_i , \dots , v_k] ,
\end{gather*}
where the element $\hat{v}_i$ is omitted in the right-hand side. This operator can be used to define the chain complex
\begin{gather*}
\cdots \longrightarrow C_{k+1} \longrightarrow C_{k} \longrightarrow C_{k-1} \longrightarrow \cdots
\end{gather*}
and out of this the homology of $S$. To this end we introduce the spaces of $k$-cycles $Z_k (S) = \ker {\partial_k}$ and of $k$-boundaries $B_k (S) = \im {\partial_{k+1}}$. Then the $k$-th homology group $H_k (S)$ is defined as the quotient $Z_k (S) / B_k (S)$. The Betti numbers are the ranks of the homology groups, $b_k = \dim H_k (S)= \dim Z_k (S) - \dim B_k (S)$, and measure the number of $k$-cycles which are not $k$-boundaries.

A very important feature of this construction is its functoriality. A simplicial map $f$ between two simplicial complexes $S_1$ and $S_2$, is a map between the corresponding vertex sets such that a~simplex $\sigma$ of $S_1$ is mapped into a simplex $f (\sigma)$ of~$S_2$. A simplicial map takes a~$p$-simplex into a $k$-simplex, with $k \le p$. A simplicial map $f \colon S_1 \longrightarrow S_2$ induces a map between the vector spaces of $p$-chains, $C_p (f)\colon C_p (S_1) \longrightarrow C_p (S_2)$. Collecting all the induced maps we form the chain map~$C_{\bullet} (f)$, that is a collection of maps
\begin{gather*}
\xymatrix@C=8mm{ \cdots \ar[r] & C_p (S_1) \ar[r]^{\partial_p^{S_1}} \ar[d]^{C_p (f)} & C_{p-1} (S_1) \ar[r] \ar[d]^{C_{p-1} (f)} & \cdots \\
 \cdots \ar[r] & C_p (S_1) \ar[r]^{\partial_p^{S_2}} & C_{p-1} (S_1) \ar[r] & \cdots
 }
\end{gather*}
such that
\begin{gather*}
C_{p-1} (f) \circ \partial_{p}^{S_1} = \partial_{p}^{S_2} \circ C_p (f) .
\end{gather*}
In particular a chain map $C_\bullet (f)$ induced by the simplicial map $f$, induces a map between homology groups
\begin{gather*}
f_\star \colon \ H_i (S_1) \longrightarrow H_i (S_2) .
\end{gather*}
It will be important for us the case when the simplicial map $f$ is the inclusion. Then by functoriality $f_\star$ keeps track of the individual homology classes of $H_i (S_1)$ inside $H_i (S_2)$, that is it contains the information whether an homology class remains non-trivial or not.

\subsection{Persistent homology and barcodes}

In order to define persistent homology we need to introduce a few preliminary notions. Consider a~field $\mathbb{K}$. An $\mathbb{N}$-\textit{persistence} $\mathbb{K}$-vector space (or module) is a collection of vector spaces $\{ V_n \}_{n \in \mathbb{N}}$ over~$\mathbb{K}$ indexed by a natural number $n \in \mathbb{N}$ together with a collection of morphisms $\rho_{i,j} \colon V_i \longrightarrow V_j$ for every $i$ and $j$ so that $i \le j$. We further require a compatibility condition, that is $\rho_{i,k} \cdot \rho_{k,j} = \rho_{i,j}$ whenever $i \le k \le j$. Morphisms between $\mathbb{N}$-persistence vector spaces $\{ V_n \}$ and $\{ W_m \}$ are naturally defined as a collection of maps $f_i \colon V_i \longrightarrow W_i$ such that the diagrams
\begin{gather*}
\xymatrix@C=30mm{
V_i \ar[d]^{f_i} \ar[r]^{\rho_{i,j}} & V_j \ar[d]^{f_j} \\
W_i \ar[r]^{\tau_{i,j}} & W_j
}
\end{gather*}
commute. The same construction can be generalized to abelian groups, simplicial complexes, chain complexes and so on. More formally a version of these arguments can be applied to any category $\mathsf{Cat}$ to define the category of $\mathbb{N}$-persistence objects $\mathbb{N}_{\rm pers} (\mathsf{Cat})$.

There is a particular class of $\mathbb{N}$-persistence modules which are of interest in topological data analysis. A persistence module $\{ V_i \}_{i \in \mathbb{N}}$ is called \textit{tame} if: i)~each $V_i$ is finite dimensional, and ii)~$\rho_{n,n+1} \colon V_n \longrightarrow V_{n+1}$ is an isomorphism for large enough $n$. The reason such modules are interesting is that $\mathbb{N}$-persistence tame $\mathbb{K}$-modules are in one to one correspondence with finitely generated modules over the graded ring~$\mathbb{K}[t]$. The fact that the latter are finitely generated allows for a classification theorem for persistence modules, in terms of their barcodes~\cite{ZC05}.

In order to explain this classification theorem, we define the $\mathbb{N}$-persistence module $\mathbb{K} (m,n)$ as
\begin{gather*}
\mathbb{K} (m,n) =
\begin{cases}
0, &{\rm if} \ i < m \ \text{or} \ i > n, \\
\mathbb{K}, & \text{otherwise},
\end{cases}
\end{gather*}
where $m \le n$ are two integers, $m$ is non-negative and $n$ can be infinity. The morphism $\rho$ is simply $\rho_{i,j} = \mathrm{id}_\mathbb{K}$ for $m \le i < j \le n$. $\mathbb{K} (m,n)$ is also known as an interval module, which assigns a non-trivial vector space only to a certain interval.

The classification theorem then states that a given tame $\mathbb{N}$-persistence $\mathbb{K}$-module admits the unique (up to ordering of factors) decomposition
\begin{gather*}
\{ V_i \}_i \simeq \bigoplus_{j=0}^N \mathbb{K} (m_j , n_j) .
\end{gather*}
A tame persistence module is therefore completely specified by a collection of $N$ intervals, for a certain $N \in \mathbb{N}$, to which we assign a non-trivial vector space. An important consequence of this theorem is that we can completely specify a persistence module by its \textit{barcode}. A barcode is simply the collection of non-negative integers $(m_i , n_i)$, where $0 \le m \le n$ and eventually $n$ can be $+\infty$, which specify when the persistence module is non-trivial. This classification result is a generalization of the well known fact from elementary algebra that ordinary vector spaces are classified up to isomorphisms by their dimension. In a similar fashion persistence modules are characterized by a sequence of intervals. We will represent graphically such a collection of intervals by drawing a series of bars (hence the name barcode).

The reason these facts are important for us is that the persistent homology of a point cloud (or of any topological space) gives a persistent module. Then a barcode becomes a very effective tool to summarize and visualize homological features.

Consider a point cloud $\mathsf{X}$, a collection of points in $\mathbb{R}^N$. We denote by $\mathsf{X}_\epsilon$ the point cloud $\mathsf{X}$ where every point $\mathsf{x} \in \mathsf{X}$ has been replaced by a ball $B_\epsilon (\mathsf{x})$ of radius $\epsilon$ centered at $\mathsf{x}$. We regard $\mathsf{X}_\epsilon$ as a continuous family of topological spaces indexed by the real variable $\epsilon \in \mathbb{R}_{\ge0}$, with $\mathsf{X}_0 = \mathsf{X}$. For any fixed collection of values $0 = \epsilon_0 < \epsilon_1 < \epsilon_2 < \cdots$, we have the sequence of inclusions
\begin{gather*}
\mathsf{X}_{\epsilon_0} \hookrightarrow \mathsf{X}_{\epsilon_1} \hookrightarrow \mathsf{X}_{\epsilon_2} \hookrightarrow \cdots .
\end{gather*}
Similarly for each $\mathsf{X}_\epsilon$ we construct the associated Vietoris--Rips complex $\mathsf{VR}_{\epsilon/2} (\mathsf{X})$. Again this is a continuous family of simplicial complexes parametrized by $\epsilon$. On the other hand only for certain values of $\epsilon$ the simplicial complexes will be distinct, and again for those parameters we have a sequence of inclusions
\begin{gather} \label{VRfiltration}
\mathsf{VR}_{\epsilon_0} (\mathsf{X}) \hookrightarrow \mathsf{VR}_{\epsilon_1} (\mathsf{X}) \hookrightarrow \mathsf{VR}_{\epsilon_2} (\mathsf{X}) \hookrightarrow \cdots .
\end{gather}
For each $\mathsf{VR}_{\epsilon} (\mathsf{X})$ we construct the associated chain complex with coefficients in $\mathbb{Z}_p$, with~$p$ a~prime. Then passing to the $i$-th homology gives the $\mathbb{N}$-persistence module
\begin{gather} \label{Npersistence}
H_i (\mathsf{VR}_{\epsilon_0} (\mathsf{X}) ; \mathbb{Z}_p) \hookrightarrow H_i (\mathsf{VR}_{\epsilon_1} (\mathsf{X}) ; \mathbb{Z}_p) \hookrightarrow H_i (\mathsf{VR}_{\epsilon_2} (\mathsf{X}) ; \mathbb{Z}_p) \hookrightarrow \cdots .
\end{gather}
Technically both $\mathsf{VR}_{\epsilon} (\mathsf{X})$ and the $H_i (\mathsf{VR}_{\epsilon} (\mathsf{X}) ; \mathbb{Z}_p)$ are really $\mathbb{R}$-persistence modules, indexed by the real variable $\epsilon$. However only for a finite number of $\epsilon$'s these complexes are really distinct, and we can therefore talk of $\mathbb{N}$-persistence modules. More formally we pick any order preserving map $\mathbb{N} \longrightarrow \mathbb{R}$ and construct the $\mathbb{N}$-persistence modules out of the $\mathbb{R}$-persistence modules we have just defined, as explained more in detail in \cite{C09}.

Note that all this construction relies essentially on the functoriality of homology. The maps in \eqref{Npersistence} are those induced by the inclusions of \eqref{VRfiltration}. Without these maps \eqref{Npersistence} would just be a collection of vector spaces. It is immediate to see, and follows just from the definition of homology and from its functoriality, that these maps obey all the required properties to define a persistence module.

In particular this means that for every $i$ the $\mathbb{N}$-persistence module in \eqref{Npersistence} is completely characterized by a collection of barcodes. These barcodes capture topological features of the point cloud $\mathsf{X}$.

In this case we can also give a perhaps more direct description of the barcodes. The inclusions between topological spaces lift to maps between the homology groups; we call this map
\begin{gather*}
\mathsf{f}^{a,b}_i \colon \ H_i (\mathsf{VR}_{\epsilon_a} (\mathsf{X}) ; \mathbb{Z}_p) \longrightarrow H_i (\mathsf{VR}_{\epsilon_b} (\mathsf{X}) ; \mathbb{Z}_p)
\end{gather*}
for $\epsilon_a \le \epsilon_b$. Then we define the \textit{i-th persistent homology group} $\mathsf{H}^{a,b}_i = \im \mathsf{f}^{a,b}_i$, or more explicitly
\begin{gather*}
\mathsf{H}^{a,b}_i = \frac{Z_i (\mathsf{VR}_{\epsilon_a} (\mathsf{X}) )}{B_i (\mathsf{VR}_{\epsilon_b} (\mathsf{X}) ) \cap Z_i (\mathsf{VR}_{\epsilon_a} (\mathsf{X}) )} ,
\end{gather*}
where for simplicity we have not written down the inclusions. The \textit{i-th persistent Betti number} is naturally defined as $\beta_i^{a,b} = \rank \mathsf{H}^{a,b}_i$. The persistent Betti number $\beta_i^{a,b}$ is given by the number of barcodes of $H_i (\mathsf{VR}_{\epsilon} (\mathsf{X}) ; \mathbb{Z}_p) $ which span the whole interval $[ \epsilon_a , \epsilon_b ]$.

\subsection{Barcodes and topological features}

Let us expand a bit on the interpretation of the barcodes. Barcodes are a visual device which represent the number of persistent generators in the $i$-th homology group $H_i ( \mathsf{VR}_\epsilon (\mathsf{X}) ; \mathbb{Z}_p)$. Given two values of the proximity parameter $\epsilon_1 < \epsilon_2$, a persistent homology class along the interval $[\epsilon_1 , \epsilon_2]$ is a non-trivial homology class in $H_i ( \mathsf{VR}_{\epsilon_1} (\mathsf{X}) ; \mathbb{Z}_p)$ which is mapped into a non-trivial homology class in $H_i ( \mathsf{VR}_{\epsilon_2} (\mathsf{X}) ; \mathbb{Z}_p)$.

Consider the barcodes at a fixed value of $\epsilon$. This means that we are looking at the point cloud $\mathsf{X}$ at a certain characteristic scale given by $\epsilon$. At this scale the generators of $H_i ( \mathsf{VR}_\epsilon (\mathsf{X}) ; \mathbb{Z}_p)$ capture topological features of the data set: they represent $i$-dimensional configurations of points shaped as cycles which are not boundaries, meaning delimiting ``holes'' which are not filled up by other points. The persistence of these generators is a measure of how long these holes last as the value of the proximity parameter $\epsilon$ increases. Intuitively the clearer is the topological feature, for example if a certain hole contains none or very few points in its interior, the longer the persistent homology class lasts. A non-trivial long-lived persistent class in $H_i ( \mathsf{VR}_\epsilon (\mathsf{X}) ; \mathbb{Z}_p)$ indicates that the points in the data set cluster around a $i$-cycle without ``filling it up''.

For example persistent homology classes in $H_0 ( \mathsf{VR}_\epsilon (\mathsf{X}) ; \mathbb{Z}_p)$ measure the number of connected components in the point cloud as a function of the length scale. At very small values of $\epsilon$ each barcode correspond to a point in the original point cloud $\mathsf{X}$. As $\epsilon$ increases, neighbor points will form a single connected component. Long-lived barcodes are evidence for clustering of data into different regions. This could indicate for example that the data have a tendency to accumulate towards a certain point or area. A clear division into connected areas with similar behavior, for example the repetition of the same pattern in the barcodes, could also be regarded as evidence of an existing symmetry in the underlying physical problem.

A similar reasoning holds for higher homology groups. A persistent generator of $H_1 ( \mathsf{VR}_\epsilon (\mathsf{X}) ;\allowbreak \mathbb{Z}_p)$ which is long-lived implies that at different length scales a certain area of the point cloud~$\mathsf{X}$ is naturally well approximated by a one-dimensional manifold with the topology of an empty circle $S^1$. Similarly non-trivial elements of $H_i ( \mathsf{VR}_\epsilon (\mathsf{X}) ; \mathbb{Z}_p)$ will generically suggest that a certain region of the point cloud $\mathsf{X}$ can be approximated by a higher dimensional manifold with a given topology. The fact that data live on a certain shape can for example suggest a good coordinate system to approximate the point cloud; in general the presence of non-trivial persistent topologies in $\mathsf{X}$ is a hint of the existence of correlations between the data points, for example in the form of a set of equations which constrain regions of $\mathsf{X}$.

When discussing the barcode distribution for $H_i ( \mathsf{VR}_\epsilon (\mathsf{X}) ; \mathbb{Z}_p)$, we will loosely use the terminology: barcodes at Betti number $i$, barcodes in degree $i$, or barcodes for $H_i$.

\subsection{An example}

\begin{figure}[htbp]\centering
\begin{minipage}[b]{0.42\textwidth}\centering
\includegraphics[width=1\textwidth]{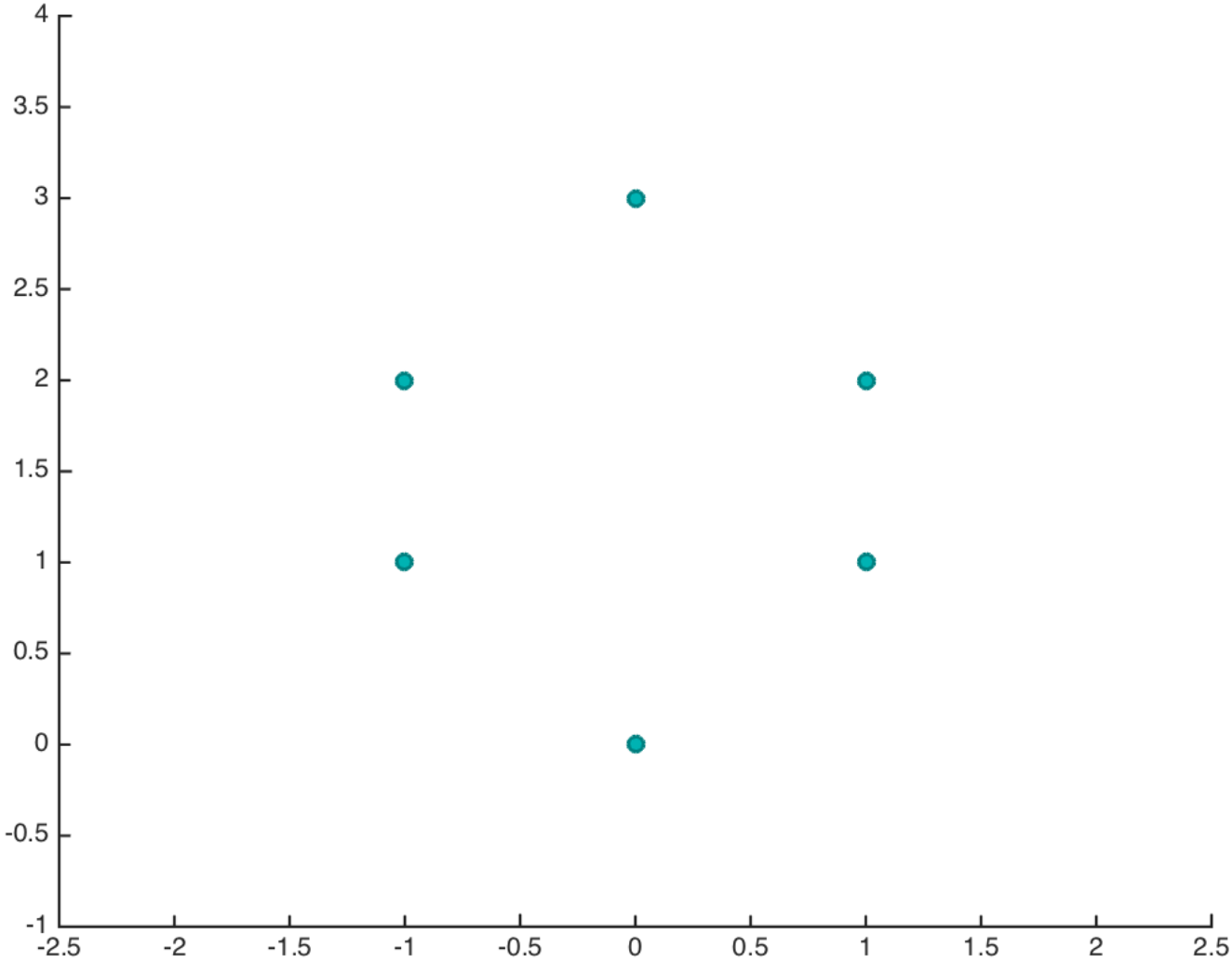}
\end{minipage}
\qquad
\begin{minipage}[b]{0.42\textwidth}\centering
\includegraphics[width=1\textwidth]{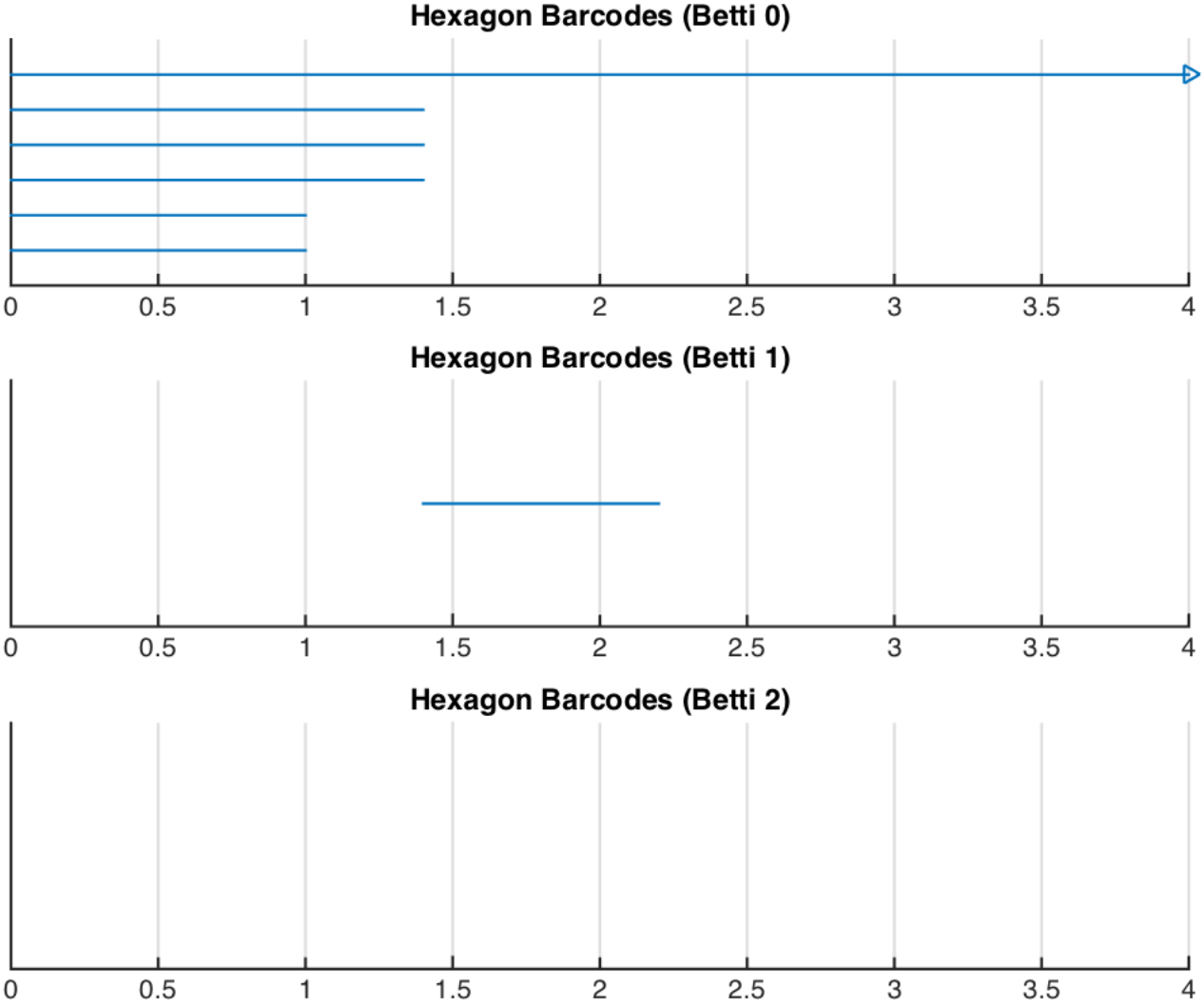}
\end{minipage}
\caption{\textit{Left}. A configuration of points which form the vertices of an hexagon. \textit{Right.} The corresponding barcodes computed from the Vietoris--Rips complex. The non-trivial barcode at Betti number~1 makes precise the statement that the six points look like an hexagon, which is topologically a circle.}\label{Hexagon}
\end{figure}

Before we proceed to apply these techniques to the study of BPS states, let us go through a~simple example. We take for our point cloud $\mathsf{X}$ the simple configuration of points shown in Fig.~\ref{Hexagon}. We want to understand its topological features using persistent homology, and in the process explain how to apply the relevant techniques step by step.

The configuration of points in Fig.~\ref{Hexagon} has a clear hexagonal shape. From a topological perspective this is equivalent to say that the points are distributed along a circle. While this is clear just by looking at the Figure, we would like to abstract this information in a collection of barcodes. What we gain in this abstraction will be clear in the following sections, where we will have to confront higher dimensional point clouds where no simple visualization tool is available.
We draw the relevant barcodes in Fig.~\ref{Hexagon} on the right. As with all the persistent homology computations in this paper, to obtain the barcodes we wrote a \textsc{matlab} program available in~\cite{programs}. Let us follow the formation and demise of persistent homology classes in ``time'' $\epsilon$.

\begin{figure}[htbp]\centering
\includegraphics[width=0.73\textwidth]{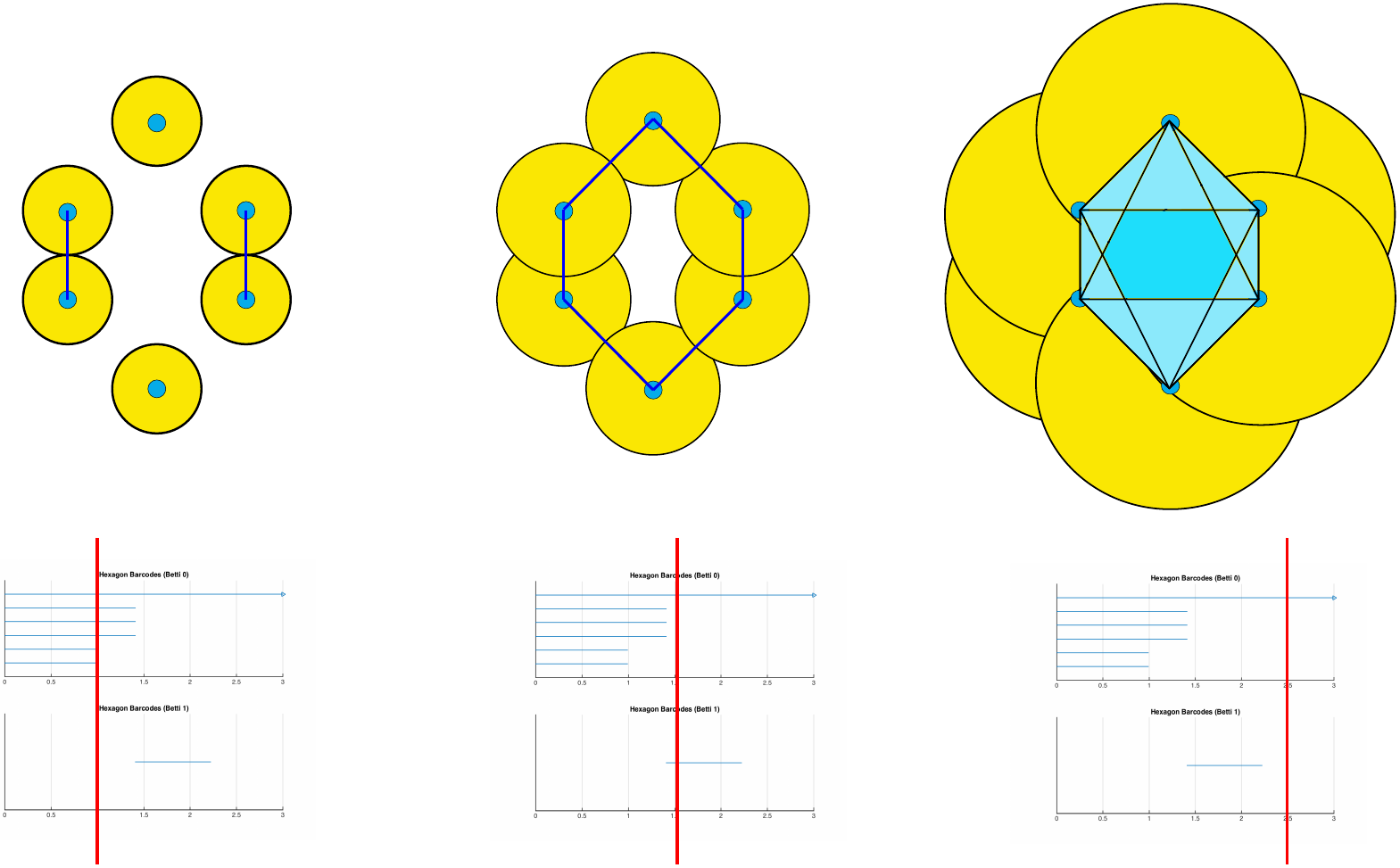}
\caption{The Vietoris--Rips complex (in blue) at various values of the proximity parameter $\epsilon$ as shown by the red line in the barcodes' plot. In yellow the balls around the points of the point cloud $\mathsf{X}$, whose radius $\epsilon/2$ is determined by the proximity parameter. At $\epsilon=1$ we see the formation of two edges and at $\epsilon=1.5$ a complete one-cycle. As we increase $\epsilon$ new simplexes form and the one-cycle becomes the boundary of a face in the Vietoris--Rips complex, thus disappearing from the homology.}
\label{HexagonInk}
\end{figure}

We draw different stages of the $\epsilon$ evolution in Fig.~\ref{HexagonInk}. At $\epsilon=0$ we just have the original six points and there is nothing worth noticing: the homology obviously gives six distinct connected components and no further feature. At $\epsilon=1$ two edges form, as shown in Fig.~\ref{HexagonInk}. As a~consequence the first homology of the Vietoris--Rips complex should capture the four connected components of $\mathsf{X}_{\epsilon=1}$. Indeed we see that precisely at~$\epsilon=1$ two homology classes disappear, corresponding to the formation of two edges connecting the respective vertices. Therefore immediately before $\epsilon=1$ there are six persistent homology classes in degree~0 corresponding to six barcodes; immediately after two of these classes have become trivial in homology (two vertices are the boundary of an edge) and only four barcodes are left.

At $\epsilon=1.5$ we see that a 1-cycle form has formed. Any point has an edge connecting it to its two nearest neighbors. By now each individual zeroth homology class but one has died by merging into a single class, and the only barcode left for $H_0$ signals the only connected component of the simplex. On the other hand we see a non trivial barcode for $H_1$, corresponding to the non-trivial cycle in the Vietoris--Rips complex.

When we reach $\epsilon=2.5$ each one of the six original points of the hexagon is now at a distance less then $\epsilon$ from each other point. As a consequence a face of the simplex has formed. The persistent homology class of $H_1$ has already disappeared, corresponding to the fact that the non-trivial one cycle we saw at $\epsilon=1.5$ is now the boundary of the face shown in Fig.~\ref{HexagonInk}.

In this way the persistent homology of the point cloud $\mathsf{X}$ captures its essential topological features: the number of points is determined by the $H_0$ barcodes at small values of the proximity parameters; the existence of a long-lived persistent homology class at $H_1$ is tantamount to the statement that the point cloud $\mathsf{X}$ has ``the shape of a circle''; finally the number of $H_0$ barcodes present at large values of $\epsilon$, in this case just one, is an information on the number of clusters in the distribution of points.

Note that whether we call a class long-lived or short-lived depends somewhat on the context. In this case we could identify the barcode at $H_1$ directly as an interesting topological feature of the hexagonal point cloud $\mathsf{X}$ because we knew of its shape. In general one needs some physical input from the problem at hand, as we will see repeatedly in the following sections.

\section[BPS spectra in theories of class $\protect{\mathcal{S} [A_{K}]}$]{BPS spectra in theories of class $\boldsymbol{\mathcal{S} [A_{K}]}$} \label{classS}

We have discussed techniques to extract topological information out of a distribution of points. In this section we will apply these techniques to BPS states in supersymmetric quantum field theory. We will consider $\mathcal{N}=2$ ${\rm SU}(3)$ super Yang--Mills and construct two point clouds $\mathsf{X}$ out of the distribution of BPS states in two different chambers. Then we will use the tools of persistent homology to compare their features. We will also learn how to extract physical information out of the barcodes.

Most of the recent progress in understanding BPS spectra has been focussed on theories of class $\mathcal{S} [A_{K}]$. These theories can be seen as the low energy limit of the compactification on $\mathbb{R}^{3,1} \times \mathcal{C}$ of the six dimensional $\mathcal{N}=(2,0)$ superconformal theory. This perspective provides an alternative description of many physical quantities in terms of the geometry of the curve $\mathcal{C}$.

In this context the moduli space of quantum vacua is the Coulomb branch $\mathcal{B}$ which para\-met\-ri\-zes tuples $u = (\phi_2 , \dots , \phi_K)$ of meromorphic differentials with prescribed singularities. The Wilsonian effective action of the theory in the Coulomb branch is completely determined by a~family of curves $\Sigma_u$, which are $K$-fold branched coverings of~$\mathcal{C}$. $\Sigma_u$ inherits a natural holomorphic one form $\lambda_u$ which descends from the Liouville one-form on~$\mathcal{C}$.

The lattice of electric and magnetic charges $\Gamma$ is identified with a quotient of $H_1 (\Sigma_u ; \mathbb{Z})$ by the lattice of flavor charges. For any state of charge $\gamma \in \Gamma$, the central charge is given by
\begin{gather*}
\mathcal{Z}_\gamma (u) = \frac{1}{\pi} \int_\gamma \lambda_u ,
\end{gather*}
where we have identified the charge $\gamma$ with an homology class in $H_1 (\Sigma_u ; \mathbb{Z})$. Given the pair $(\Sigma_u , \lambda_u)$ the central charge is in principle known at any point $u \in \mathcal{B}$.

The BPS spectra of theories of class $\mathcal{S} [A_K]$ have striking and unexpected features for $K > 2$. Firstly these theories will generically have higher spin BPS supermultiplets at generic points in their Coulomb branch. Secondly, and more surprisingly, they have \textit{wild spectra}: chambers where the number of BPS states with mass less or equal to a given mass $M$ grows exponentially with~$M$. These features were demonstrated explicitly for~$\mathrm{SU}(3)$ super Yang--Mills with $\mathcal{N}=2$ in~\cite{Galakhov:2013oja} and are believed to hold generically. This phenomenon gives a striking example of the type of understanding of quantum field theory that we gain by studying the wall-crossing behavior of BPS states. With these techniques available to perform controlled computations at various values of the coupling constant, even a relatively simple quantum field theory such as $\mathrm{SU}(3)$ Yang--Mills with $\mathcal{N}=2$ is full of surprises.

We will now apply the formalism outlined in Section~\ref{persistence} to this theory. In particular we will discuss the topological features of the BPS spectrum in a weak coupling chamber and study their behavior as we cross a wall of marginal stability into a wild chamber. We will see that the differences are quite striking, even at the topological level.

In this particular case, the Seiberg--Witten curve $\Sigma$ is a three sheeted covering of $\mathcal{C}$, which has the topology of a cylinder, with six ramification points:
\begin{gather*}
\lambda^3 - \frac{u_2}{z^2} \lambda + \left( \frac{1}{z^2} + \frac{u_3}{z^3} + \frac{1}{z^4} \right) = 0 .
\end{gather*}
One way to compute the BPS spectra is to start from a chamber where the spectrum is known and then apply the wall-crossing formula~\eqref{KSWCF}. The strong coupling region is characterized by small values of the moduli $u_2$ and $u_3$ and has a finite spectrum consisting of six hypermultiplets. The spectrum generator decomposes as
\begin{gather*} 
\mathcal{K}_{\gamma_4} \mathcal{K}_{\gamma_3} \mathcal{K}_{\gamma_2 + \gamma_4} \mathcal{K}_{\gamma_1 + \gamma_3} \mathcal{K}_{\gamma_2} \mathcal{K}_{\gamma_1} .
\end{gather*}
For this theory the rank of the charge lattice $\Gamma$ is four and we have picked a basis $\{ \gamma_i \}$, with $i=1,2,3,4$, so that $\langle \gamma_1 , \gamma_2 \rangle = \langle \gamma_2 , \gamma_3 \rangle = \langle \gamma_3 , \gamma_4 \rangle = -2$ and $\langle \gamma_1 , \gamma_3 \rangle = \langle \gamma_2 , \gamma_4 \rangle = 1$ and all the other parings are vanishing.

\subsection{Wild wall crossing}

To obtain the relevant BPS spectra we will borrow several results from~\cite{Galakhov:2013oja}. The strategy is to start in the strong coupling chamber and follow a path $p$ in the Coulomb branch $\mathcal{B}$. The path is defined by crossing the following walls of marginal stability in this order: $\mathsf{MS} (\gamma_1 + \gamma_3 , \gamma_2 + \gamma_4)$, $\mathsf{MS} (\gamma_1 , \gamma_2)$, $\mathsf{MS} (\gamma_3 , \gamma_4)$ and $\mathsf{MS} (\gamma_1 , \gamma_2 + \gamma_4)$. It ends within a chamber that we will call $\mathscr{C}^{{\rm SU}(3)}_1$. For each of these walls the two charges whose central charge becomes parallel have pairing equal to two: each time the situation is a direct analog of the transition from strong to weak coupling in pure ${\rm SU}(2)$ super Yang--Mills, and each time a similar spectrum is generated consisting of a~vector multiplet shrouded by an infinite cloud of hypermultiplets. In this chamber the spectrum can be written down explicitly. Following~\cite{Galakhov:2013oja} we introduce the notation
\begin{gather*}
\Pi^{n,m} (\gamma_a,\gamma_b) = \left( \prod_{k \nearrow n}^{\infty} \mathcal{K}_{(k+1) \gamma_a + k \gamma_b} \right) \mathcal{K}_{\gamma_a + \gamma_b}^{-2} \left( \prod_{k \searrow m}^{\infty} \mathcal{K}_{k \gamma_a + (k+1) \gamma_b} \right) ,
\end{gather*}
where $k \nearrow n$ means that the product is taken in order so that the value of $k$ increases from left to right starting from $n$, and similarly for $k \searrow m$. The spectrum in the chamber $\mathscr{C}^{{\rm SU}(3)}_1$ was computed explicitly in \cite{Galakhov:2013oja} and is given by
\begin{gather*}
\Pi^{(0,0)} (\gamma_3 , \gamma_4) \Pi^{(0,1)} (\gamma_1 + \gamma_3 , \gamma_2 + \gamma_4) \Pi^{(0,0)} (\gamma_1 , \gamma_2 + \gamma_4) \Pi^{(1,0)} (\gamma_1 , \gamma_2) .
\end{gather*}
This spectrum contains four vectormultiplets and an infinite series of dyonic stable hypermultiplets.

\begin{figure}[t]\centering
\begin{minipage}[b]{0.43\textwidth}\centering
\includegraphics[width=0.9\textwidth]{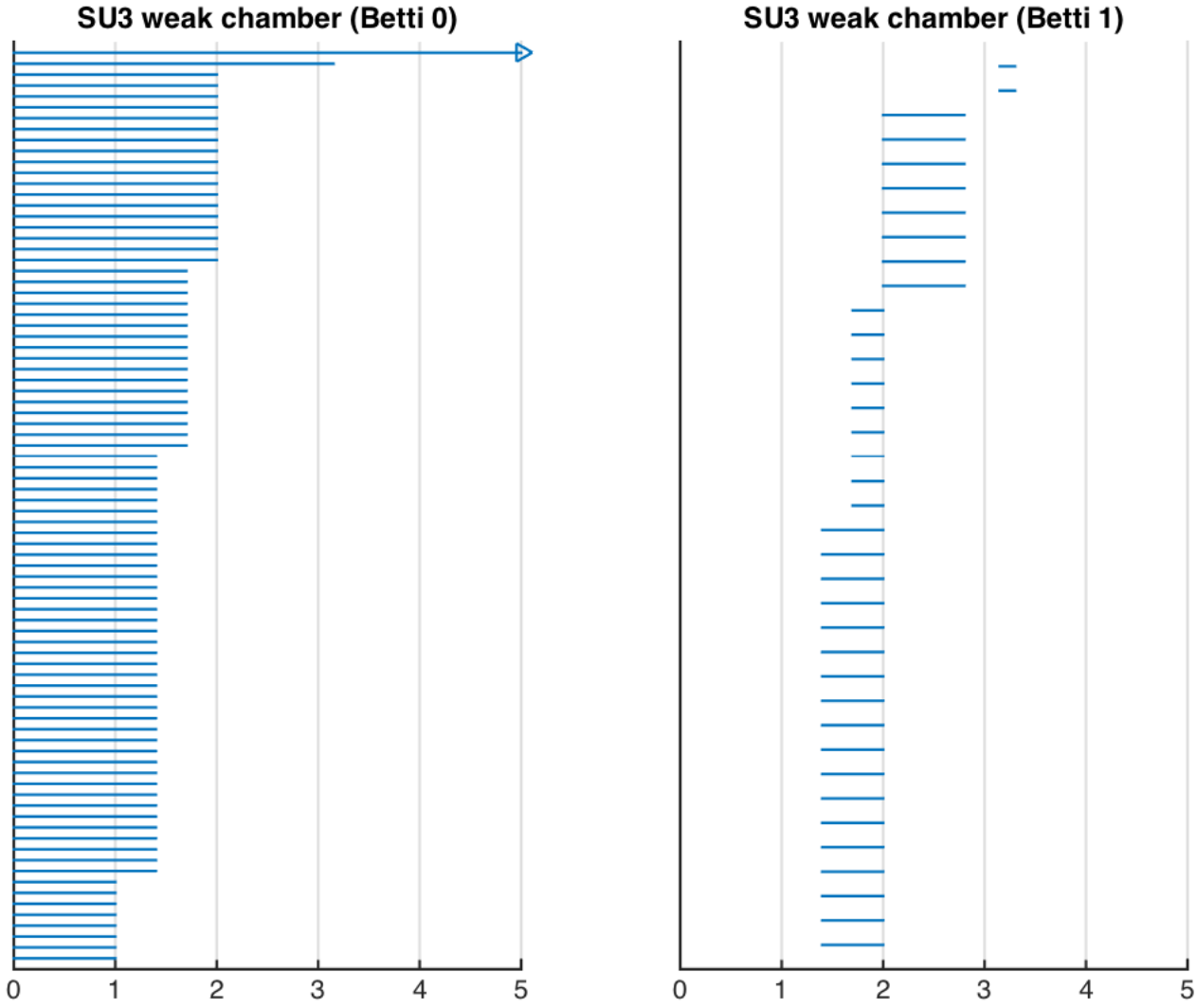}
\end{minipage}
\qquad
\begin{minipage}[b]{0.43\textwidth}\centering
\includegraphics[width=0.9\textwidth]{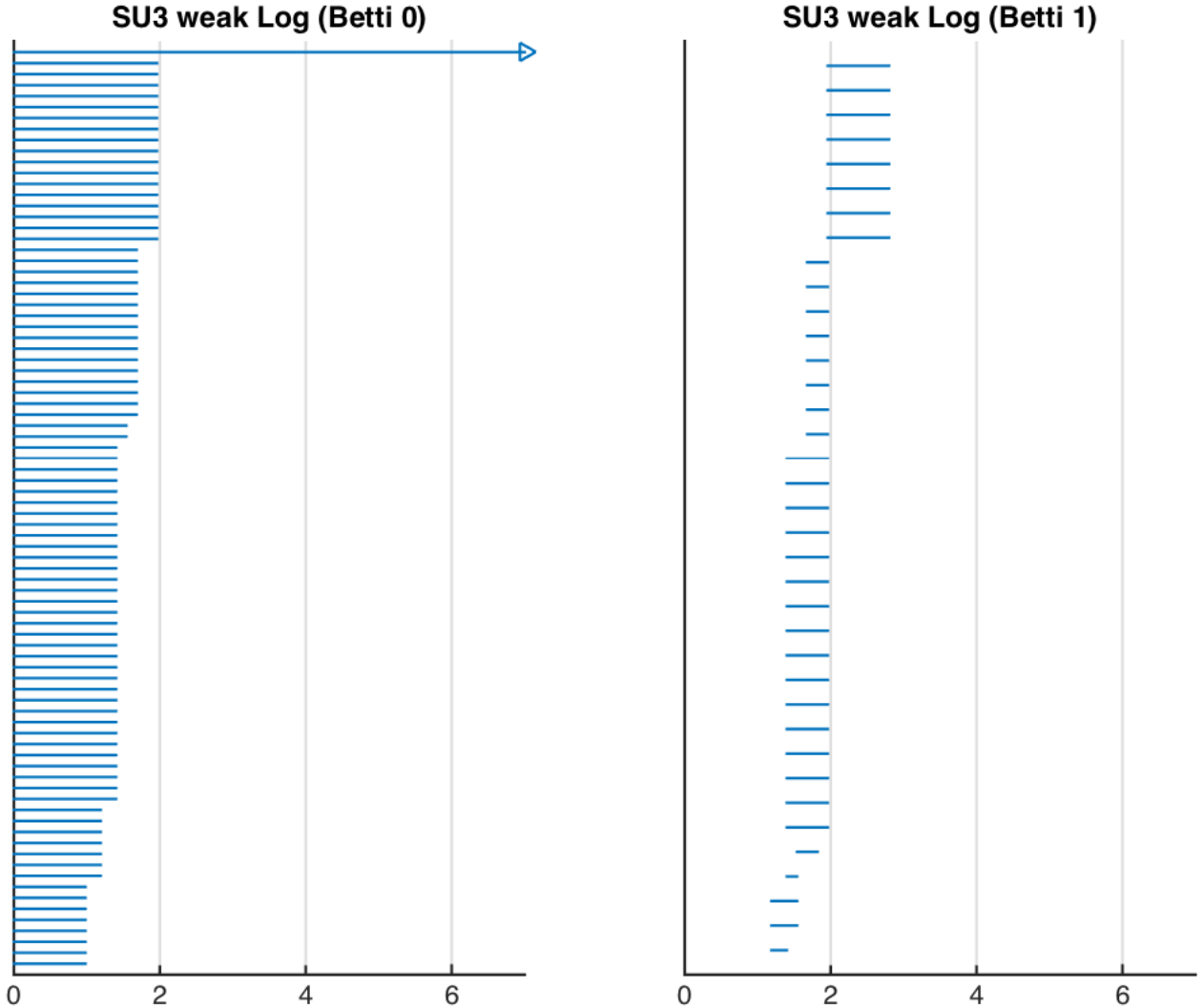}
\end{minipage}
\caption{Barcodes corresponding to the BPS spectrum in the chamber $\mathscr{C}^{{\rm SU}(3)}_1$. \textit{Left.} The point cloud $\mathsf{X}$ is constructed from vectors of the form $\mathsf{x} = (d_1 , d_2 , d_3 , d_4 , \Omega (\gamma ; u))$. \textit{Right.} The logarithm of the degeneracies $\log |\Omega (\gamma ; u)|$ is now used in the point cloud.}
\label{SU3DTweakLogBarcodes}
\label{SU3DTweakBarcodes}
\end{figure}

Now we look at the topological structures of this spectrum, using the Rips--Vietoris complex. That is we consider a point cloud $\mathsf{X}$ in $\mathbb{R}^5$ where each vector has the form $\mathsf{x} = (d_1 , d_2 , d_3 , d_4 , \allowbreak\Omega (\gamma ; u))$, where $\gamma = \sum_i^4 d_i \gamma_i$ is the charge of a stable particle and $u \in \mathscr{C}^{{\rm SU}(3)}_1$. After constructing the filtered Rips--Vietoris complex as explained in Section~\ref{persistence}, we pass to the homology over $\mathbb{Z}_2$ and compute the barcodes. They are shown in Fig.~\ref{SU3DTweakBarcodes}. The point cloud consists of 84 states and the construction of the filtered Vietoris--Rips complex involves a total of 5348 simplices.\footnote{This number is a function of the proximity parameter $\epsilon$, as well as of how many homology groups $H_k$ are included in the computation, in the sense that if one decides to truncate the computation at some~$H_k$, higher dimensional simplices can be neglected. In the following we will by convention only show barcodes up to a value of the proximity parameter $\epsilon$ and of the Betti numbers, which contain interesting topological features, or at least the topological features we want to discuss. We will use the terminology ``number of simplices'' in a similar way, referring only to the simplices used in the homology computations under these conditions.} Non-trivial persistent homology classes are present only for $H_0$ and~$H_1$.

For future reference we show in Fig.~\ref{SU3DTweakLogBarcodes} on the right the barcodes with the logarithm of the (abso\-lute value of the) degeneracies, that is obtained from the point cloud $(d_1 , d_2 , d_3 , d_4 ,\log \,|\allowbreak \Omega (\gamma ; u)|)$. Note that this change does not really modify the BPS point cloud substantially, since all the non vanishing degeneracies are $1$ or $-2$ for hypermultiplets and vector multiplets respectively. On the other hand the relative distance between points are now different and as a consequence the total number of simplices changes, in this case increases to 13708. The distribution of barcodes does not deviate significantly between Fig.~\ref{SU3DTweakLogBarcodes} on the left and on the right, as expected.

\looseness=-1 Now we cross the wall $\mathsf{MS} (2 \gamma_1 + \gamma_2 , \gamma_2 + \gamma_4 )$ to enter into the wild chamber $\mathscr{C}^{{\rm SU}(3)}_2$. Note that $\langle 2 \gamma_1 + \gamma_2 , \gamma_2 + \gamma_4 \rangle = 3$, which implies that in this chamber the BPS quiver has a representative in its mutation class which contains the 3-Kronecker quiver as a subquiver. In crossing the wall a~plethora of higher spin multiplets is generated with wild degeneracies. The spectrum is given by
\begin{gather*} 
\Pi^{(0,0)} (\gamma_3 , \gamma_4) \Pi^{(0,1)} (\gamma_1 + \gamma_3 , \gamma_2 + \gamma_4) \Pi^{(0,1)} (\gamma_1 , \gamma_2 + \gamma_4) \Xi (2 \gamma_1 + \gamma_2 , \gamma_2 + \gamma_4) \Pi^{(2,0)} (\gamma_1 , \gamma_2) ,
\end{gather*}
where $\Xi (2 \gamma_1 + \gamma_2 , \gamma_2 + \gamma_4) $ is a so-called $3$ cohort in \cite{Galakhov:2013oja}.
\begin{figure}[th!]\centering
\begin{minipage}[b]{0.43\textwidth}\centering
\includegraphics[width=0.9\textwidth]{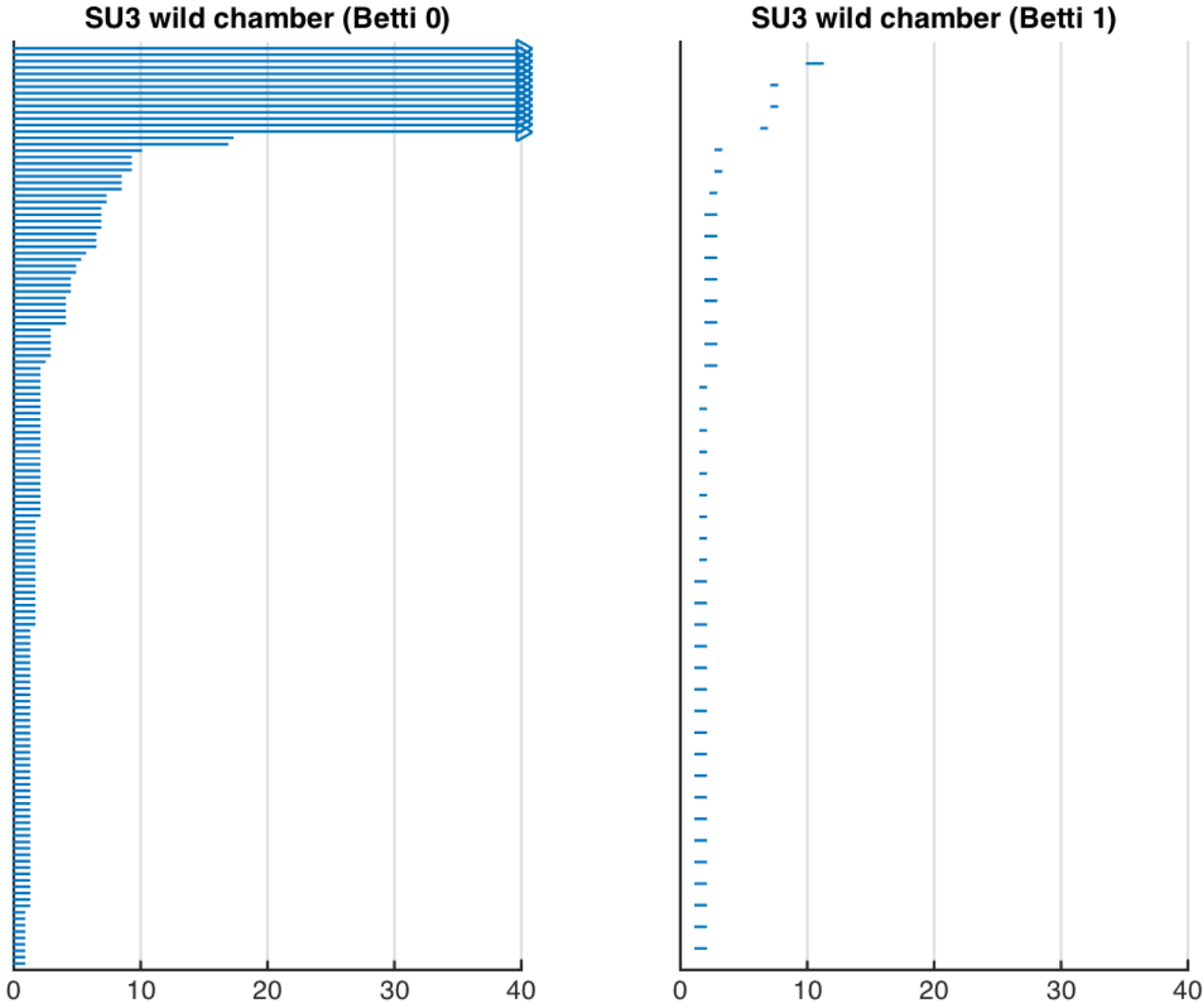}
\end{minipage}
\qquad
\begin{minipage}[b]{0.43\textwidth}\centering
\includegraphics[width=0.9\textwidth]{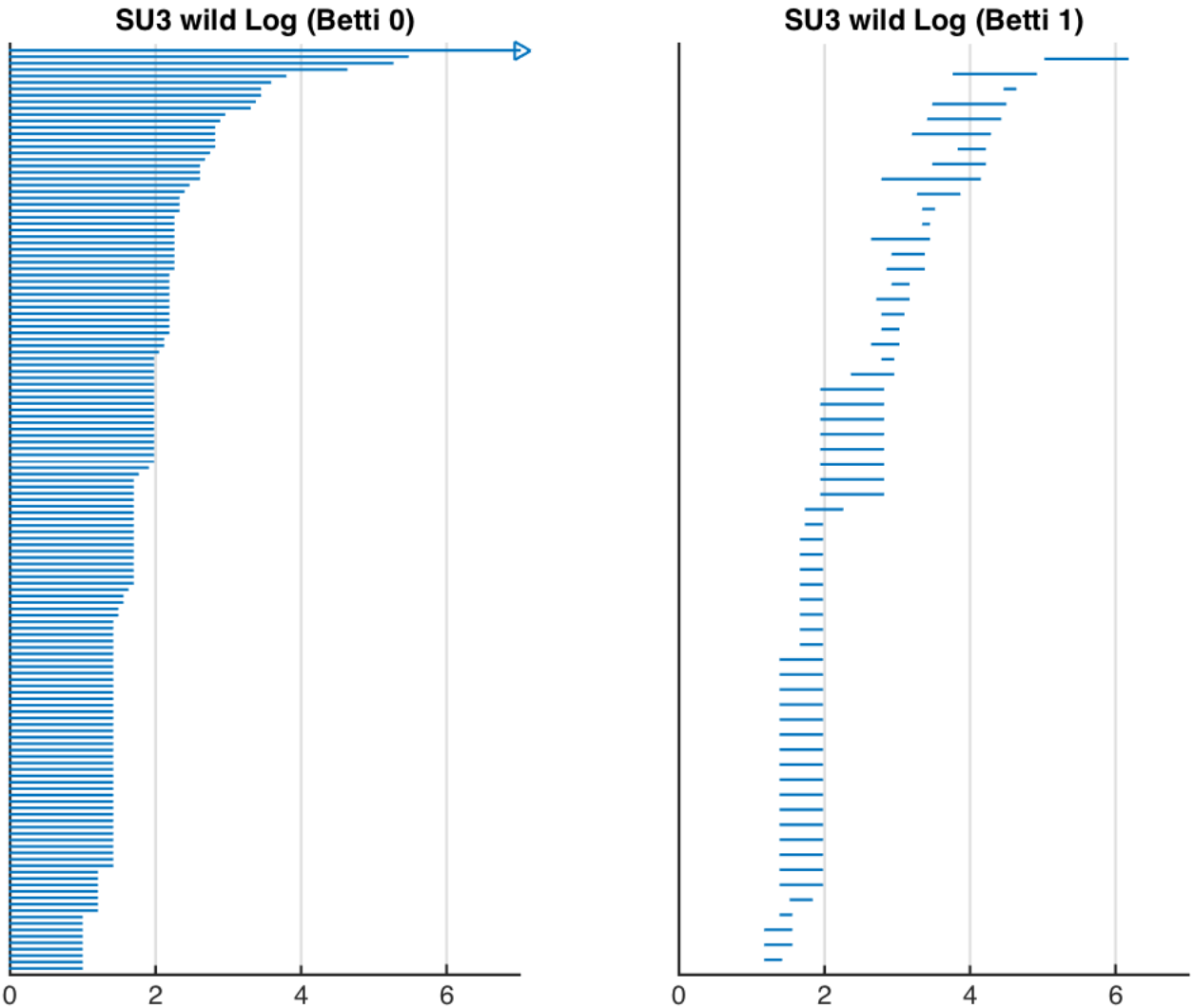}
\end{minipage}
\caption{Barcodes for the BPS spectrum in the wild chamber $\mathscr{C}^{{\rm SU}(3)}_2$. \textit{Left.} The point cloud is $\mathsf{X}$ constructed out of vectors of the form $(d_1 , d_2 , d_3 , d_4 , \Omega (\gamma ; u))$. \textit{Right.} Vectors in the point cloud are of the type $(d_1 , d_2 , d_3 , d_4 ,\log |\Omega (\gamma ; u)|)$.}
\label{SU3DTwildBarcodes}
\label{SU3DTwildLogBarcodes}
\end{figure}
The $3$ cohort does not have a closed form expression and its BPS degeneracies have been computed in \cite{Galakhov:2013oja} up to a total charge $\gamma$ of $15$. The explicit results of the computations can be found in \cite[Appendix~A.2]{Galakhov:2013oja}. Using these results we construct a sample of 144 points corresponding to as many BPS states. The Vietoris--Rips complex runs over a total of 210156 simplices and its persistent homology classes (again over~$\mathbb{Z}_2$) are shown in the left in Fig.~\ref{SU3DTwildBarcodes}. The exponential growth of the degeneracies shows up as the presence of many very long-lived barcodes in degree~0. This is easy to understand; due to the exponential growth of the point cloud the homology of the Vietoris--Rips complex sees many points as individual connected components for very large values of the proximity parameter $\epsilon$.

Since the degeneracies are exponentially growing it is useful to consider also the point cloud obtained by taking the logarithm of the modulus of the degeneracies. This is merely a trick to simplify the computations of the simplices. This significantly reduces the lifespan of long-lived homology classes, also altering the number of simplices. On the other hand the features of the barcode distribution are easier to visualize, and shown in Fig.~\ref{SU3DTwildBarcodes} on the right. The total number of simplices for which there are interesting topological features is now reduced to 27766.

Across the wall we see distinctively a transition at the level of the topology of the point cloud. This is clear for example in Fig.~\ref{SU3DTwildBarcodes} (\textit{left}) where the barcodes for $H_0$ are very long-lived. If we compare with Fig.~\ref{SU3DTweakBarcodes} (\textit{left}) we see that the wild chamber has a much larger number of connected component at very large scales $\epsilon$. This is indeed a consequence of the exponential growth in the number of states: since the degeneracies $\Omega (\gamma ; u)$ grow exponentially the points are further and further apart and therefore at the same scale $\epsilon$ at which the points in Fig.~\ref{SU3DTweakBarcodes} are already grouped in a single cluster, the degeneracies in Fig.~\ref{SU3DTwildBarcodes} still look like many connected components. In other words at a certain threshold the values of the degeneracies begin to grow too far apart for the Vietoris--Rips complex to form edges and the homology sees each point as an individual connected component.

To have a more meaningful comparison, we turn to the logarithm of the degeneracies, so to damp the exponential growth. We will see in the next sections that there are more refined methods, but for the moment this will be enough. We now compare the results in Figs.~\ref{SU3DTweakLogBarcodes} and~\ref{SU3DTwildLogBarcodes} (\textit{right}). The transition between the two chambers is still very clear: the barcodes in the chamber $\mathscr{C}^{{\rm SU}(3)}_1$ are very regular and the dependence on the scale $\epsilon$ rather mild. On the other hand we see in Fig.~\ref{SU3DTwildLogBarcodes} how persistent homology captures the wilderness of the BPS spectrum. The homological features are very irregular, especially for $H_1$ and non-trivial 1-cycles persist at every $\epsilon$-scale. This features terminate in Fig.~\ref{SU3DTwildLogBarcodes} at a certain value of $\epsilon$ only because we are using a finite sample, with roughly the same number of points as in the chamber $\mathscr{C}^{{\rm SU}(3)}_1$, but would continue indefinitely were we to increase the number of degeneracies computed via the wall-crossing formula in \cite{Galakhov:2013oja}.

This is an example of the kind of information we can get on the BPS spectra using the forma\-lism of persistent homology. In this section we have just compared the topological features of the BPS spectrum in two different chambers. Topology gives a clear meaning to the statement that these chambers are qualitatively different. The topological features of the two distributions are captured by the basic topological invariants of the $\mathbb{N}$-persistence modules, namely the barcodes. It is remarkable that upon crossing the wall of marginal stability, the transition between the two chambers is very sharp even at the topological level.

We will now discuss a similar problem in a string theory compactification, namely the BPS spectra in various chambers of the conifold. But before that we need to refine our techniques in the next section.

\section{Witness and lazy witness} \label{witness}

We have shown how to use the Vietoris--Rips complex $\mathsf{VR}_\epsilon$ to extract topological information from the BPS spectra in the form of barcodes. The computation of the barcodes is done exactly by evaluating the homology of the simplicial complex for any value of the proximity parameter $\epsilon$. However in many cases the point cloud $\mathsf{X}$ is too large for a direct computation. In this cases certain approximation schemes are available, which we will now describe. These approximation schemes are suited to deal efficiently with large sets of data. The main idea is to use certain criteria to select only a certain subset~$\mathsf{L}$ of the point cloud $\mathsf{X}$, which is then used to form a~simplicial complex which approximates the Vietoris--Rips complex~\cite{C09}.

The operator which performs such a choice is a \textit{landmark selector}. The most used landmark selectors are:
\begin{enumerate}\itemsep=0pt
\item \textit{Random selector.} The most straightforward way to choose a subset $\mathsf{L} \subset \mathsf{X}$ is by picking a number of points at random. This procedure is quite useful, although in practice it is better to choose various random subsets $\mathsf{L}$ and perform homological computations for each one separately. Indeed it is quite possible that a random selection would miss essential features of a point cloud $\mathsf{X}$. The limitations of random selection are well known from Monte Carlo algorithms and will not be repeated here.
\item \textit{Minmax selector.} This operator selects a collection of points $\mathsf{L}$ which, in a very specific sense, is as spread out as possible. The minmax algorithm works inductively by maximizing the distance of a point from a previously chosen set. More in detail, the algorithm starts from a randomly chosen point $\mathsf{x}_0$. Let $d ( \, , \, ) \colon \mathsf{X}^2 \longrightarrow \mathbb{R}$ be the standard distance function between points (which is inherited from the ambient metric of $\mathbb{R}^N$ where the point cloud is embedded). The choice of the remaining points in the set $\mathsf{L}$ proceeds by induction. Denote by $\mathsf{L}_i$ the minmax selection of $i$ points in $\mathsf{X}$. Then $\mathsf{L}_{i+1}$ consist of the same set to which a single point $\mathsf{z}$ is added, in such a way that the function $d (\mathsf{x}_j , \mathsf{z})$ is maximized for each $\mathsf{x}_j \in \mathsf{L}_i$. Because of this inductive definition, the landmark selected set $\mathsf{L}$ will consists of points which are spread apart from each other as much as possible. Therefore in general we expect this selector to capture features of a point cloud $\mathsf{X}$ better than a random selector. On the other hand one has to keep in mind that there are cases where this expectation will fail: for example since the minmax selector will generically pick out outlier points, a~random selector might work better with very dense sets.
\end{enumerate}

Having chosen a landmark set $\mathsf{L}$, we can define a version of the Vietoris--Rips complex which is based on it. We shall use two such simplicial complexes, the \textit{witness complex} $\mathsf{W} (\mathsf{X},\mathsf{L},\epsilon)$ and the \textit{lazy witness complex} $\mathsf{LW}_{\nu} (\mathsf{X} , \mathsf{L} , \epsilon)$. In both cases the vertex set is given by $\mathsf{L}$; what changes is the definition of the simplices. These are defined as follows:
\begin{enumerate}\itemsep=0pt
\item \textit{Witness complex $\mathsf{W} (\mathsf{X},\mathsf{L},\epsilon)$.} Let $\mathsf{x}$ be a point in $\mathsf{X}$. Denote by $d_k (\mathsf{x})$, for $k>0$, the distance between $\mathsf{x}$ and its $(k+1)$-th closest landmark point. Now we declare that a collection of vertices $\mathsf{l}_i \in \mathsf{L}$ for $i=0,\dots , k$ form the simplex $[\mathsf{l}_0 , \dots , \mathsf{l}_k]$ if all of its faces are in $\mathsf{W} (\mathsf{X},\mathsf{L},\epsilon)$ and there exists a witness point $\mathsf{x} \in \mathsf{X}$ so that the following condition holds
\begin{gather*}
\max \{ d (\mathsf{l}_0 , \mathsf{x}) , d (\mathsf{l}_1 , \mathsf{x}) , \dots , d (\mathsf{l}_k , \mathsf{x}) \} \le d_k (\mathsf{x}) + \epsilon .
\end{gather*}
\item \textit{Lazy Witness complex $\mathsf{LW}_{\nu} (\mathsf{X} , \mathsf{L} , \epsilon)$.} Let $\nu \in \mathbb{N}$, and let $d_\nu (\mathsf{x})$ be the distance between $\mathsf{x}$ and the $\nu$-th closest landmark point (with $d_\nu (\mathsf{x}) \equiv 0$ if $\nu = 0$). Then for $\mathsf{l}_1$ and $\mathsf{l}_2$ points of~$\mathsf{L}$, $[\mathsf{l}_1 , \mathsf{l_2}]$ is an edge in $\mathsf{LW}_{\nu} (\mathsf{X} , \mathsf{L} , \epsilon)$ if there exists a witness point $\mathsf{x} \in \mathsf{X}$ so that the following condition holds
\begin{gather*}
\max \{ d (\mathsf{l}_1 , \mathsf{x}) , d (\mathsf{l}_2 , \mathsf{x}) \} \le d_\nu (\mathsf{x}) + \epsilon .
\end{gather*}
A higher dimensional simplex defines an element of $\mathsf{LW}_{\nu} (\mathsf{X} , \mathsf{L} , \epsilon)$ if and only if all of its edges are in $\mathsf{LW}_{\nu} (\mathsf{X} , \mathsf{L} , \epsilon)$
\end{enumerate}
These definitions are those implemented in \textsc{javaplex}. It is easy to see that if $\epsilon_1 < \epsilon_2$, then we have $\mathsf{W} (\mathsf{X} , \mathsf{L} , \epsilon_1) \subset \mathsf{W} (\mathsf{X} , \mathsf{L} , \epsilon_2)$ and $\mathsf{LW}_{\nu} (\mathsf{X} , \mathsf{L} , \epsilon_1) \subset \mathsf{LW}_{\nu} (\mathsf{X} , \mathsf{L} , \epsilon_2)$.

We can now use the same arguments of Section~\ref{persistence} to argue that by letting $\epsilon$ vary we induce filtrations of witness and lazy witness complexes and that taking the $i$-th homology of any such sequence as a function of $\epsilon$ defines a $\mathbb{N}$-persistence module. Therefore we can easily define persistent homology groups and barcodes and use the complexes $\mathsf{W} (\mathsf{X} , \mathsf{L} , \epsilon)$ and $\mathsf{LW}_{\nu} (\mathsf{X} , \mathsf{L} , \epsilon)$ to study persistent topological features of distributions of BPS states. Unless specified otherwise, when using the lazy witness complex we will always set $\nu=1$ and omit it from the notation.

\section{BPS invariants on the conifold} \label{conifold}

In our second example we consider BPS states in a local string theory compactification, the resolved conifold. The geometry is given by the total space of the fibration $X=\mathcal{O} (-1) \oplus \mathcal{O} (-1) \longrightarrow \mathbb{P}^1$ and has one K\"ahler modulus $t$, the complexified size of the base $\mathbb{P}^1$. We are interested in a particular class of BPS states which are given by bound states of a gas of light D0 branes and D2 branes wrapping the $\mathbb{P}^1$, with a single D6 brane wrapping the full non compact total space. Such invariants have been computed in quite some detail at any point of the moduli space \cite{Andriyash:2010yf,Jafferis:2008uf,nakanagao,szendroi}.

Since the geometry is non-compact, to properly define the charges it is necessary to embed it into a compact Calabi--Yau and then take a local limit. This is done by considering a compact Calabi--Yau $\tilde{X}$ and taking the limit where the K\"ahler moduli of all the homology classes become large, with the sole exception of the K\"ahler class of a rigid rational curve.

Since we are taking a local limit, it is enough to consider the compact Calabi--Yau $\tilde{X}$ in the large radius approximation. BPS states are labeled by charge vectors $\gamma \in \Gamma$, where
\begin{gather*}
\gamma \in \Gamma = \Gamma^{\rm m} \oplus \Gamma^{\rm e} = \big( H^0 \big(\tilde{X} ,
 \mathbb{Z}\big) \oplus H^2 \big(\tilde{X} , \mathbb{Z}\big) \big) \oplus \big( H^4 \big(\tilde{X} , \mathbb{Z}\big)
 \oplus H^6 \big(\tilde{X} , \mathbb{Z}\big) \big) .
\end{gather*}
By Poincar\'e duality D-branes wrapping $p$-cycles correspond to charges as
\begin{gather*}
{\rm D}p \longleftrightarrow H^{6-p} \big(\tilde{X} , \mathbb{Z}\big) = H_{p} \big(\tilde{X} , \mathbb{Z}\big) , \qquad p = 0,2,4,6 .
\end{gather*}
The Dirac--Schwinger--Zwanziger pairing has the geometric definition
\begin{gather*}
 \langle\gamma , \gamma' \rangle_{\Gamma} = \int_{\tilde{X}} \gamma\wedge (-1)^{{\rm deg}/2} \gamma' .
\end{gather*}
In the large radius limit the central charge of a BPS state with charge $\gamma$ is given by the integral
\begin{gather*}
\mathcal{Z}_{\tilde{X}} \big( \gamma;\tilde{t} \big) = -\int_{\tilde{X}} \gamma \wedge \e^{-\tilde{t}},
\end{gather*}
where $\tilde{t} = B+ \ii J$ is the complexified K\"ahler form given by the background supergravity two-form $B$-field and the K\"ahler $(1,1)$-form $J$ of~$\tilde{X}$.

Now we take the local limit, following \cite{Jafferis:2008uf}. We parametrize the K\"ahler form as
\begin{gather*}
\tilde{t} = t_{\mathbb{P}^1} + L \e^{\ii \varphi} t_{\perp},
\end{gather*}
where $t_{\mathbb{P}^1}$ and $t_{\perp}$ are classes in $H^2 \big(\tilde{X} ; \mathbb{C}\big)$ such that
\begin{gather*}
\int_{\mathbb{P}^1} t_{\mathbb{P}^1} = z , \qquad \int_{\mathbb{P}^1} t_{\perp} = 0 .
\end{gather*}
Now we take the local limit by sending $L \longrightarrow + \infty$.

In this limit the relevant BPS configurations are multi-centered solutions with core charge $\gamma_c = (1,0,0,0)$ and halo charge $\gamma_h = (0,0,- \beta , n)$. Conventionally we write the latter as $\gamma_h = (0,0, -m , n)$, where $m$ denotes the number of times the class $\beta$ wraps\footnote{We use a different notation from \cite{Jafferis:2008uf}: we call $\beta$ the homology class of a curve, while in \cite{Jafferis:2008uf} $\beta$ is the 4-form Poincar\'e dual to the $\mathbb{P}^1$.} the $\mathbb{P}^1$, $\beta = m [\mathbb{P}^1]$. Walls of marginal stability can be computed explicitly as a function of $m$ and $n$. In the local limit the central charges retain some dependence on the parameter $\varphi$. Therefore the walls of marginal stability can be parametrized on the space $(z , \varphi)$, where the K\"ahler moduli space is extended by the parameter $\varphi$ due to the local limit. Physically $\varphi$ represent the density of the components of the B-field normal to the exceptional locus.

The walls of marginal stability are \cite{Andriyash:2010yf,Jafferis:2008uf,nakanagao}{\samepage
\begin{gather*}
\mathsf{MS}_n^m = \left\{ (z , \varphi) \colon \varphi = \frac13 \arg \left(z + \frac{n}{m}\right) + \frac{\pi}{3} \right\} ,\\
\mathsf{MS}_n^{-m} = \left\{ (z , \varphi) \colon \varphi = \frac13 \arg \left(z - \frac{n}{m}\right) \right\} ,\\
\mathsf{MS}_{-n}^{-m} = \left\{ (z , \varphi) \colon \varphi = \frac13 \arg \left(z + \frac{n}{m}\right) \right\} ,
\end{gather*}
with $n \ge 0$ and $m \ge 0$. We will denote a chamber between two walls as $\mathscr{C} \big[\mathsf{MS}_{n_1}^{m_1} , \mathsf{MS}_{n_2}^{m_2}\big]$.}

Now we will study the topological features of the BPS spectra in various chambers and compare them. From a physical perspective this is similar to what we have done in Section~\ref{classS}, except that the BPS spectra count states in string theory and not in quantum field theory (albeit we have rendered gravity non-dynamical by taking a local limit).

This problem is also interesting from a mathematical perspective. Indeed a proper treatment of BPS particles in string theory requires to work at the level of a derived category, in this case $\mathcal{D} (\mathsf{coh} (X))$. However for practical computations one typically chooses to work in a simpler abelian category. Indeed choosing a stability condition on $\mathcal{D} (\mathsf{coh} (X))$ corresponds to the choice of an abelian category $\mathsf{A}$ and a stability condition on $\mathsf{A}$. In this case we will discuss, for example, BPS states computed from the moduli space of ideal sheaves at large radius, or from the moduli space of cyclic modules of a quiver in a certain chamber. These objects are mathematically quite different, yet describe the same physical setting up to wall-crossings. We will provide a~new viewpoint using persistent topology.

\subsection{Chamber structure}

We begin by considering the chambers of the form $\mathscr{C} \big[\mathsf{MS}_{n+1}^{1} , \mathsf{MS}_{n}^{1}\big]$. These chambers are directly connected to the core region, where only the D6 brane exists as a stable state, situated near the wall with $\varphi = \frac13 \arg z + \frac{\pi}{3}$. The partition function in each chamber can be obtained by applying the KSWCF \eqref{KSWCF}. Because of the particularly simple charge configuration of the bound states, the wall-crossing formula reduces to the semi-primitive version \cite{Denef:2007vg}, which amounts in multiplying the partition function by the partition function of a halo of D0 and D2 branes. After crossing the appropriate walls of marginal stability, the core D6 brane will be bound to a halo of charge $\gamma_h = (0,0, -m , n)$. In decreasing $\varphi$ from the core region, the first walls encountered are of the form $\mathsf{MS}_{n}^{1}$.

We introduce the standard notation $q = \e^{\lambda}$ where $\lambda$ is the topological string coupling, and $Q = \e^{-z}$. The resulting partition function in the chambers $\mathscr{C} \big[\mathsf{MS}_{n+1}^{1} , \mathsf{MS}_{n}^{1}\big]$ is
\begin{gather} \label{Zcorechambers}
Z_n (q, Q) = \prod_{k=1}^n \big( 1 - (-q)^k Q \big)^k .
\end{gather}
In these chambers the physical BPS spectrum is particularly simple and can be obtained simply by expanding~\eqref{Zcorechambers} at any fixed $n \ge 0$.

At large radius the effective dynamics of the bound states is captured by a six dimensional topological field theory, which arises as the topological twist of the $\mathcal{N}=2$ supersymmetric abelian theory obtained as low energy approximation to the DBI action on the D6 brane worldvolume \cite{Cirafici:2008sn, Iqbal:2003ds}. To reach this chamber one has to cross all the chambers of the form $\mathscr{C} \big[\mathsf{MS}_{n+1}^{-1} , \mathsf{MS}_{n}^{-1}\big]$ by taking smaller values for $\varphi$, for $n \longrightarrow \infty$. A notable feature of the large radius chamber, as well as other chambers in local threefolds, is that walls of marginal stability accumulate towards it, leaving a certain ambiguity to the statement ``crossing all the walls''. The behavior is similar to what happens in quantum field theory with BPS rays, which accumulate to higher spin fields. Luckily in this situation the limit can be taken analytically. The computation of the BPS spectrum can also be done directly by other means, confirming the wall crossing prediction. The gauge theory partition function
\begin{gather*} 
Z^X_\mathrm{gauge} (q , Q ) = \sum_{k \beta} q^k Q^{\beta} {\tt DT}_{k,\beta}(X)
\end{gather*}
can be evaluated explicitly by equivariant localization. In our notation $Q^{\beta} = \e^{- m z}$. Such a~partition function receives contributions from point like instantons located at the toric fixed points, say at the north and south pole of the base $\mathbb{P}^1$, as well as extended instantons fibered over the $\mathbb{P}^1$. In the case of the conifold the result can be written in closed form as~\cite{Iqbal:2003ds}
\begin{gather} \label{conifoldLR}
Z^X_{\mathrm{gauge}} (q , Q) = M (-q)^2 \prod_{k=1}^{\infty} \big(1 - (-q)^k Q\big)^k .
\end{gather}
We have used the partition function \eqref{conifoldLR} to compute a large number of enumerative invariants and used them to construct a point cloud.

In the noncommutative crepant resolution (NCCR) chamber, geometrical concepts are replaced by algebraic ones. We reach this chamber by taking smaller and smaller values of $\varphi$, crossing all walls of the form $\mathsf{MS}_n^{-1}$ until we reach the region $\frac13 \arg z < \varphi < \frac13 \arg (z-1)$. The conifold point lies at the boundary of this region. Indeed in a noncommutative crepant resolution ordinary geometrical notions don't hold anymore and geometry is replaced by a certain algebra. Such a phenomenon is generic in toric threefolds. This algebra is the Jacobian algebra of an appropriate quiver \cite{szendroi}. The center of this algebra is the equation of the singular limit of the threefold; in this sense the Jacobian algebra is a ``resolution'' of the singularity. The physics of such resolutions can be understood in terms of a certain deformation of a cohomological TQFT, called ``stacky gauge theory'', which has been discussed at length in \cite{Cirafici:2008sn,Cirafici:2010bd,Cirafici:2012qc}. We refer the interested reader to the review \cite{framedquivers} for more details.
The result can be summarized in the partition function \cite{szendroi}
\begin{gather} \label{ZNCCR}
Z_{\rm NCCR} (q , Q) = M (-q)^2 \prod_{k=1}^{\infty} \big( 1 - (-q)^k Q \big)^k \big( 1 - (-q)^k Q^{-1} \big)^k .
\end{gather}
Again we can compute the BPS degeneracies by a direct expansion and we will momentarily use persistent homology to study their distribution.

Now we move to the study of the chambers $\mathscr{C} \big[\mathsf{MS}_{-n}^{-1} , \mathsf{MS}_{-n-1}^{-1}\big]$. In those chambers, a different qualitative phenomenon happens, namely the crossing of certain conjugation walls $\mathscr{S}$ after which the core charge of the bound states, identified with the D6 brane at large radius, changes \cite{Andriyash:2010yf}. These conjugation walls are associated with particles becoming massless. As a consequence there is a monodromy around the locus where the state becomes massless in the moduli space. Upon crossing such walls the core charge changes according to the monodromy, while an extra halo of particles appears to ensure charge conservation.

The partition function can be computed using the wall-crossing formula, giving~\cite{Andriyash:2010yf}
\begin{gather} \label{Zconjchambers}
\mathcal{Z}_n (q , Q) = \left( \prod_{k=1}^n \big(q^k Q\big)^k \right) M (-q)^2 \prod_{k=1}^{\infty} \big( 1 - (-q)^k Q^{-1} \big)^k \prod_{k=n+1}^{\infty} \big( 1 - (-q)^k Q \big)^k .
\end{gather}
Here the presence of the factor $\Big( \prod\limits_{k=1}^n \big(q^k Q\big)^k \Big)$ is a consequence of the monodromy
\begin{gather*}
\Gamma_0 \longrightarrow \Gamma_0 - \sum_{l=1}^{n} \langle \Gamma_0 , \gamma_{1,l} \rangle \gamma_{1,l} ,
\end{gather*}
which a core charge $\Gamma_0$ undergoes after crossing the conjugation walls $\mathscr{S} (\gamma_{1,l} , \Gamma_0)$ for $l=1,\dots,n$. Each time a conjugation wall $\mathscr{S} (\gamma_{1,l} , \Gamma_0)$ is crossed, the core charge changes accordingly to the above monodromy, and an halo appears, made of a filled Fermi sea of particles with charge $\gamma_{1,l}$. Therefore the nature of the bound states changes in a qualitative fashion, and we are interested in seeing how does this change affects the distribution of BPS invariants.

\subsection{Topological analysis}

As we have seen the structure of BPS states on the conifold has very distinctive features in each chamber. We want now to understand at the qualitative level offered by the topological analysis how the nature of the BPS bound states differs from one chamber to another. We use the partition functions we have described above to generate a rather large number of BPS invariants, as the coefficients of the expansion
\begin{gather*}
Z_{\mathscr{C}} (q,Q) = \sum_{k , m \in \mathbb{Z}} \Omega \left( \gamma = (1 , 0 , -m , k) ; \mathscr{C} = (z ,\varphi) \right) q^k Q^m .
\end{gather*}
In our case this partition function is given by \eqref{Zcorechambers}, \eqref{conifoldLR}, \eqref{ZNCCR} and \eqref{Zconjchambers} respectively. We have shown explicitly the chamber dependence as well as the charge vector of a state.

\begin{figure}[hbtp]\centering
\begin{minipage}[b]{0.45\textwidth}\centering
\includegraphics[width=1\textwidth]{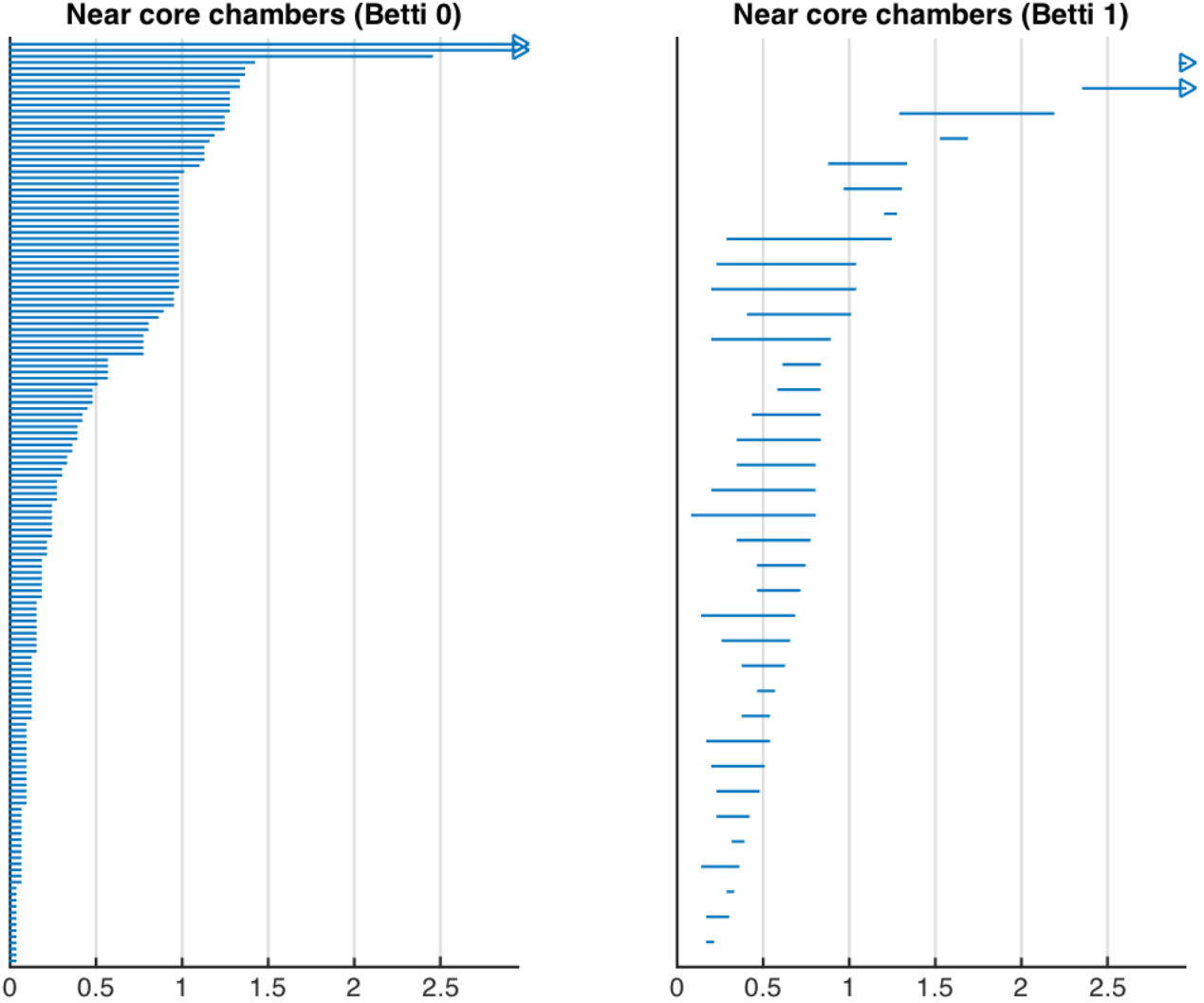}
\end{minipage}
\qquad
\begin{minipage}[b]{0.45\textwidth}
\centering
\includegraphics[width=1\textwidth]{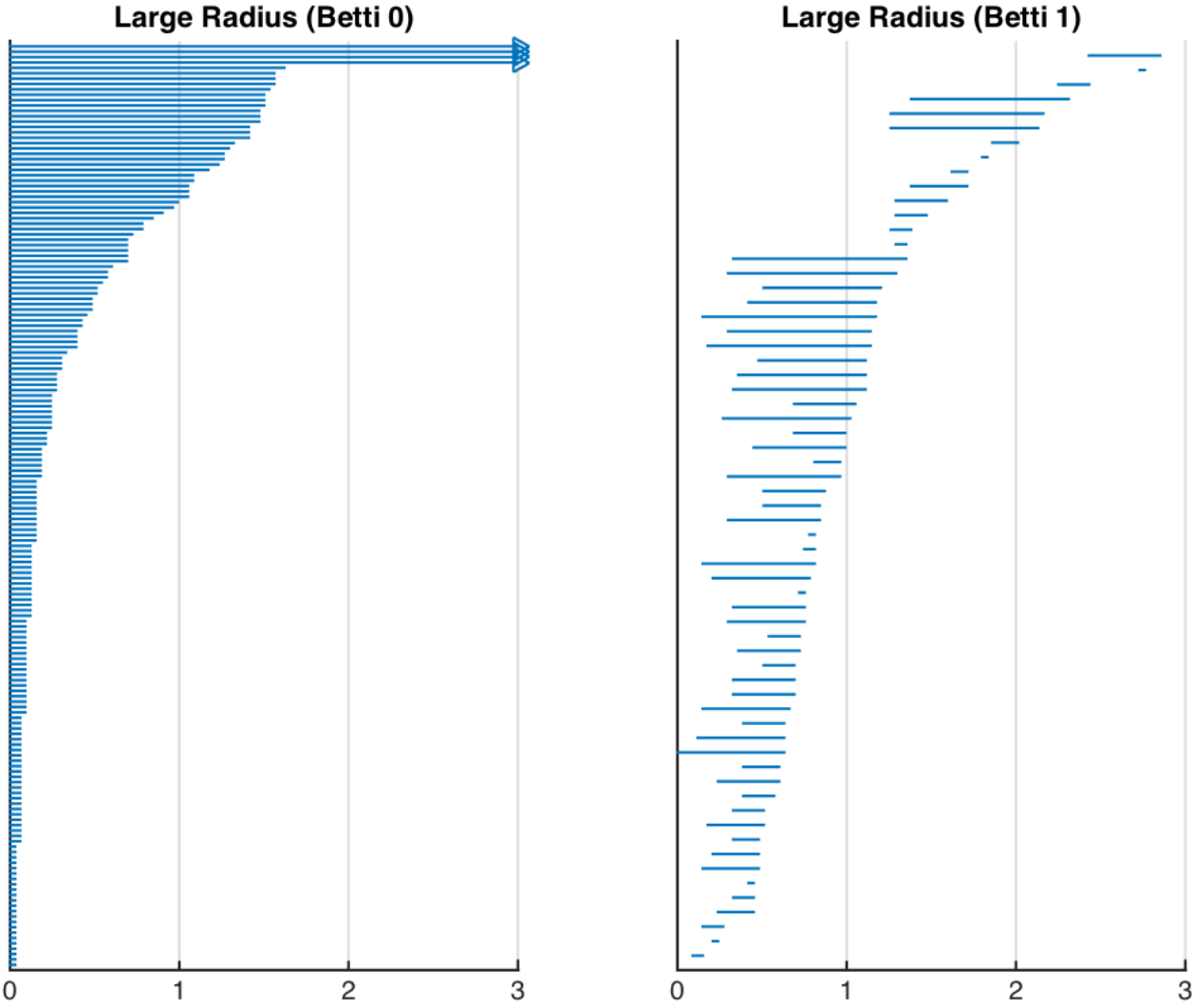}
\end{minipage}
\caption{Spectra of BPS states for the conifold. The barcodes have been obtained using the lazy witness complex $\mathsf{LW} (\mathsf{X} , \mathsf{L} , \epsilon)$ on a point cloud of the form $\mathsf{x} = ( k , m , \log |\Omega(\gamma;\mathscr{C})|) $. \textit{Left.} Chambers of the form $\mathscr{C} [\mathsf{MS}_{n+1}^{1} , \mathsf{MS}_{n}^{1}]$ for $n=90$. The point cloud is made out of 1093 BPS invariants and the filtered homology computation involves 3457 simplices. \textit{Right}. The large radius chamber. The point cloud is constructed with 1183 invariants and the filtered homology computation involves 3632 simplices.}\label{Conifold1}
\end{figure}

Since for Calabi--Yau manifolds the degeneracies of BPS states grow rather fast, we take the log of $\Omega (\gamma ; \mathscr{C})$. Therefore in all the cases our point cloud $\mathsf{X}$ contains vectors of the form $\mathsf{x} = ( k , m , \log |\Omega (\gamma ; \mathscr{C})|)$, where the chamber $\mathscr{C}$ is specified by the value of the moduli $(z,\varphi)$.

\begin{figure}[htbp]\centering
\begin{minipage}[b]{0.45\textwidth}\centering
\includegraphics[width=1\textwidth]{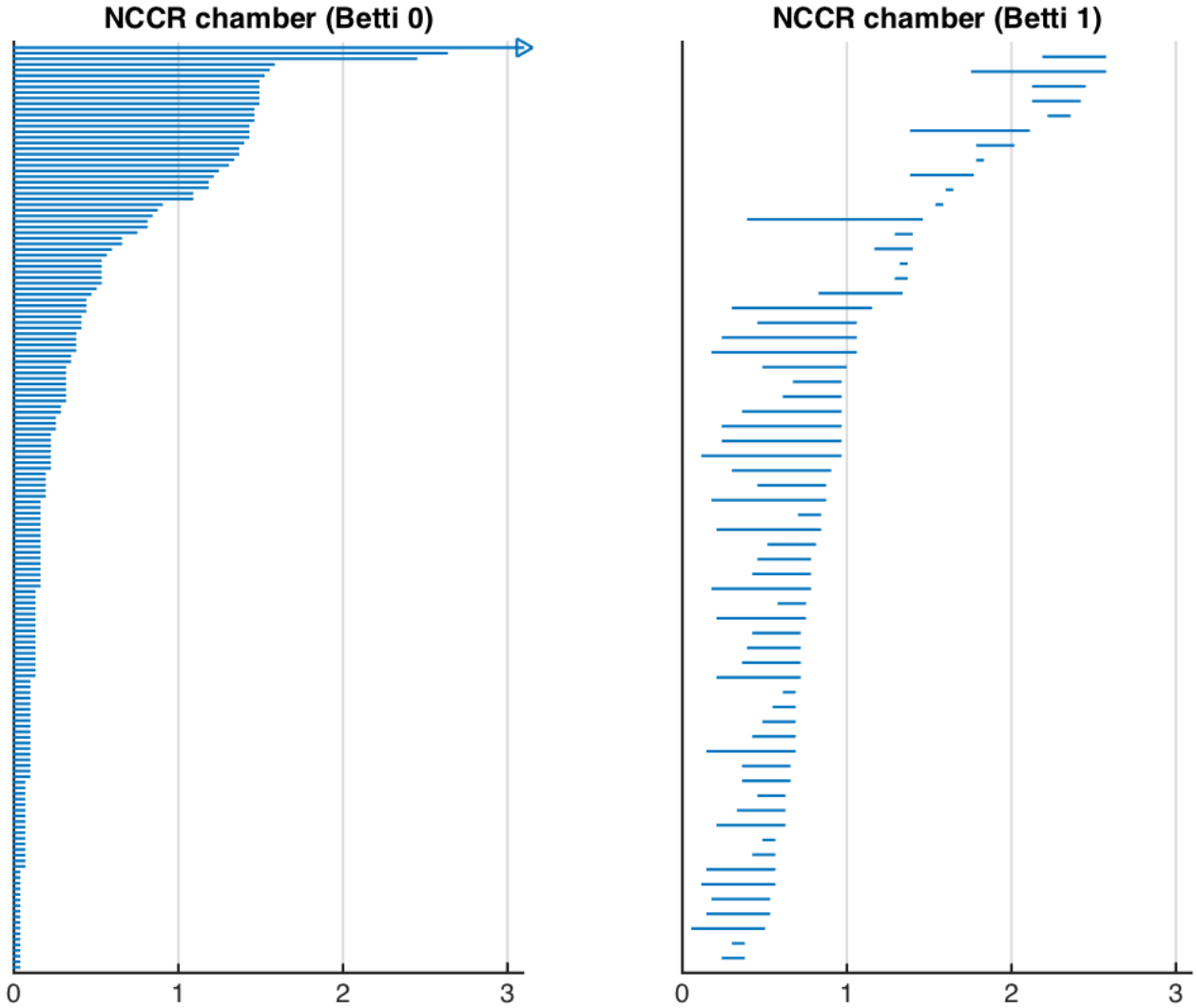}
\end{minipage}
\qquad
\begin{minipage}[b]{0.45\textwidth}\centering
\includegraphics[width=1\textwidth]{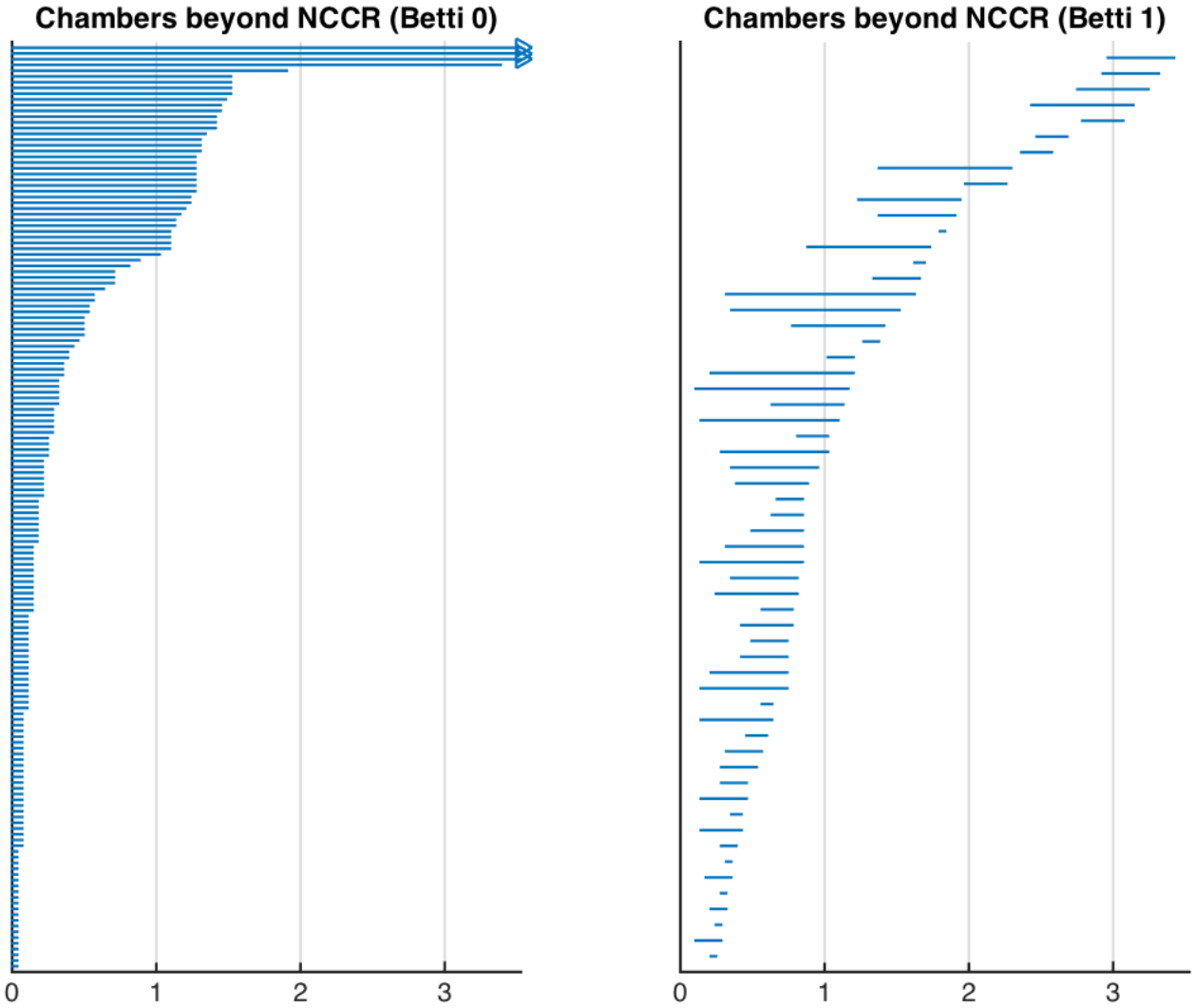}
\end{minipage}
\caption{Spectra of BPS states for the conifold. The barcodes have been obtained using the lazy witness complex $\mathsf{LW} (\mathsf{X} , \mathsf{L} , \epsilon)$ on a point cloud of the form $\mathsf{x} = (k , m , \log |\Omega(\gamma;\mathscr{C}) |) $. \textit{Left.} The noncommutative crepant resolution chamber. The point cloud is made of 1165 invariants and the number of simplices is 4361. \textit{Right}. Chambers of the kind $\mathscr{C} [\mathsf{MS}_{-n}^{-1} , \mathsf{MS}_{-n-1}^{-1}]$ for $n=45$. The point cloud has 1230 points and the number of simplices is 5095.}\label{Conifold2}
\end{figure}

\begin{figure}[th!]\centering
\includegraphics[width=0.46\textwidth]{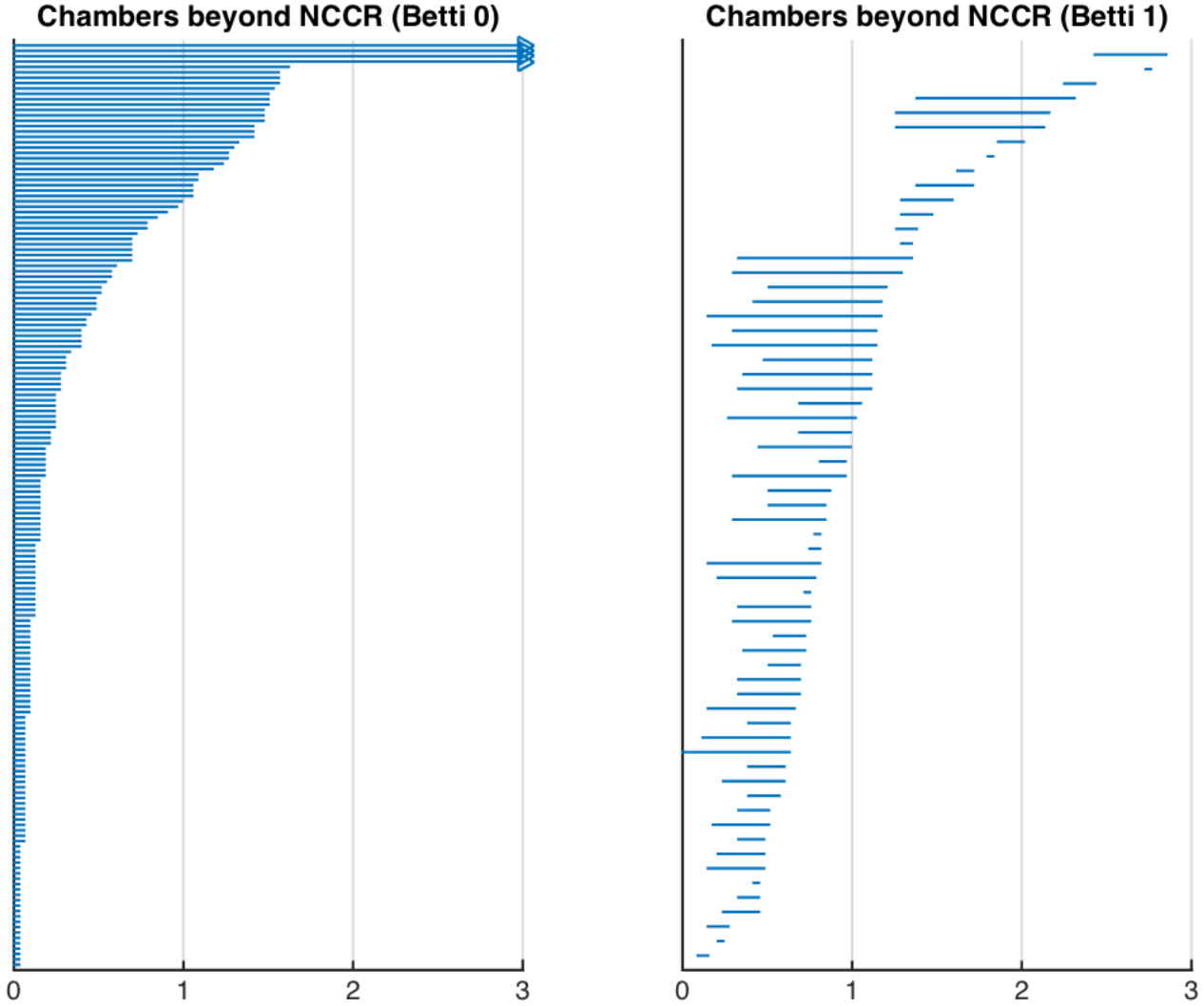}
\caption{Chambers of the kind $\mathscr{C} [\mathsf{MS}_{-n}^{-1} , \mathsf{MS}_{-n-1}^{-1}]$ for $n=90$. The point cloud has 1183 points and the number of simplices is 3632.}\label{Conifold3}
\end{figure}

For each of the chambers under consideration we take around a thousands of BPS invariants. For those partition functions in chambers $\mathscr{C}$ which depend on a choice of an integer~$n$, namely~\eqref{Zcorechambers} and~\eqref{Zconjchambers}, we chose this integer high enough to provide enough invariants. Due to the large number of states, we employ the lazy witness complex $\mathsf{LW} (\mathsf{X} , \mathsf{L} , \epsilon)$, with~200 landmark points. The computation of persistent homology leads to the barcode distributions shown in Figs.~\ref{Conifold1},~\ref{Conifold2} and~\ref{Conifold3}.

Let us understand what kind of topological information we can infer. Consider firstly Fig.~\ref{Conifold1}. On the left we have the chambers close to the core region, while on the right the large radius chamber. The transition is very clear and by looking at the homology in degree one, we see that the large radius chamber presents many more topological features. In a sense moving towards large radius there is an increase in complexity as more cycles form in an irregular pattern. We interpret this as evidence that the stability condition which identifies the large radius chamber offers more possibilities to construct BPS states, while the enumerative problem in chambers nearby the core region is comparatively poorer. Interestingly the length of the long-lived persistent classes is roughly comparable. Looking at the zeroth Betti number in both cases most homology classes have died by $\epsilon \simeq 1.5$.

Also quite interesting are the similarities between the large radius chamber, Fig.~\ref{Conifold1} on the right, where BPS invariants count ideal sheaves on the resolved conifold, and the noncommutative crepant resolution chamber, Fig.~\ref{Conifold2} on the left, where BPS invariants count cyclic modules over the path algebra of the framed conifold quiver. Despite the difference between the enumerative problems the shapes of the barcode diagrams are roughly comparable, the main dissimilarity being the presence of more homology classes in degree zero which persist over $\epsilon \simeq 3$ in the large radius region. Of course while we are discussing the distribution of barcodes, the actual homology classes have a different behavior in both cases, and it would be very interesting to try to interpret each interval, especially in degree one. The abundance of non-trivial 1-cycles in both cases points out that, if we try to interpret the point clouds as a discretization of an underlying geometry, at a generic point within this geometry there should exist a canonical set of circle coordinates.\footnote{Canonical in the sense of being associated to the 1-form dual to the 1-cycle. More precisely, assuming there is no torsion, one can lift cohomology from $\mathbb{Z}_2$ to $\mathbb{R}$, and then pick the harmonic representative of the 1-form, which locally integrates to a smooth coordinate function.}

Another intriguing phenomenon happens when continuing beyond the noncommutative cre\-pant resolution chamber, to chambers of the form $\mathscr{C} \big[\mathsf{MS}_{-n}^{-1} , \mathsf{MS}_{-n-1}^{-1}\big]$. We have computed the persistent homology in Fig.~\ref{Conifold2} on the right and Fig.~\ref{Conifold3} for two values of the chamber label~$n$. As~$n$ increases the barcode distribution interpolates between the noncommutative chamber and the large radius chamber. In Fig.~\ref{Conifold3} the similarities with Fig.~\ref{Conifold1} on the right, are striking, despite the actual generating functions being distinct. A physical interpretation of these chambers is non straightforward; as explained in \cite{Andriyash:2010yf}, proceeding by keeping $z$ fixed while sending $\varphi$ to zero, makes the large radius approximation to the periods used to compute the central charges, less reliable. What we seem to be finding is that getting further and further from the noncommutative chamber, one reproduces a similar structure to large radius. Indeed this is reminiscent of the result of \cite{szendroi} which relates the partition function of large radius Donaldson--Thomas invariants, and the partition function of noncommutative Donaldson--Thomas invariants, with the large radius partition function of the flipped conifold, where a topology changing transition has replaced the exceptional divisor with a topologically distinct $\mathbb{P}^1$. According to this interpretation, the shape of the barcode distribution in Fig.~\ref{Conifold3} seem to imply that the only difference with the large radius BPS states is the core charge, which has changed due to crossing a large number of conjugation walls~$\mathscr{S}$. It would be very interesting to have an independent check in the low energy effective theory.

\section{Quantum geometry of compact threefolds} \label{qgeom}

Now we want to discuss BPS invariants for a few compact Calabi--Yau varieties. So far we have fixed an underlying compactification and we have studied how the BPS spectra are affected by crossing walls of marginal stability. We will now take a somewhat different perspective, by comparing BPS spectra in different compactifications but in the ``same'' chamber, at large radius. What we mean by the same chamber is as follows. The topological string on a compact Calabi--Yau $X$ captures a particular collection of BPS states which play a role in the microscopic description of certain five dimensional black holes~\cite{Katz:1999xq}, the Gopakumar--Vafa invariants \cite{Gopakumar:1998ii,Gopakumar:1998jq}. It is generically believed, and in some cases proven, that the generating function of Gopakumar--Vafa invariants coincides with the generating function of Donaldson--Thomas invariants in a~certain chamber, upon a change of variables. This chamber captures physically BPS spectra of bound states of D0/D2 branes with a single D6 brane. Having a single D6 brane means that the relevant moduli space is the moduli space of ideal sheaves on~$X$. Implicit in the definition of an ideal sheaf, is a notion of a chamber, for which ideal sheaves represent stable objects. We will compare BPS spectra in such a chamber for different Calabi--Yaus.

We have chosen certain one parameter varieties for which a great deal is known about Gopakumar--Vafa invariants, and the topological string explicitly solved up to a certain ge\-nus~\cite{Huang:2006hq}. Thanks to Castelnuovo's theory of curves, these results allow for the computation of several Donaldson--Thomas invariants \cite{Huang:2007sb}.

\subsection{Topological strings on one parameter CY}

We will discuss the distributions of BPS invariants on certain one-parameters models where the topological string amplitudes were computed explicitly in \cite{Huang:2006hq} to high genus. These models have all the form of complete intersection Calabi--Yaus in weighted projective spaces. Complete intersection manifolds are constructed as the zero locus of a finite number of homogeneous polynomials in a product of projective spaces. If we denote a weighted projective space by $\mathbb{P}^{n-1} (w_1 , \dots , w_n)$, then a complete intersection CY of degree $( d_1 , \dots , d_k )$ will be denoted by $X_{d_1, \dots , d_k} (w_1, \dots , w_n)$ and $w^m$ will denote the $m$ times repetition of the weight $w$. For example the quintic threefold $X_5 (1)$ is obtained as the zero locus of a single degree 5 polynomial in $\mathbb{P}^4$.

A collection of 13 such threefolds was discussed in \cite{Huang:2006hq}. All these one-parameter models have the property that their mirror has a Picard--Fuchs system with three regular singular points. The moduli spaces have the form $\mathcal{M} (X) = \mathbb{P}^1 \setminus \{ 0 , 1 , \infty \} $ where the point denoted by $\infty$ denotes the point of maximal monodromy at large radius, where we will discuss BPS spectra of D-branes bound states. The other two points correspond to a conifold point and a point with rational branching, for example an orbifold point corresponding to a Gepner model. The constraints imposed by modularity on the topological string amplitudes implies that these can be written as polynomials over a ring generated by certain modular objects.

In particular the holomorphic ambiguity in the topological amplitude is fixed by the specific form of the amplitude near the conifold and orbifold points; for example at the conifold points, the ``gap condition'' determines the form of the poles in the amplitude via knowledge of the light BPS states, and this structure is universal~\cite{Huang:2006hq}.

\subsection{BPS invariants}

To the Calabi--Yaus $X$ under consideration one can associate the partition function
\begin{gather*}
Z_{\GV} (X ; q , t) = \prod_{d=1}^{\infty} \left[ \prod_{r=1}^{\infty} \big(1 - q^r \e^{-d t}\big)^{r \GV_{0,d}} \prod_{g=1}^{\infty} \prod_{l=0}^{2g-2} \big( 1 - q^{g-l-1} \e^{- d t} \big)^{(-1)^{g+l} \left(\begin{smallmatrix} 2 g-2 \\ l \end{smallmatrix}\right) \GV_{g,d} }\right] .
\end{gather*}
The topological string free energy $F (X ; q, t) = \log Z_{\GV} (X ; q , t)$ was computed in \cite{Huang:2006hq} using the techniques outline above, up to high genus. Here $t$ represents the K\"ahler modulus of the Calabi--Yau, while $q= \e^{\ii \lambda}$, with $\lambda$ the topological string coupling. The integers $\GV_{g,d}$ are the Gopakumar--Vafa invariants of \cite{Gopakumar:1998ii,Gopakumar:1998jq}. The free energy $F (X ; q, t)$ has an interpretation as a certain coupling in the low energy $\mathcal{N}=2$ supergravity. This term can be obtained by a one-loop computation in a~constant graviphoton background, where it receives contribution from BPS particles engineered by an M2 brane wrapping a degree $d$ curve, in the M-theory limit of the type IIA string. The Gopakumar--Vafa invariant $\GV_{g,d}$ is a twisted supersymmetric index which captures the contribution of these particles to the effective coupling. The quantum numbers of these particles can be labelled by the representation of the little group ${\rm SO}(4) \sim {\rm SU}(2)_L \otimes {\rm SU}(2)_R$. It turns out that only supersymmetric particles with quantum number $g$ associated with representations of ${\rm SU}(2)_L$ of the form
\begin{gather*}
\left[ 2 (\mathbf{0}) + \left( \mathbf{\frac12}\right) \right]^{\otimes g}
\end{gather*}
contribute to the index.

These couplings and the invariants $\GV_{g,d}$ play a very important role in the microscopical description of black holes in quantum gravity. A five dimensional black hole can be engineered via an M-theory compactification on a Calabi--Yau, by an M2 brane wrapping a curve $\beta = d t \in H_2 (X , \mathbb{Z})$. The microscopic degeneracies of such a black hole, with charge $\beta$ and angular momentum $m$ in ${\rm SU}(2)_L$ can be computed as functions of the invariants $\GV_{g,d}$, and reproduce precisely the macroscopic entropy~\cite{Katz:1999xq}.

In this section we will be more interested in certain D-brane bound states which are dual to the above description. We will study numerical Donaldson--Thomas invariants of the moduli space of ideal sheaves of $X$ which physically represent the degeneracies of BPS bound states of a D6 brane wrapping $X$ with a gas of D0/D2 branes. Conjecturally the duality has the form
\begin{gather} \label{DTandGV}
Z_{\DT} (X ; q , t) = \sum_{n,d} \DT_{n,d} q^n \e^{-d t} = Z_{\GV} (X ; q , t) M (-q)^{\chi (X)} ,
\end{gather}
where
\begin{gather*}
M(q) = \prod_{n=1}^{\infty} \big( 1 - q^n \big)^{-n}
\end{gather*}
is the MacMahon function. Using the knowledge of the topological string amplitude computed in \cite{Huang:2006hq} we will follow the approach of~\cite{Huang:2007sb} to compute the BPS invariants $\DT_{n,d}$ up to a~certain~$d$. By expanding \eqref{DTandGV} one sees that the invariant $\DT_{n,d}$ with fixed~$d$, receives contributions from all the invariants $\GV_{g,d'}$ with $d' \le d$, but arbitrary $g$. Therefore in principle one should know the full right hand side of \eqref{DTandGV} to compute the invariants $\DT_{g,d}$. However for the models at hand there is an enormous simplification coming from Castelnuovo theory. Roughly speaking Castelnuovo's theory provides a bound on the genus $g$ of a degree $d$ curve in a projective space, and certain generalizations thereof. For the models we are studying, the Castelnuovo bound is either known or estimated in~\cite{Huang:2007sb}. Due to the Castelnuovo bound, given a degree $d$, there is a~$g_{\max}$ so that for $g > g_{\max}$ all the $\GV_{g,d}$ are vanishing. For example for the quintic $X_5 (1)$
\begin{gather*}
g_{\max} \le \frac{1}{10} \big( 10 + 5 d + d^2 \big) .
\end{gather*}
Taking as inputs the invariants $\GV_{g,d}$ of \cite{Huang:2007sb, Huang:2006hq} and using Castelnuovo's theory, we have computed all the BPS degeneracies $\DT_{n,d}$ up to $d=9$ and $n=9$ for the complete intersection varieties $X_5 (1)$, $X_{3,3} \big(1^6\big)$, $X_{4,2} \big(1^6\big)$, $X_{3,2,2} \big(1^7\big)$, $X_{2,2,2,2} \big(1^8\big)$ and $X_{4,3} \big(1^5,2\big)$, to have an homogenous set of data to use as a point cloud. We have similar computations for the remaining of the~13 one parameter models, but with less data; the results are in line with what we will discuss momentarily but we will not include them here.

\subsection{Topological analysis}

We collect and discuss here our results. Out of our physical settings we construct a number of point clouds $\mathsf{X}$ which have the form $\mathsf{x} = (d , n , \log |\DT_{n,d}|)$ for $\mathsf{x} \in \mathsf{X}$. Again we are taking the logarithm of the degeneracies. As we have already stressed this should be handled with care, since the logarithm is likely to wash out more subtle topological features in the BPS spectra. However as already explained this is not a problem in our case, since we are not using persistent homology to determine the properties of a BPS spectrum on its own, but to compare various spectra. From a purely computational perspective, taking the logarithm is almost a necessity. The BPS degeneracies that we are discussing correspond to black hole microstates and on general grounds in quantum gravity these numbers grow exponentially.

Let us apply our formalism. We construct the point cloud $\mathsf{X}$ with BPS spectra consisting of 108--135 states, depending on the geometry. These are all the non-vanishing Donaldson--Thomas invariants $\DT_{n,d}$ of the compact threefold up to degree $d=9$, and $n=9$ (chosen in order to have point clouds of roughly the same order). Out of $\mathsf{X}$ we construct the family of topological spaces~$\mathsf{X}_\epsilon$ and the Vietoris--Rips complex $\mathsf{VR}_\epsilon (\mathsf{X})$. Then we pass to the homology $H_i (\mathsf{VR}_\epsilon (\mathsf{X}) ; \mathbb{Z}_2)$ and look at the topological features of the $\mathbb{N}$-persistence modules.

\begin{figure}[th!]\centering
\begin{minipage}[b]{0.45\textwidth}\centering
\includegraphics[width=1\textwidth]{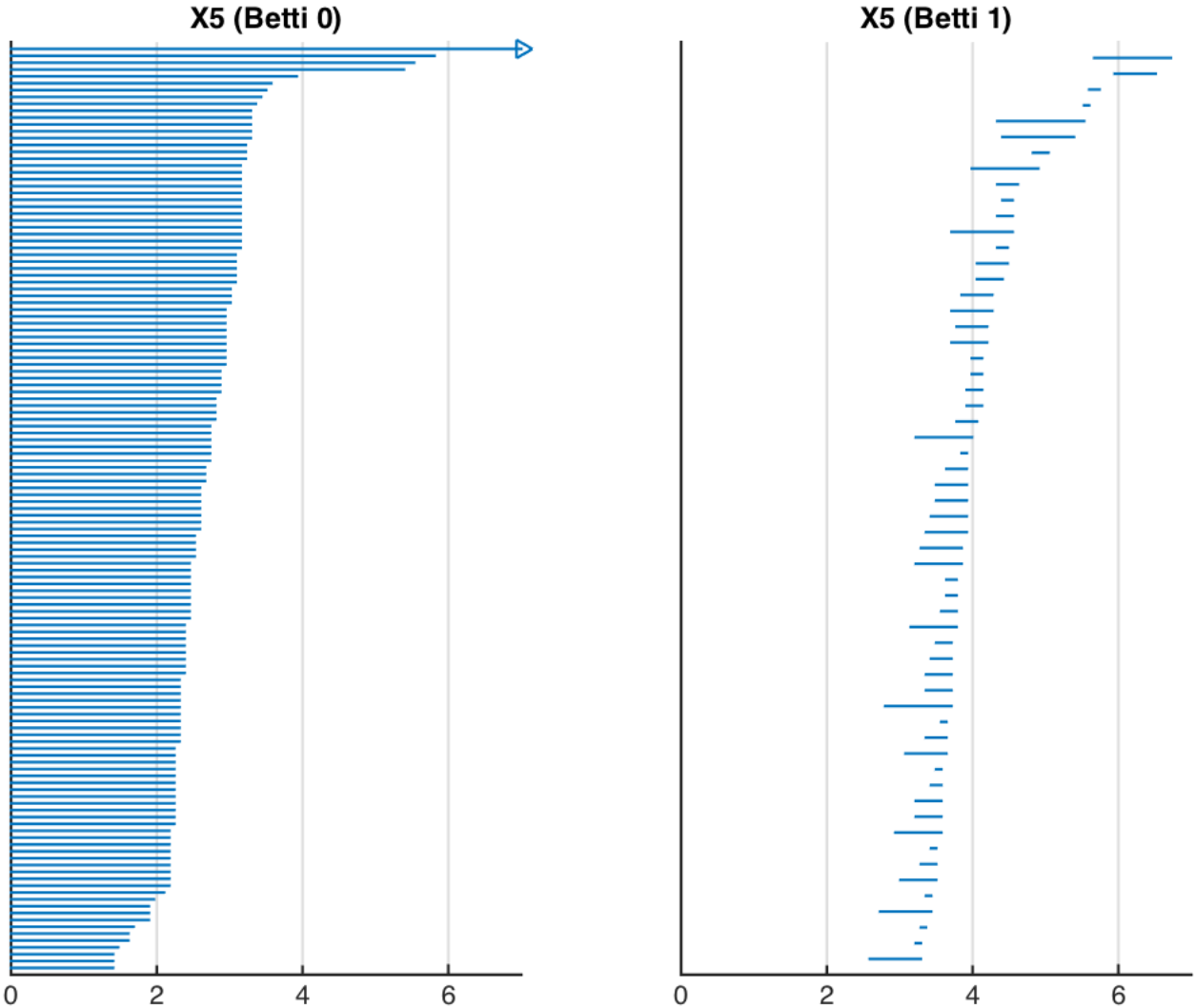}
\end{minipage}
\qquad
\begin{minipage}[b]{0.45\textwidth}\centering
\includegraphics[width=1\textwidth]{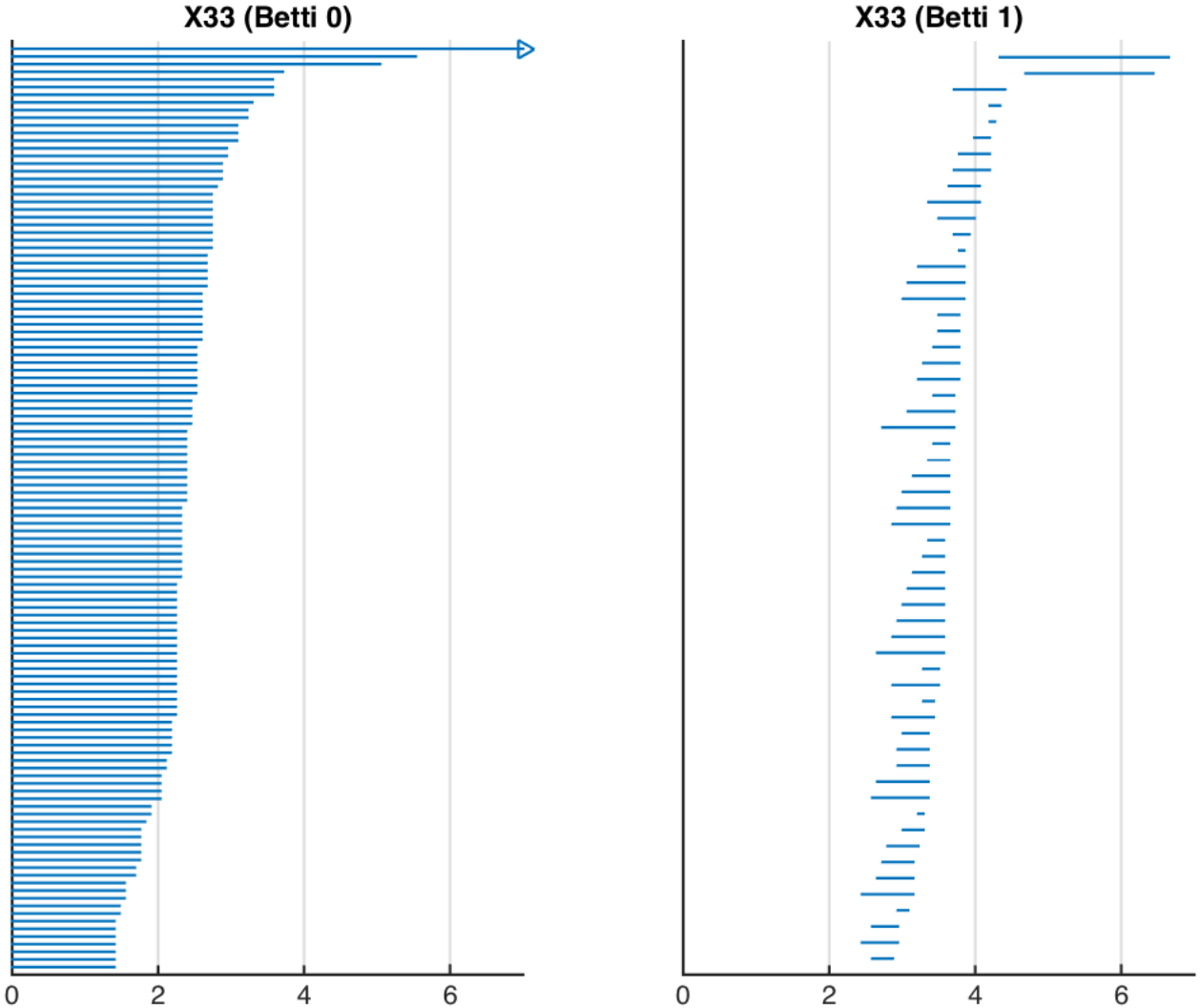}
\end{minipage}
\caption{Barcodes for the distribution of Donaldson--Thomas invariants on compact Calabi--Yaus. The computation uses the Vietoris--Rips complex $\mathsf{VR}_\epsilon (\mathsf{X})$ build on a point cloud $\mathsf{X}$ whose generic element is $\mathsf{x} = (d , n , \log |\DT_{n,d}|)$. We give explicitly the number of simplices used in the computation. \textit{Left.} The quintic $X_5 (1)$, with 135 BPS invariants, from 2662 simplices. \textit{Right.} The variety $X_{3,3} (1^6)$, with 121 BPS invariants, from 3472 simplices.}\label{compact1}
\end{figure}

\begin{figure}[th!]\centering
\begin{minipage}[b]{0.45\textwidth}\centering
\includegraphics[width=1\textwidth]{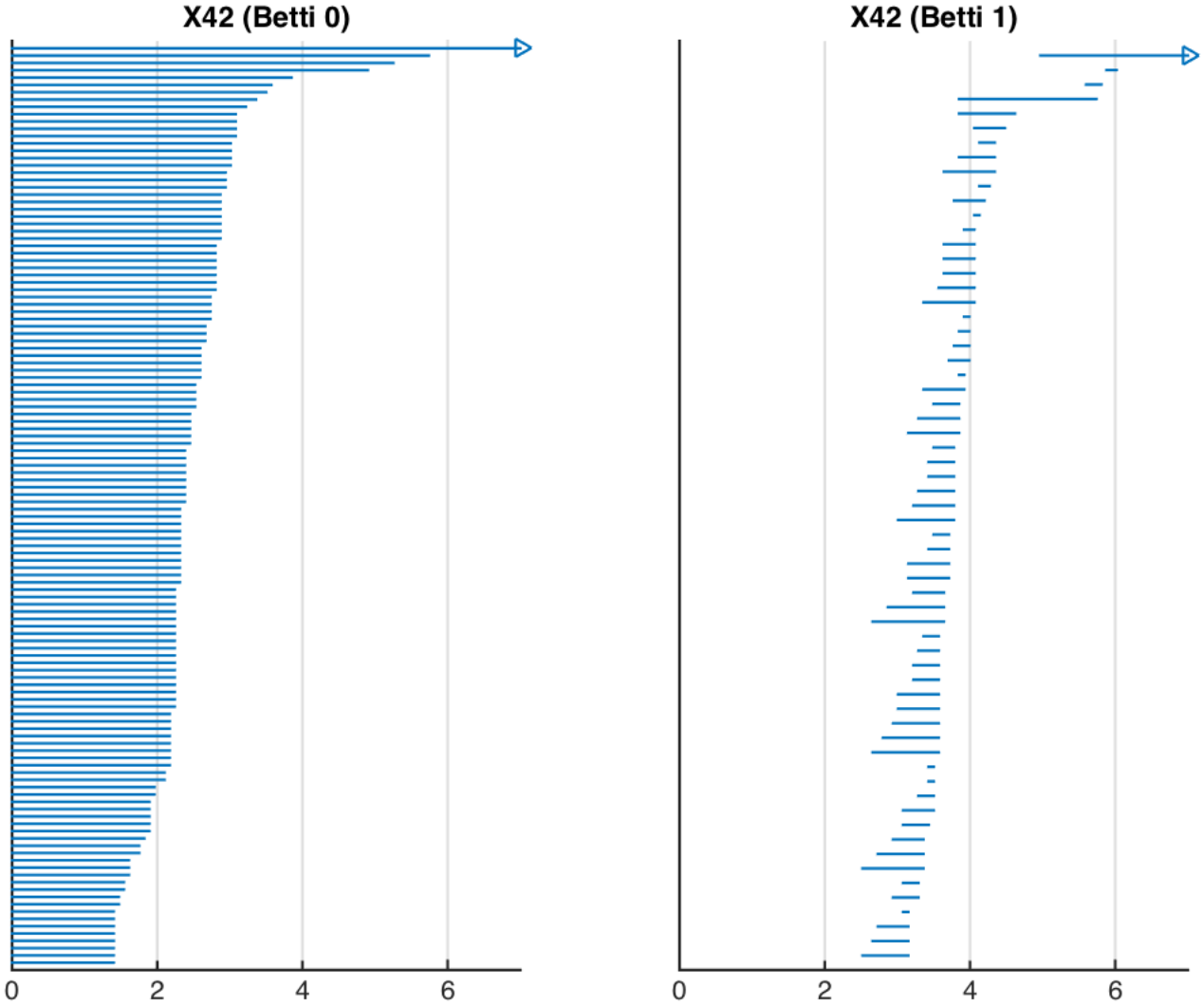}
\end{minipage}
\qquad
\begin{minipage}[b]{0.45\textwidth}\centering
\includegraphics[width=1\textwidth]{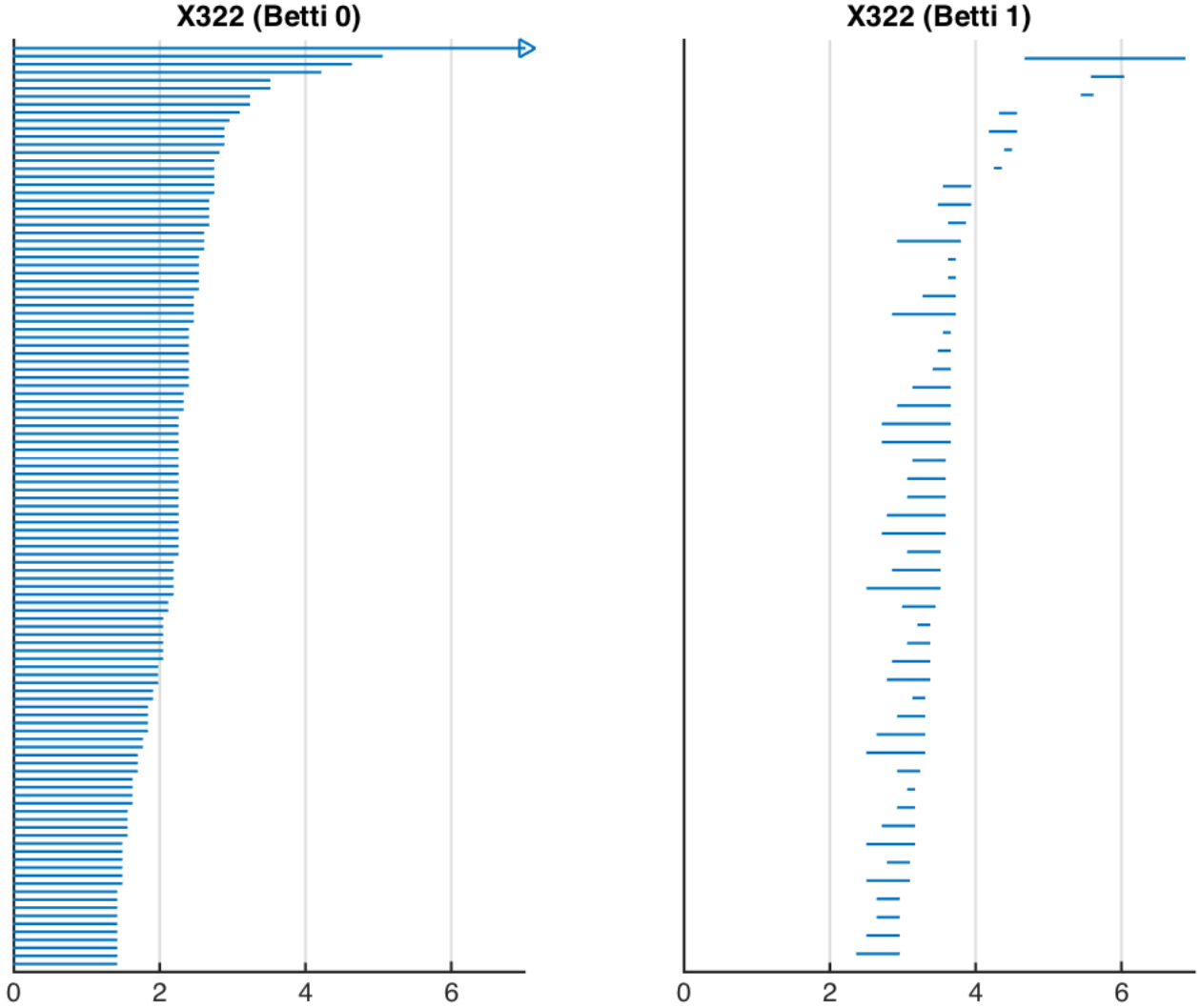}
\end{minipage}
\caption{Barcode computation, same as in Fig.~\ref{compact1}. \textit{Left.} The variety $X_{4,2} (1^6)$, with 126 BPS invariants, from 3200 simplices. \textit{Right.} The variety $X_{3,2,2} (1^7)$, with 115 BPS invariants, from 3795 simplices.}\label{compact2}
\end{figure}

\begin{figure}[htbp]\centering
\begin{minipage}[b]{0.45\textwidth}\centering
\includegraphics[width=1\textwidth]{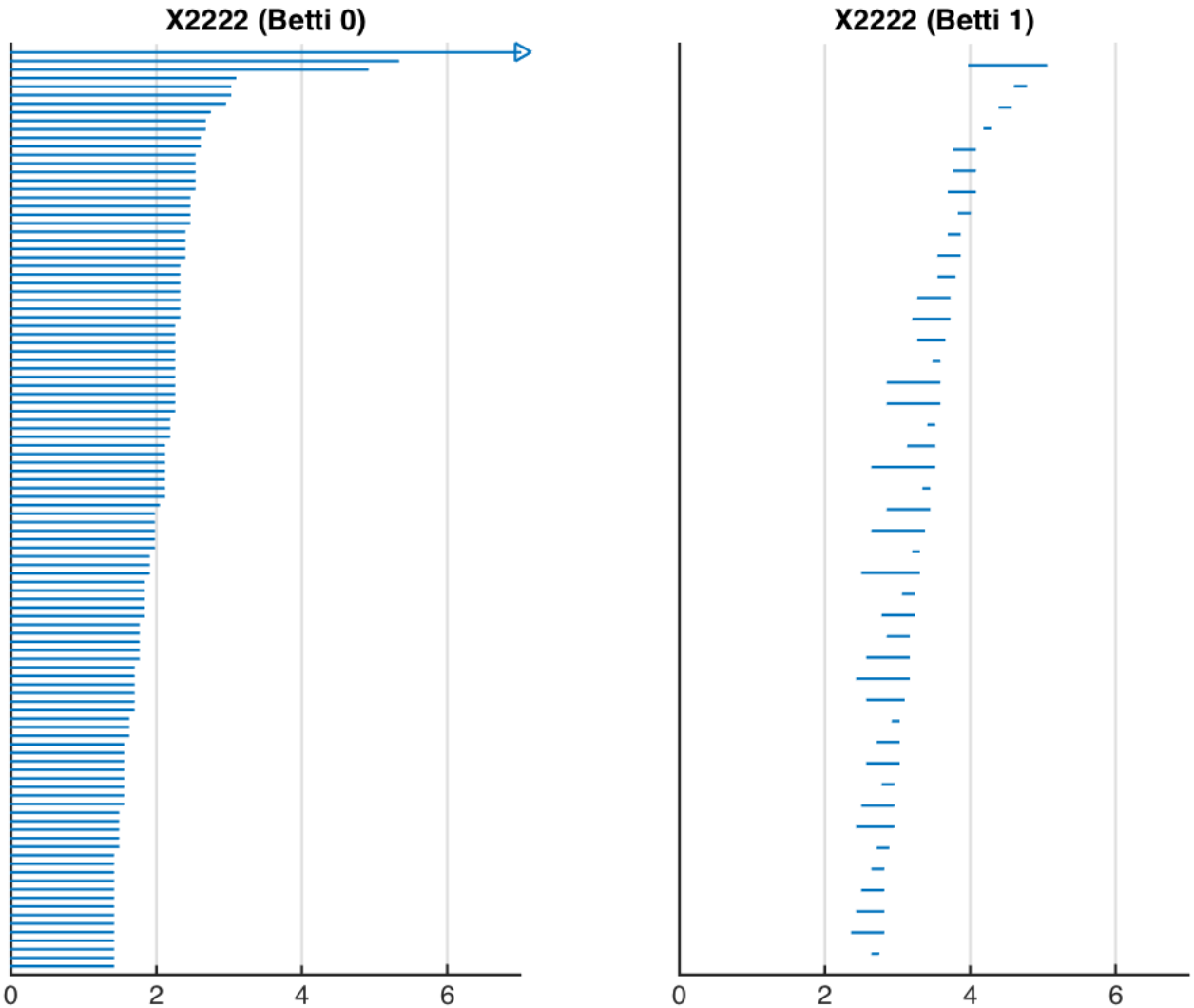}
\end{minipage}
\qquad
\begin{minipage}[b]{0.45\textwidth}\centering
\includegraphics[width=1\textwidth]{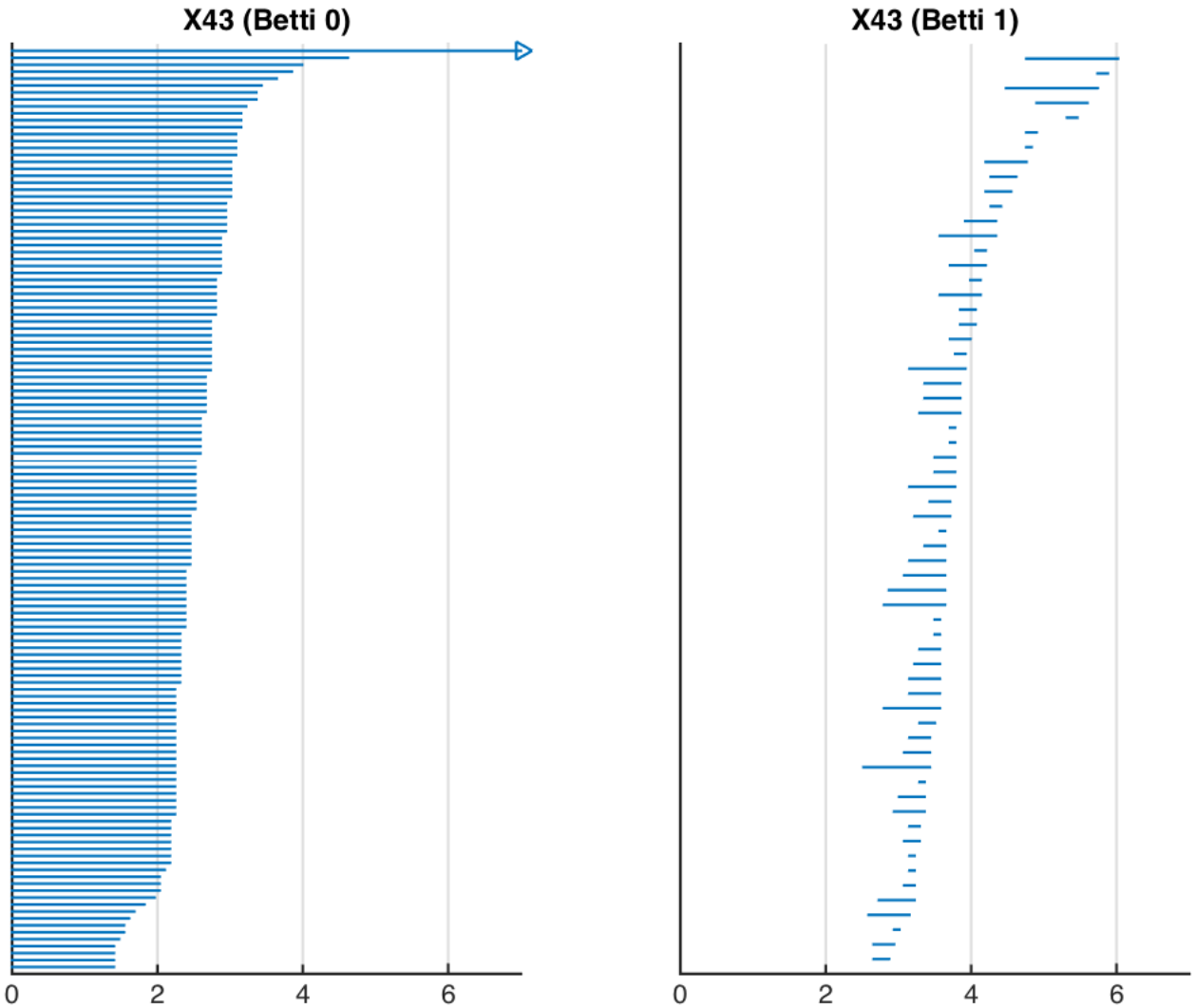}
\end{minipage}
\caption{Barcode computation, same as in Fig.~\ref{compact1}. \textit{Left.} The variety $X_{2,2,2,2} (1^8)$, with 108 BPS invariants, from 4163 simplices. \textit{Right.} The variety $X_{4,3} (1^5,2)$, with 133 BPS invariants, from 2988 simplices.}\label{compact3}
\end{figure}

One immediate feature of Figs.~\ref{compact1}--\ref{compact3} is that they all look rather similar. The differences between the distributions of barcodes are minimal. For certain geometries the $H_0$ barcodes are slightly more long-lived and for others the $H_1$ distribution is a bit more regular. However these differences, despite the damping due to the logarithmic scale, are still very small and it is natural to think that there is a physical mechanism behind this.

Of course there is a natural candidate for the problem at hand: the universality expected for black hole physics. The degeneracies of black hole microstates must be such as to recover the macroscopic entropy, given by the area law: for the large charges the entropy scales with the area of the black hole horizon. Furthermore the area law receives an infinite series of higher derivative corrections~\cite{LopesCardoso:1998wt} which are expected to be determined by the topological string \cite{Denef:2007vg,LopesCardoso:2006bg,Cardoso:2014kwa, Ooguri:2004zv}.

What we are seeing here is that also the topological features of the BPS spectra (but \textit{not} the actual values of the BPS invariants, which fluctuate wildly between different geometries) appear to be universal. Note that the usual ideas about black hole universality refers to the emergence of universal features in the large charges expansion. Here we are dealing with the BPS spectrum without any limit (although it should be noted that from a similar partial knowledge, the authors of \cite{Huang:2007sb} could extrapolate a large order result which agrees with the macroscopic predictions).

It is instructive to compare this behaviour with the results of Section~\ref{conifold}, as seen in the Figs.~\ref{Conifold1}, \ref{Conifold2} and \ref{Conifold3}. Consider first the $H_0$ barcodes. In the case of compact threefolds they are evenly distributed, not too distant from an average lifespan at around $\epsilon \sim 2$. On the other hand in the case of the conifold, we see in all chambers a~tendency to cluster rather rapidly, with many barcodes disappearing almost immediately; here the main topological feature is a~smaller number of connected components. Similarly in the conifold case the behaviour of the $H_1$ barcodes is very irregular, with relatively long lived cycles appearing at every length scale. This is contrasted with the case of compact threefolds. Indeed from Figs.~\ref{compact1}, \ref{compact2} and \ref{compact3} we see that the appearance of non trivial cycles is mostly contained within a certain length scale, around $\epsilon \sim 3$. These topological features imply that in the case of compact threefolds the distribution of BPS degeneracies appears more regular and uniform, when compared with the case of the conifold where the BPS degeneracies cluster in fewer connected components.

It is natural to believe that what we are seeing in the case of compact threefolds is how universal aspects of black hole physics impact the topological features of the distribution of BPS states. In particular it is not just the shape of the barcode distribution, but the actual lengths and positions of the barcodes which are comparable, both in degree zero and one. We stress again that the values of the BPS invariants and the corresponding point clouds are very different from case to case. It is natural to conjecture that this phenomenon would become more pronounced were we to increase the number of BPS invariants in the point clouds. To be more precise we put forward the following conjecture, which we expect to be valid at least for one-parameter Calabi--Yaus: as the number of BPS invariants increases, approaching the large charge limit, the $\mathbb{N}$-persistence modules associated with $H_0$ and $H_1$ tend to universal $\mathbb{N}$-persistence modules. Note that if we interpret the point clouds as a discretization of an underlying surface, we can rephrase the above statement in terms of a single universal homotopy class of surfaces, governing the distribution of BPS states.

Assuming that such a conjecture is true, it would seem natural to imagine that such a~homotopy class of surfaces is determined by the attractor mechanism. Unfortunately we do not now how such an explicit link could be realized.

It is clear that we cannot draw any conclusive statement from the limited spectra we have analyzed here. To check our conjecture directly would require a better knowledge of the BPS spectra to higher values of the curves' degree, and possibly for a larger sample of Calabi--Yaus.

\section[$\mathcal{N}=4$ dyons and quantum black holes]{$\boldsymbol{\mathcal{N}=4}$ dyons and quantum black holes} \label{dyons}

As next example we now consider dyons in an $\mathcal{N}=4$ compactification on $\rm{K3} \times \torus^2$. The reason to consider this example is that contrary to the cases discussed in the previous section, the BPS invariants are directly constrained by modularity and appear as coefficients of certain number theoretical functions. Indeed in this case the high degree of supersymmetry and the action of the modular group ${\rm SL}(2;\mathbb{Z}) \times {\rm SO}(22,6;\mathbb{Z})$ of S- and T-dualities allows for explicit counting functions for BPS states~\cite{Dijkgraaf:1996it}. The number of $1/4$ BPS states can be written as coefficients of the Fourier expansion of $1/\Phi_{10}$, where $\Phi_{10}$ is the Igusa cusp form of weight $10$. These coefficients represent indices of BPS bound states computed at weak string coupling; as the string coupling grows the geometry backreacts and a black hole can form. Certain Fourier coefficients of this expansion then represent degeneracies of black holes microstates which can then be compared with the macroscopic thermodynamical quantities \cite{Dabholkar:2005dt, LopesCardoso:2004xf,LopesCardoso:2006bg,Shih:2005he}. We will be interested in the interplay between such counting functions and modularity; the relevant counting functions are meromorphic Jacobi forms and the associated mock Jacobi forms are interpreted as partition functions of single-centered black holes~\cite{Dabholkar:2012nd}. We want to discuss how (mock) modularity of the microscopic degeneracies impacts the topological features of the distribution of BPS states. For a macroscopic supergravity perspective on how modularity affects the counting functions, see~\cite{Cardoso:2013ysa, Cardoso:2014hma}.

The charges of the BPS states take values in the lattice of electric-magnetic charges $\Gamma_e^{6,22} \oplus \Gamma_m^{6,22}$, where each factor is isomorphic to the lattice $H_2 (\rm{K3} ; \mathbb{Z}) \oplus 3 \Gamma^{1,1}$, and $\Gamma^{1,1}$ is an hyperbolic lattice. The microscopic degeneracies can be labelled by three T-duality invariants $(n,l,m)$. They can be extracted from the expansion of the Siegel modular form of weight 10, $\Phi_{10} (\tau , z , v)$, which is itself a function of the chemical potentials for the T-duality invariants \cite{Dijkgraaf:1996it}. By writing $w = \e^{2 \pi \ii v}$, we can expand
\begin{gather*}
\frac{1}{\Phi_{10}} = \sum_{m \ge 1} \psi_m (\tau ,z) w^m ,
\end{gather*}
where the functions $\psi_m (\tau , z)$ are meromorphic Jacobi forms of index $m$ \cite{Dabholkar:2012nd}. Similarly we will also introduce the notation $q= \e^{2 \pi \ii \tau}$ and $\zeta = \e^{2 \pi \ii z}$. A meromorphic Jacobi form $\psi (\tau , z)$ of weight $k$ and index $m$ is holomorphic in $\tau$ and meromorphic in $z$ and transforms as
\begin{gather} \label{jacobitransf}
\psi \left( \frac{a \tau +b}{c \tau + d} , \frac{z}{c \tau + d} \right) = (c \tau + d)^k \e^{\frac{2 \pi \ii m c z^2}{c \tau + d}} \psi (\tau , z) , \qquad \text{with} \quad \left( \begin{matrix} a & b \\ c & d \end{matrix} \right) \in \mathrm{SL} (2 , \mathbb{Z}) ,
\end{gather}
under the modular group, and enjoys the elliptic transformation
\begin{gather} \label{elliptictransf}
\psi (\tau , z + \lambda \tau + \mu) = \e^{- 2 \pi \ii m (\lambda^2 \tau + 2 \lambda z)} \psi (\tau , z) ,
\end{gather}
under translations of $z$ by $\mathbb{Z} \tau + \mathbb{Z}$, where $\lambda , \mu \in \mathbb{Z}$.

All the expansions we shall consider in this section are meaningful for a particular choice of the compactification moduli which correspond to the attractor region. Explicitly we can write the meromorphic Jacobi forms $\psi_m (\tau , z)$ in terms of the elliptic genus of a symmetric product of~K3 surfaces \cite{Dabholkar:2012nd}
\begin{gather} \label{defpsim}
\psi_m (\tau , z) = \frac{1}{A (\tau , z)} \frac{1}{\eta (\tau)^{24}} \mathscr{E} \big(\tau , z ; \mathrm{Sym}^{m+1} (\rm{K3})\big) .
\end{gather}
We have introduced the standard notation
\begin{gather*}
A (\tau , z) = \frac{\theta_1^2 (\tau , z)}{\eta^6 (\tau)} , \\
\eta (\tau) = q^{\frac{1}{24}} \prod_{n \ge 1} (1 - q^n) \\
B (\tau , z) = 4 \left(\frac{\theta_2^2 (\tau , z)}{\theta_2^2 (\tau)} +\frac{\theta_3^2 (\tau , z)}{\theta_3^2 (\tau)} +\frac{\theta_4^2 (\tau , z)}{\theta_4^2 (\tau)}\right) .
\end{gather*}
and the last function will be used momentarily. The first few $\psi_m$ functions can be written down explicitly \cite{Dabholkar:2012nd}
\begin{gather*}
\psi_1 = \frac{1}{4 \eta^{24}} \big( 9 A^{-1} B^2+ 3 E_4 A \big) , \\
\psi_2 = \frac{1}{27 \eta^{24}} \big( 50 A^{-1} B^3 + 48 E_4 A B + 10 E_6 A^2 \big) , \\
\psi_3 = \frac{1}{384 \eta^{24}} \big( 475 A^{-1} B^4 + 886 E_4 A B^2 + 360 E_6 A^2 B + 199 E_4^2 A^3 \big) , \\
\psi_4 = \frac{1}{72 \eta^{24}} \big( 51 A^{-1} B^5 + 155 E_4 A B^3 + 93 E_6 A^2 B^2 + 102 E_4^2 A^3 B + 31 E_4 E_6 A^4 \big) .
\end{gather*}
We have introduced the Eisenstein series $E_k$ of weight $k$, with $k \ge 2$,
\begin{gather*}
E_k (\tau) = 1 - \frac{2 k}{B_k} \sum_{n \ge 1} \sigma_{k-1} (n) q^n
\end{gather*}
with $\sigma_{k-1} (n) = \sum_{d \vert n} d^{k-1}$ and $B_k$ the $k$-th Bernoulli number. We will use the explicit form of these functions to compute the degeneracies of single centered black holes and study their distributions.

Meromorphic Jacobi forms are associated to mock Jacobi forms. This in practice means that the functions $\psi_m$ can be written as a sum of a finite part and a polar part
\begin{gather*}
\psi_m = \psi_m^F + \psi_m^P .
\end{gather*}
Both are holomorphic in $\tau$, but while the finite part is holomorphic also in $z$ the polar part is not and is indeed completely determined by the poles of $\psi_m$. The finite part $\psi_m^F$ does not transform as a Jacobi form, but~\eqref{jacobitransf} is recovered upon adding the polar part $\psi_m^P$. Both $\psi_m^F$ and $\psi_m^P$ still enjoy the elliptic transformation~\eqref{elliptictransf}.

Physically this decomposition corresponds to the fact that the Fourier coefficients of $\psi_m^F$ capture the degeneracies of single centered black holes, while the polar part $\psi_m^P$ determines the jump in the degeneracies due to decay into two centered black holes at walls of marginal stability~\cite{Dabholkar:2012nd}. More complicated decays are forbidden by $\mathcal{N}=4$ supersymmetry (in the sense that too many supercharges realized non-linearly produce too many fermionic zero modes to contribute to the relevant indices). The presence of the polar part is necessary by consistency with wall-crossing, but its contribution has to be subtracted to determine the degeneracies of single centered black holes \cite{Dabholkar:2012nd}. In particular the polar part of $\psi_m$ is known for all~$m$
\begin{gather*}
\psi_m^P (\tau , z) = \frac{p_{24} (m+1)}{\eta^{24} (\tau)} \sum_{s \in \mathbb{Z}} \frac{q^{m s^2 + s} \zeta^{2 m s+1}}{(1-\zeta q^s)^2} .
\end{gather*}
The function $p_{n} (m)$ counts integer partitions of $m$ with $n$ available slots. The finite part $\psi_m^F$ is a mock Jacobi form of index $m$, which means that it can be completed by $\psi_m^P$ to give the Jacobi form $\psi_m$. If we expand
\begin{gather*}
\psi_m^F = \sum_{n,l} c(n,r) q^n \zeta^l ,
\end{gather*}
then the precise physical statement is that the microscopic degeneracies $d (n,l,m)$ corresponding to single centered black holes are related to the coefficients of $\psi_m$ as $d (n,l,m) = (-1)^{l+1} c (n,l)$ for $n \ge m$. Furthermore on can restrict to the range $0 \le l \le m$ due to elliptic invariance of~$\psi_m$.

\begin{figure}[htbp]\centering
\includegraphics[width=0.6\textwidth]{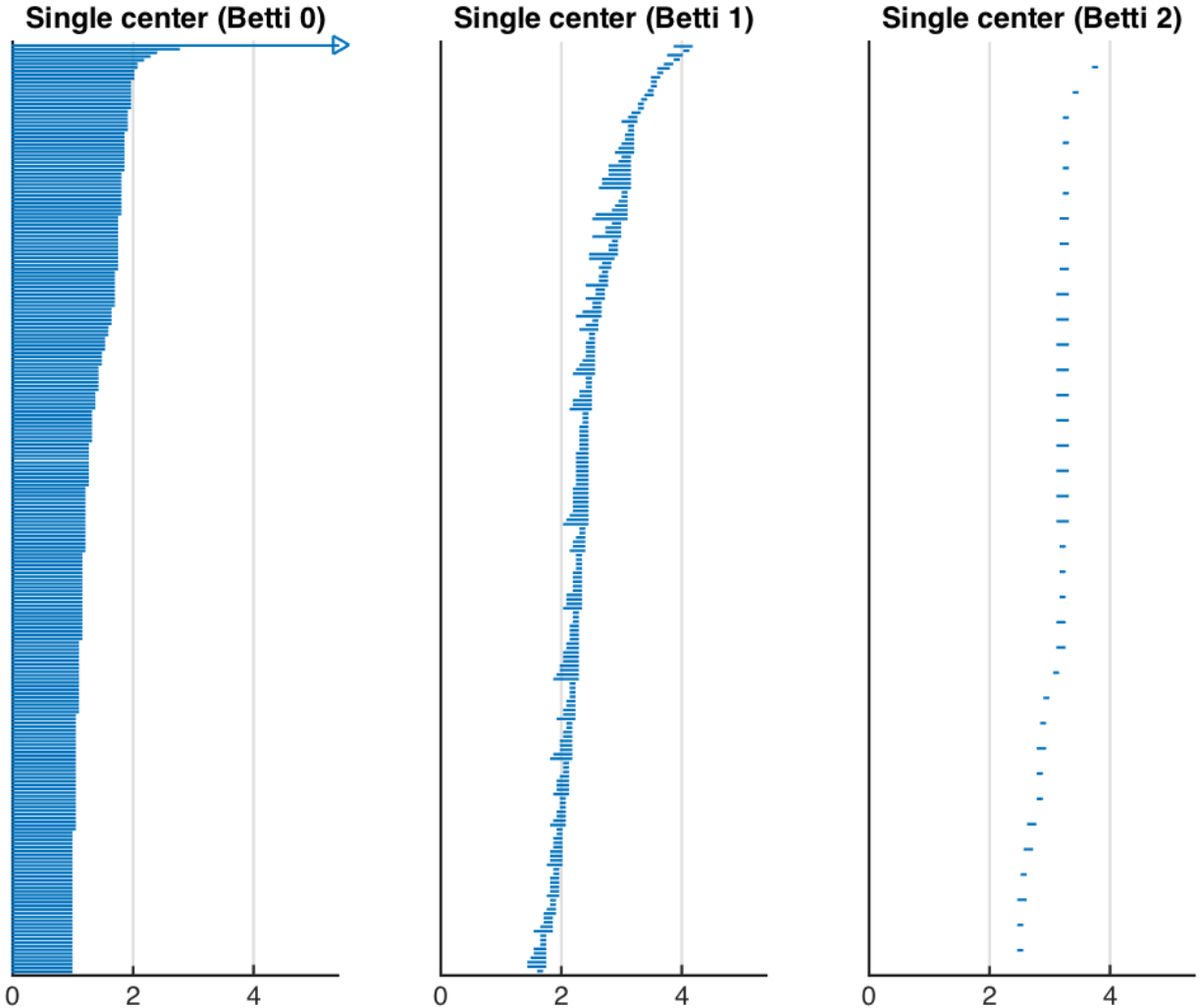}
\caption{Barcodes for the spectrum of single-centered BPS $\mathcal{N}=4$ dyons. The point cloud is constructed out of vectors of the form $\mathsf{x} = (\log | d(n,l,m) | , m , l , n)$. The degeneracies are obtained from the coefficients of $\psi_m^F$, having subtracted the polar part from $\psi_m$. To compute the persistent homology we use the Vietoris--Rips complex $\mathsf{VR}_\epsilon (\mathsf{X})$ with 254 BPS states, running over 254073 simplices.}\label{dyons1}
\end{figure}

\begin{figure}[htbp]\centering
\begin{minipage}[b]{0.45\textwidth}\centering
\includegraphics[width=1\textwidth]{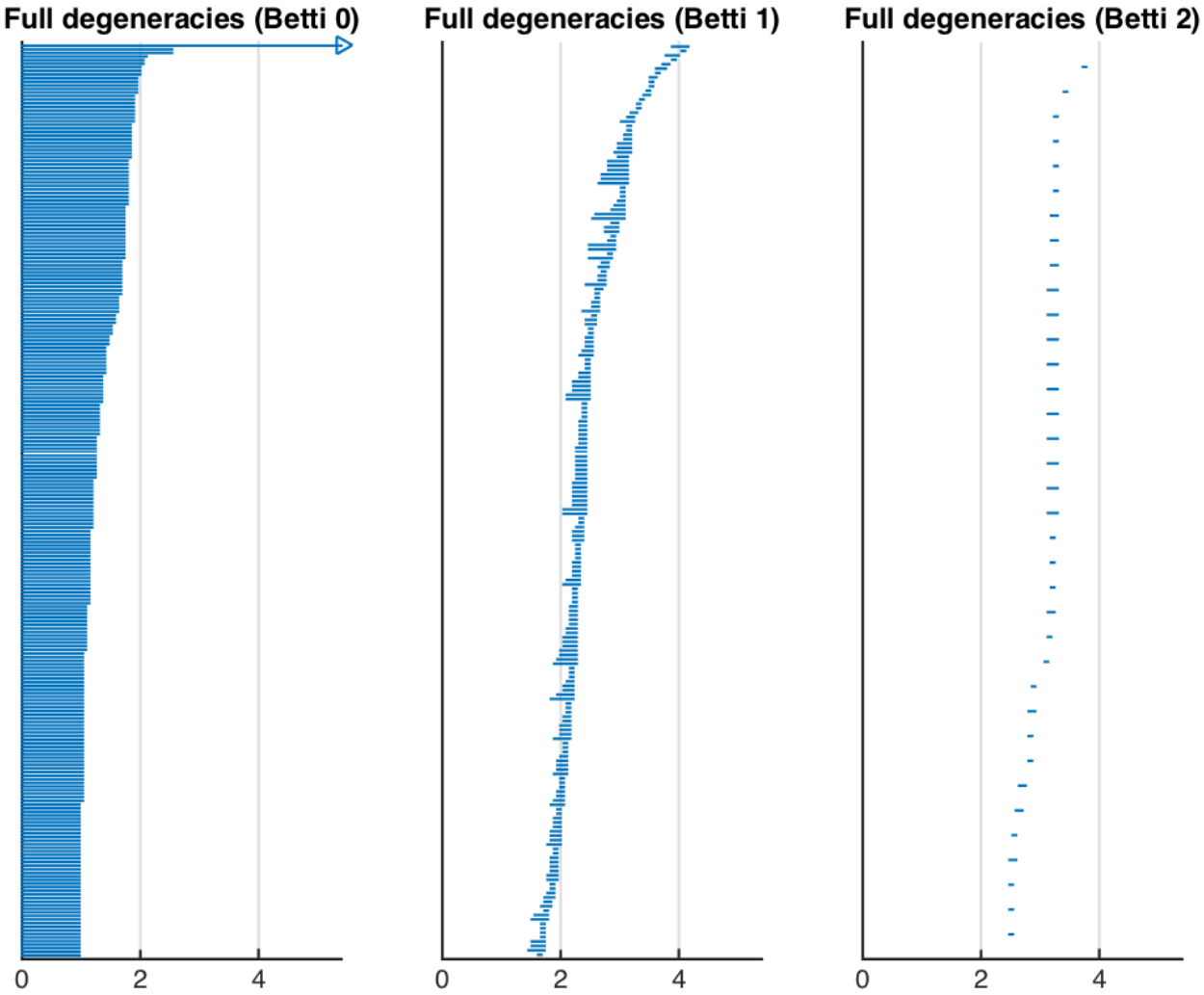}
\end{minipage}
\qquad
\begin{minipage}[b]{0.45\textwidth}\centering
\includegraphics[width=1\textwidth]{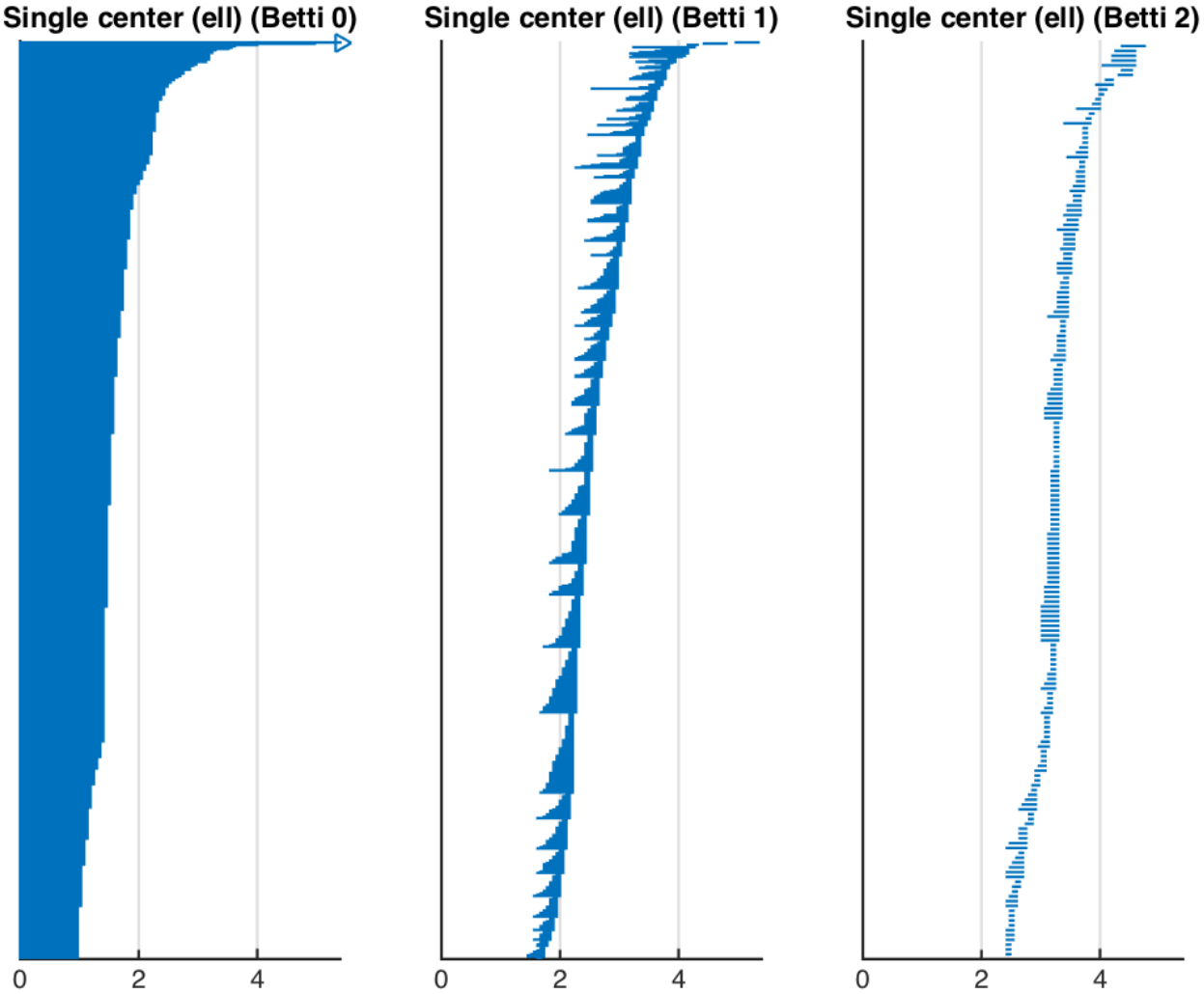}
\end{minipage}
\caption{Barcodes for the spectrum of $\mathcal{N}=4$ dyons, the point cloud has the same form as in Fig.~\ref{dyons1}. \textit{Left.} The degeneracies include the contribution of two-centered states, and are obtained from the expansion of $\psi_m$. The point cloud contains 254 BPS states and the persistent homology computation runs over a total of 255260 simplices. \textit{Right.} Single centered degeneracies, but now relaxing the condition $0 \le l \le m$, that is including modular elliptic images. The point cloud contains 707 states and the homological computation runs over 1318999 simplices.}\label{dyons2}
\end{figure}

We have computed the barcodes for the coefficients of $\psi_m^F$ and $\psi_m$ for $m=1,2,3,4$, up to a~certain value of $n$. We want to understand the impact on the degeneracies of subtracting the two centered contribution. We have also computed the persistent homology for $\psi_m^F$ relaxing the condition $0 \le l \le m$ to see explicitly the constraint imposed by elliptic invariance on the data set. The results are presented in Figs.~\ref{dyons1} and \ref{dyons2}. Let us now discuss them.

Consider Fig.~\ref{dyons1}. The point cloud is constructed out of degeneracies $d (n,l,m)$ of microstates of single centered black holes, obtained from subtracting the polar part from $\psi_m$. Differently from all the cases we have seen in the previous sections, the persistent homology of $\mathcal{N}=4$ single centered black hole degeneracies has non-trivial classes in degree 2. These are not much long-lived, albeit their lifespan is comparable with those we see in the classes for the first Betti number. These classes represent 2-cycles regularly distributed within the point cloud. One interesting feature is that their lifespans are mostly distributed around the same scale $\epsilon \sim 3$. Around this scale a number of 2-cycles appear simultaneously and with the same size, since all the lifespans have comparable lengths. This pattern is somewhat reminiscent of the ``voids'' found in the distribution of flux vacua in rigid Calabi--Yaus in \cite{Denef:2004ze}. In those class of type IIB string flux compactification the only complex structure parameter in the flux superpotential is the axion-dilaton $\tau_{\rm ad}$. When one studies the distribution of minima of the flux superpotential in a fundamental domain of $\mathrm{SL}(2 ; \mathbb{Z})$ in the $\tau_{\rm ad}$-plane, one discovers empty regions of the form of circles of various sizes, with a big degeneracy of vacua at the center. This pattern is due to the~$\mathrm{SL}(2 ; \mathbb{Z})$ symmetry. It is natural to suspect that the similar structure we see in Fig.~\ref{dyons1} is due to the (mock) modularity of the degeneracy partition function. Partial evidence comes from a~persistent homology study of the flux vacua of~\cite{Denef:2004ze}, discussed in~\cite{TDAvacua}, where a similar pattern in the persistent homology is reproduced.

In Fig.~\ref{dyons2} on the left, we show the results of the same computation, now done including also two-centered black holes. The generating function is now the full $\psi_m$ and not $\psi^F_m$. The barcode distributions are essentially identical to those in Fig.~\ref{dyons1} (again we stress that the actual numbers in the point cloud are in general different). In other words the presence of two-centered solutions has no effect whatsoever at the level of persistent homology. One should note however that the expansions we are using are valid in a region of the moduli space corresponding to the attractor region. It is therefore natural that the contribution of single centered black holes is dominating. The situation in other regions of the moduli space might be rather different.

Finally Fig.~\ref{dyons2} on the right, shows again the persistent homology of single centered black holes, but relaxing the condition $0 \le l \le m$. We are therefore computing the coefficients of $\psi_m^F$ this time including the images under elliptic transformations. Perhaps unsurprisingly the barcode distribution is again basically the same, where however every feature is more pronounced.\footnote{In the comparison one should take into account that this point cloud contains many more BPS invariants that in the previous cases, which alters the precise position at which features appear and disappear. Also since the figure has to accommodate all the degree zero homology classes provided by the initial points, many features are pushed up. For example the bump at around $\epsilon \sim 2$ of Fig.~\ref{dyons1} in degree zero, is now visible in Fig.~\ref{dyons2} in the first upper third of the frame.} Here we see at the level of topology the structure of a (mock) modular form: the barcodes give a visual representation of the mechanism which connects various coefficients via elliptic transformations.

\section{Mathieu moonshine} \label{mathieu}

As a final example we will apply our techniques to study the coefficients of certain number theoretical functions which play an important role in string theory compactifications on~K3 geo\-metries. Mathieu moonshine is based on the observation of Eguchi, Ooguri and Tachikawa~\cite{Eguchi:2010ej} that the coefficients of the Fourier expansion of the elliptic genus of K3 can be written as sum of dimensions of the irreducible representation of the Mathieu group $\mathbb{M}_{24}$. This observation is a~generalization of the monstrous moonshine, where the fact that sums of the dimensions of the irreducible representations of the monster group appear as coefficients in the $J$-function, has an explicit realization via certain modules in bosonic conformal field theory. These facts have been reviewed in \cite{Cheng:2012uy,Duncan:2014vfa,Gaberdiel:2012um}, to which we refer the reader for further references, and whose presentation we will follow. We shall also assume that the relevant conjecture formulated in the moonshine literature are true.

In the case at hand the monster group, the largest sporadic group, is replaced by the Mathieu group $\mathbb{M}_{24}$, which is a subgroup of the group of permutations with 24 elements which preserves a certain fixed set, known as the extended binary Golay code. The conformal field theory is a~$\mathcal{N}=4$ superconformal sigma model with target space K3. The spectrum of supersymmetric string states is encoded in the elliptic genus
\begin{gather} \label{ell}
\mathscr{E} (\tau , z ) = \tr_{\mathcal{H}_{RR}} \left( (-1)^{J_0 + \overline{J}_0} y^{J_0} q^{L_0-\frac{c}{24}} \overline{q}^{\overline{L}_0 - \frac{\overline{c}}{24}} \right),
\end{gather}
where $q = \e^{2 \pi \ii \tau}$ and $y=\e^{2 \pi \ii z}$, and $L_0$, $\overline{L}_0$, $J_0$ and $\overline{J}_0$ are the zero modes of the corresponding operators in the superconformal algebra. The trace is taken over the Ramond--Ramond sector and only the right moving ground states contribute. As a result the elliptic genus is independent of $\overline{\tau}$. Such states form a subspace of $\mathcal{H}_{RR}$ which decomposes according to the representations of the $\mathcal{N}=4$ supersymmetry algebra. As a consequence the elliptic genus can be decomposed as
\begin{gather} \label{elldecompos}
\mathscr{E} (\tau , z ) = 20 \mathcal{H}_{\frac14,0} (\tau, z) - 2 \mathcal{H}_{\frac14,\frac12} + \sum_{n=1}^\infty A_n \mathcal{H}_{\frac14+n,\frac12} (\tau, z) ,
\end{gather}
where
\begin{gather*}
\mathcal{H}_{h,j} (\tau , z) = \tr_{\mathcal{H}_{h,j}} \left( (-1)^{J_0} y^{J_0} q^{L_0 - \frac{c}{24}} \right)
\end{gather*}
are the characters over the irreducible representation spaces $\mathcal{H}_{h,j}$ labeled by the conformal dimension $h$ and the $\mathfrak{su} (2)$ spin $j$. In particular the $A_n$ are dimension of representations~$R_n$ of~$\mathbb{M}_{24}$ (the same is also true for the first two coefficients interpreted as virtual representations).

Of course this construction is closely related to the situation of Section~\ref{dyons}. In that case the partition function of $\mathcal{N}=4$ dyons in a string compactification on $\mathrm{K3} \times \torus^2$ is proportional to the generating function of the elliptic genera of the symmetric products of K3s, via~\eqref{defpsim}. In this section we will focus on elliptic genera from a different perspective.

The elliptic genus \eqref{ell} has an interesting equivariant (or twisted) generalization. Given an element $g \in \mathbb{M}_{24}$, we have
\begin{gather} \label{ellg}
\mathscr{E}_g (\tau , z ) = \tr_{\mathcal{H}_{RR}}\left( g (-1)^{J_0 + \overline{J}_0} y^{J_0} q^{L_0-\frac{c}{24}} \overline{q}^{\overline{L}_0 - \frac{\overline{c}}{24}}\right).
\end{gather}
In this case a version of \eqref{elldecompos} holds, where the dimension of the representations are replaced by $\tr_{R_n} g $ (e similarly for the virtual representations).\footnote{We have been a bit imprecise with the definitions since K3 sigma models have a moduli space. Implicitly we are always choosing a point in the CFT moduli space where the subgroup of $\mathbb{M}_{24}$ generated by $g$ is a symmetry of the Hilbert space.} In particular we can rewrite \eqref{ellg} as
\begin{gather*}
\mathscr{E}_g (\tau , z) = \frac{\theta_1^2 (\tau , z)}{\eta^3 (\tau)} \left( \chi_g \mu (\tau , z) + H_g (\tau) \right) ,
\end{gather*}
where $\chi_g = \tr_R g$ is the character of the defining representation $R_n$ of $\mathbb{M}_{24}$ and
\begin{gather*}
\mu (\tau ,z ) = \frac{\ii y^{1/2}}{\theta_1 (\tau , z)} \sum_{n \in \mathbb{Z}} (-1)^n \frac{y^n q^{n (n+1)/2}}{1-y q^n} ,
\end{gather*}
is an Appell--Lerch sum. The Mathieu McKay--Thompson series $H_g (\tau)$ are weight $\frac12$ mock modular forms, with shadow $\chi_g \eta(\tau)^3$, and can be expanded as
\begin{gather} \label{mockHg}
H_g (\tau) = q^{-\frac18} \left( -2 + \sum_{n=1}^{\infty} q^n \tr_{K_n} g \right) ,
\end{gather}
where $K = \bigoplus_{n=1}^{\infty} K_n$ is an infinite dimensional $\mathbb{M}_{24}$ module \cite{Cheng:2010pq,Eguchi:2010fg,Gaberdiel:2010ca,Gaberdiel:2010ch}. When $g = \id$, \eqref{mockHg} is the generating function of the degeneracies of massive irreducible representations of the worldsheet $\mathcal{N}=4$ conformal algebra, as they appear in \eqref{ell} and \eqref{elldecompos}.

One can see that $\mathscr{E}_g (\tau , z)$ only depends on the conjugacy class $[g]$ of $g \in \mathbb{M}_{24}$. The conjugacy classes lead to 21 distinct elliptic genera; they are labelled by $g= $1A, 2A, 2B, 3A, 3B, 4A, 4B, 4C, 5A, 6A, 6B, 7AB, 8A, 10A, 11A, 12A, 12B, 14AB, 15AB, 21AB, 23AB. The notation is standard and used in \cite{Cheng:2012uy}. For each conjugacy class $[g]$, Table 2 of \cite{Cheng:2012uy} gives $\chi_g$ and the non universal ingredients needed to construct the function $H_g (\tau)$.

The equivariant elliptic genus has the Fourier expansion
\begin{gather} \label{EquivEll}
\mathscr{E}_g (\tau , z) = \sum_{n \ge 0 , l \in \mathbb{Z}} c_g ( n , l) q^n y^l .
\end{gather}
Due to modularity one has $c_g (n,l) = c_g (4n-l)$. These coefficients satisfy $c_g (k) = \mathrm{str}_{\hat{K}_k} g$, where $\hat{K}_k$ is the degree $k$ component of a certain infinite dimensional $\mathbb{Z}$-graded $\mathbb{M}_{24}$ module, obtained from $K$, and whose precise form is known but not relevant for the present discussion. The coefficients $c_g (n)$ thus have a precise relation to the irreducible representations of $\mathbb{M}_{24}$. From the determination of the exact coefficients, their distribution and asymptotic formulae, one gathers essential informations about the representation theory of $\mathbb{M}_{24}$ and its realization as a vertex operator algebra.

Now we want to address the problem from a different angle and use our formalism to understand topological features of the distributions of the coefficients of the equivariant elliptic genera $\mathscr{E}_g (\tau , z)$.

We construct a point cloud $\mathsf{X}^{g}$ whose elements have the form $\mathsf{x} = (\log \vert c_g (n,l) \vert , n , l)$ for each conjugacy class associated to an element $g \in \mathbb{M}_{24}$. Using the results of \cite{Cheng:2012uy} we construct explicit\-ly~$\mathscr{E}_g (\tau , z)$ and compute the first non-vanishing coefficients up to $q^{20}$. Again the coefficients are growing very fast, and the logarithm of the degeneracies makes it easier to capture the essential features using persistent topology. In all cases the point cloud consists of 259 points. For each~$[g]$ listed as above, we construct the $\mathbb{N}$-persistence modules $H_i ( \mathsf{VR}_\epsilon (\mathsf{X}^{g}) ; \mathbb{Z}_2 )$ and compute their barcodes.

We collect all the results in Appendix \ref{MoonshineBar}. All the conjugacy classes present similar topological features. It is quite likely that the explanation for this fact is modularity, which as we have seen in the previous section produces strong constraints on the allowed persistent homology classes. We stress that we are not suggesting any relation whatsoever between the coefficients of different elliptic genera; apparently modularity has a subtle way of showing up in the topological characterization of point clouds. It would be very interesting to pursue this issue further, if one general grounds given the information of modularity one can predict persistent features in the distribution of barcodes. Also note that the persistence modules as discussed in Appendix~\ref{MoonshineBar}, have only non trivial~$H_0$ and~$H_1$. On the other hand we have seen in Section~\ref{dyons} that a non trivial~$H_2$ was present and we interpreted this fact as a consequence of modular invariance. Note however that in Section~\ref{dyons}, the point cloud was constructed out of the black hole degeneracies, related to the coefficients of the elliptic genus by a multiplicative lift. To properly compare the effects of modularity in the two cases, we should consider the multiplicative lift of the equivariant elliptic genera~\eqref{EquivEll}. The idea of investigating a precise relation between modularity and topological features is very interesting, but we shall leave it to the future.

Taking at face value that modularity affects the persistent homology groups, there are still interesting more subtle feature in the barcode spectra. Indeed, even at the qualitative level, the barcode distributions associated with the elliptic genera $\mathscr{E}_g (\tau , z)$, fall into two families.

The first family contains the conjugacy classes: 2B, 3B, 4A, 4C, 6B, 10A, 12A, 12B, 21AB; the second family the remaining classes. What singles out the first class is that, despite having a similar form, the barcodes for the zeroth and first homology are consistently more short-lived than those in the second family. In the Betti number zero distribution, the persistent classes all die for values of the proximity parameter $\epsilon \sim 2$. In degree one the features are less clear-cut, the barcode distribution tends to be centered around smaller values of $\epsilon$, and most persistent classes disappear before $\epsilon \sim 3$. Another striking feature is the number of simplices involved in the computation of persistent homology, which is not apparent just by looking at the figures in Appendix~\ref{MoonshineBar}. All the classes in the second family, namely 1A, 2A, 3A, 4B, 5A, 6A, 7AB, 8A, 11A, 14AB, 15AB and 23AB generate a number of simplices roughly comparable, around 65000--68000. On the other hand the number of simplices generated in the Vietoris--Rips complex for the first family fluctuates wildly, roughly in the range 38000--135000. This indications are very qualitative, but from the behavior of the persistent homology it is natural to wonder if there is any particular property that distinguish these families.

Indeed there is. The above classes in the first family are precisely the only conjugacy classes in $\mathbb{M}_{24}$ which do not have a representative in the subgroup $\mathbb{M}_{23}$ of $\mathbb{M}_{24}$. Some classes of $\mathbb{M}_{23}$ are singled out by Mukai's theorem, which states that given a finite subgroup of the group of symplectic automorphisms of a~K3 surface, it can be embedded into $\mathbb{M}_{23}$. The other classes in~$\mathbb{M}_{23}$ do not have a geometrical interpretation.

At the technical level there is a simple explanation of the appearance of these two families. The above classes in the first family all appear to have $\chi_g = 0$, which therefore acts as a relative shift of the coefficients with respect to the elliptic genera of the second family. What is however interesting here is that, with a bit of hindsight, such behavior could have been predicted just by looking at the equivariant elliptic genera, for example computed via conformal field theory, without any knowledge of the representation theory of Mathieu's groups. At a very qualitative level such features are captured by persistent homology.

\section{Discussion} \label{discussion}

In this paper we have taken the perspective of persistent homology to analyze certain enumerative BPS invariants which arise in some physical and number theoretical problems. The focus of this paper has been the comparison between the topological features which characterize different distributions. We have done so in different ways, by studying the same theory in different limits or chambers, or by studying different theories in a similar situation. The underlying theme has been that the distributions of supersymmetric states are rich in topological information, which often can be clearly traced back to physical properties of the system under consideration. This information is qualitative, and concerns the overall structure of the space of supersymmetric states. In a sense it is a measure of its topological complexity.

On a more practical level, this note has taken to the task of exemplifying the uses of new methods based on persistent homology to study physical problems in string and field theory; as well as calling the attention of the computational topology community on the wealth of enumerative and number theoretical datasets which arise in string/field theories. From a physical perspective, to have a meaningful enumerative problem it is necessary to resort to protected quantities, in this case supersymmetric states.

In the first part of the paper we have applied the tools of topological data analysis to the wall-crossing phenomenon in $\mathcal{N}=2$ theories. The indices of BPS protected quantities jump at walls of marginal stability and the effect of the jump can be in principle computed using a~wall-crossing formula. In Sections \ref{classS} and \ref{conifold} we have studied how the topological properties of the BPS spectra change upon crossing certain walls. We have done so for a quantum field theory, namely ${\rm SU}(3)$ $\mathcal{N}=2$ pure Yang--Mills, which is the simple example of a theory containing wild chambers, and for string theory on the conifold. Both theories exhibit physically interesting phenomena and we have shown their impact on the persistent homology of the BPS spectra. In~${\rm SU}(3)$ Yang--Mills we have analyzed the difference at the topological level between the spectra in the weak coupling and wild chamber, where there is an exponential growth of states and higher spin particles appear. At the level of the barcodes the transition is striking, going from an ordered simple pattern to an irregular (``wild'') distribution. This provides a qualitative measure of the complexity of the spectrum in wild chambers. In the case of the conifold the transitions are less apparent. In this case the physical features, for example the fact that physical states are realized as very different objects in the large radius geometrical chamber or in the noncommutative crepant resolution chamber, or the presence of conjugation walls where the core of the bound states changes, appear as rather small fluctuations in the BPS barcodes. On the other hand the conifold is known to be a very simple geometry, where all the BPS generating functions are known in closed form. It would be interesting to carry on this analysis for more complicated non-compact threefolds.

In this context the main result that we have obtained is the explicit verification that in certain cases one can indeed distinguish chambers by their topological features. To fully understand how general is this result, and if one can turn it into a more quantitative statement, one should undertake a systematic study of the distributions of BPS states over all chambers in several models, both in field and in string theory. Practically this consists in starting from a chamber where the spectrum is known, if available, and then using repeatedly the wall-crossing formula until all the moduli space of vacua is covered. Repeating this for several theories will generate a~large amount of data. Then a natural direction for this research program is to try to classify the typology of chambers by the topological features of the BPS distributions. Indeed we have seen examples in Section~\ref{classS} and~\ref{conifold} where the topological features between different chambers differ drastically or are almost identical. However to make this ideas more concrete one should devise more quantitative methods to compare barcodes.

\looseness=-1 In Sections \ref{qgeom}, \ref{dyons} and \ref{mathieu} we have considered BPS spectra in compact Calabi--Yaus. Interestingly the series of Donaldson--Thomas invariants corresponding to black hole microstates in the large radius limit of one parameter compact threefolds, all exhibit very similar persistent features. We interpret this fact as another sense in which the microstate counting of black holes is universal. Note that in general when talking about black hole universality, one refers to the large charge limit, where the supergravity approximation is reliable. In this limit the entropy obeys the area law and the subleading corrections are expressed in terms of certain higher derivative terms \cite{LopesCardoso:1998wt}. Here we are looking at the first few hundreds of microstates; these data are enough to extrapolate the large order behavior, which agrees with the supergravity expectations \cite{Huang:2007sb}. Therefore some universal behavior is somewhat expected; it is interesting to see it arise at the level of the barcodes.

Sections \ref{dyons} and \ref{mathieu} are concerned with another aspect of the enumerative BPS problems, more precisely in a few cases where the degeneracies of microstates arise as coefficients of known functions. In this cases the relevant functions have modular properties, and we have focused on the interplay between modularity and topology. Indeed we have seen experimentally in a~series of examples that modularity strongly constrains the shape of the barcode distributions. For example we have seen in Section~\ref{mathieu} that the barcodes associated with different elliptic genera present clear similarities. Note that these functions are very different, in the sense that there is no relation between the coefficients of the Fourier expansions of different elliptic genera. On the other hand their topological features are quite similar and we interpret this fact as a~consequence of modularity, although we don't have any a priori argument based on topology. Quite remarkably, the subtle differences in the barcode distributions can be explained in term of the different families of conjugacy classes of the Mathieu group $\mathbb{M}_{24}$.

We consider this note as a first step to understand the role that topological data analysis can play in string/field theory problems. Clearly much is left to be understood and more extensive computations of enumerative invariants are needed to put the results of this paper on firmer grounds. Also there are several other problems which seem to be amenable to a topological analysis as we have done in this paper. We list here a few which we are currently investigating:
\begin{itemize}\itemsep=0pt
\item It would be interesting to generalize this formalism to the study of string vacua. In particular a lot of compactifications of string theory are known, with $\mathcal{N}=2$ or $\mathcal{N}=1$ supersymmetry. The vacuum selection problem consists in the choice of one compactification, or a class thereof, over the others due to some particular features. It is natural to wonder if persistent homology has anything to say about this, if for examples the distributions of vacua with certain features have distinctive topological properties with respect to others. A first step in this direction appears in \cite{TDAvacua}.
\item A similar problem concerns the distribution of attractor points in the Calabi--Yau moduli space. These points corresponds to black holes via the attractor mechanism, and as argued in \cite{Moore:1998pn} they are deeply related to certain arithmetic aspects of string compactifications.
\item One of the original motivations of this paper was if there is any particular distinctive feature of BPS states in quantum field theory in the presence of defects. When a theory is modified by the presence of a defect, new BPS states appear, those which can bound to the defect. For example in six dimensional topological quantum field theories such a modification is related to the conjectural enumerative problem of Donaldson--Thomas invariants for moduli spaces of parabolic sheaves \cite{Cirafici:2013tna}. In four dimensions, it was shown in \cite{Cirafici:2013bha} that for theories in which a~line defect can be engineered via laminations on a~curve~\cite{Gaiotto:2010be}, line defects come in distinct families, which are generated by the action of a~cluster algebra. Each family is generated by the repeated action of a~certain sequence of cluster transformations and can contain an infinite number or a finite number of elements (including just one element). It is natural to wonder if the analysis we have performed in Section~\ref{classS} can be of any use in the classification problem for defects.
\item Persistent homology is very closed in spirit to Morse theory. It would be interesting to give a more physical description of persistent homology classes via the correspondence between Morse theory and supersymmetric quantum mechanics \cite{Witten:1982im}. Viceversa, generalization of this correspondence, for example along the lines of Floer theory, are likely to provide interesting variants of persistent homology.
\item The equivariant elliptic genera that we have discussed in Section~\ref{mathieu} are just a small set of functions which arise in the field of moonshine. It would be very interesting to generalize our formalism to other similar problems. For example, it is an open problem to understand the distribution of coefficients of the McKay--Thompson series for the monster moonshine modules. Partial results on the asymptotics are in \cite{Duncan:2014vfa} and it would be interesting to see what are the persistent features of these distributions.
\end{itemize}
We hope to report on these matters in the near future.

\appendix

\section{Moonshine barcodes} \label{MoonshineBar}

In this appendix we collect all of the barcodes computed in Section~\ref{mathieu}. Every point cloud $\mathsf{X}^g$ is constructed out of vectors of the form $\mathsf{x} = (\log |c_g (n,l)| , n , l)$ computed via the equivariant elliptic genus $\mathscr{E}_g (\tau , z)$. The labels of the conjugacy classes $[g]$ are indicated on top of each figure. Each point cloud consists of 259 states. For each conjugacy class $[g]$ we use the Vietoris--Rips complex to compute the persistence modules $H_i \left( \mathsf{VR}_\epsilon (\mathsf{X}^{g}) ; \mathbb{Z}_2 \right)$. This appendix contains all the associated barcodes distributions, each figure labelled by the conjugacy class.

\begin{figure}[th!]\centering
\begin{minipage}[b]{0.41\textwidth}\centering
\includegraphics[width=1\textwidth]{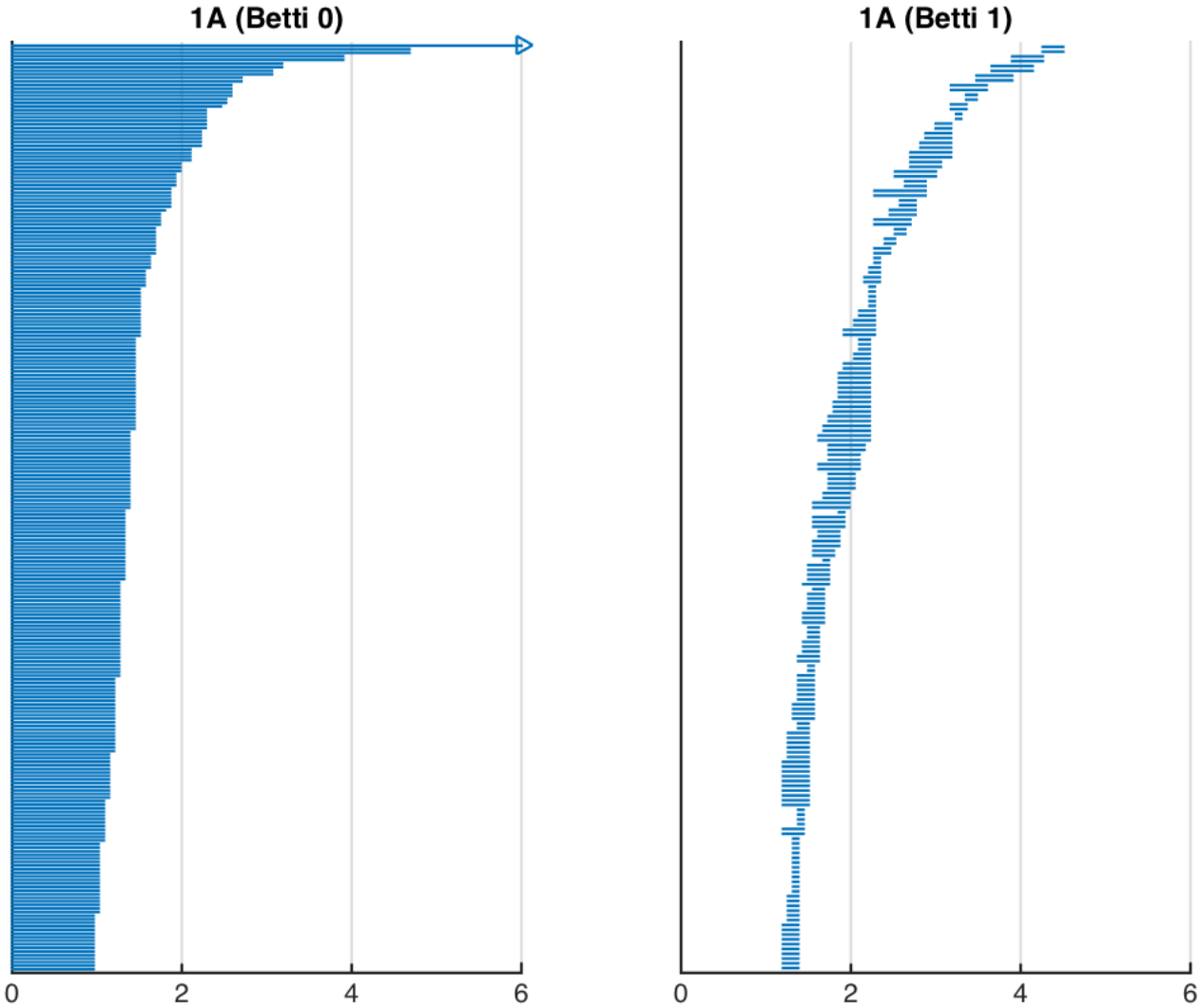}
\end{minipage}
\qquad
\begin{minipage}[b]{0.41\textwidth}\centering
\includegraphics[width=1\textwidth]{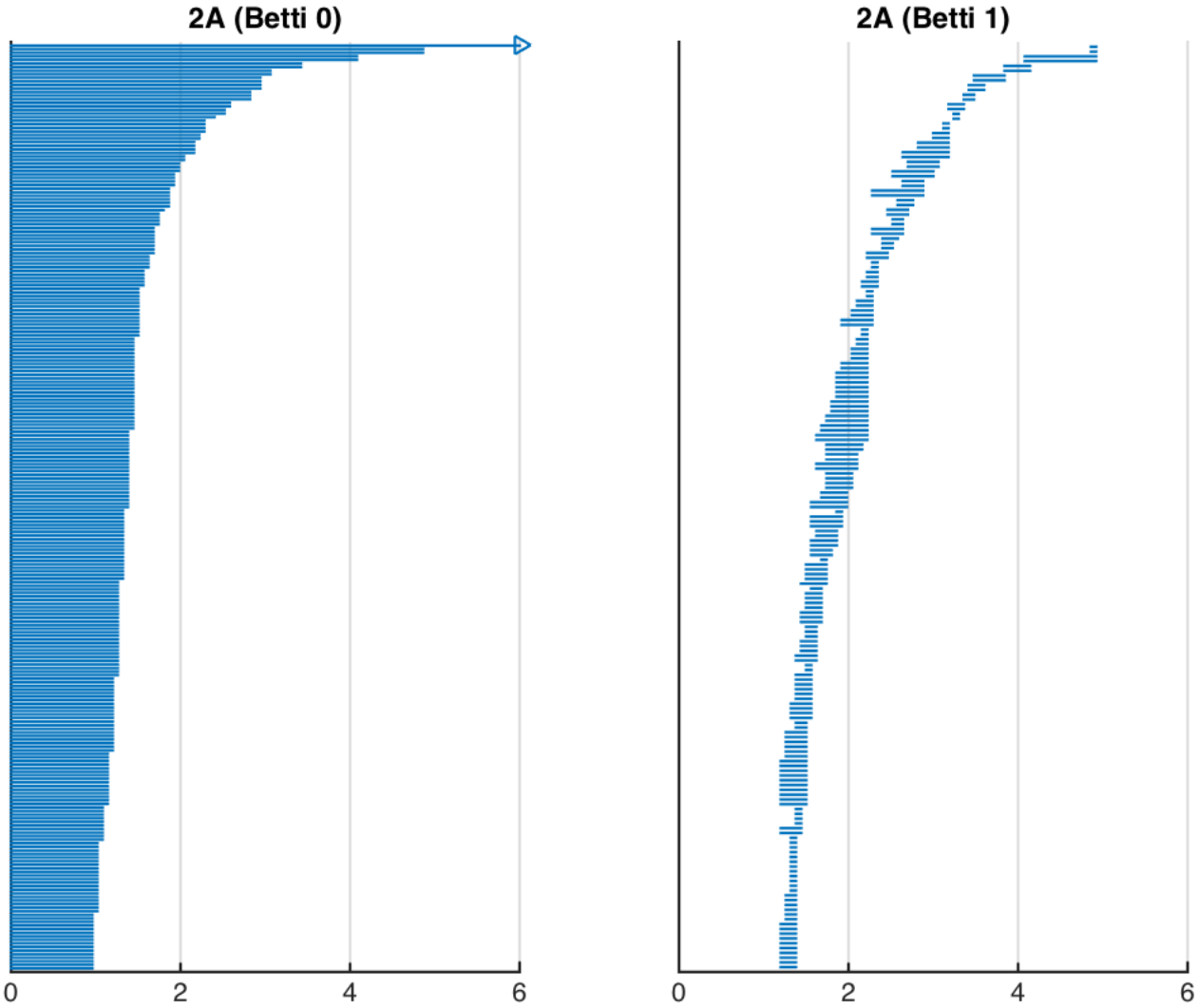}
\end{minipage}

\vspace{3mm}

\begin{minipage}[b]{0.41\textwidth}\centering
\includegraphics[width=1\textwidth]{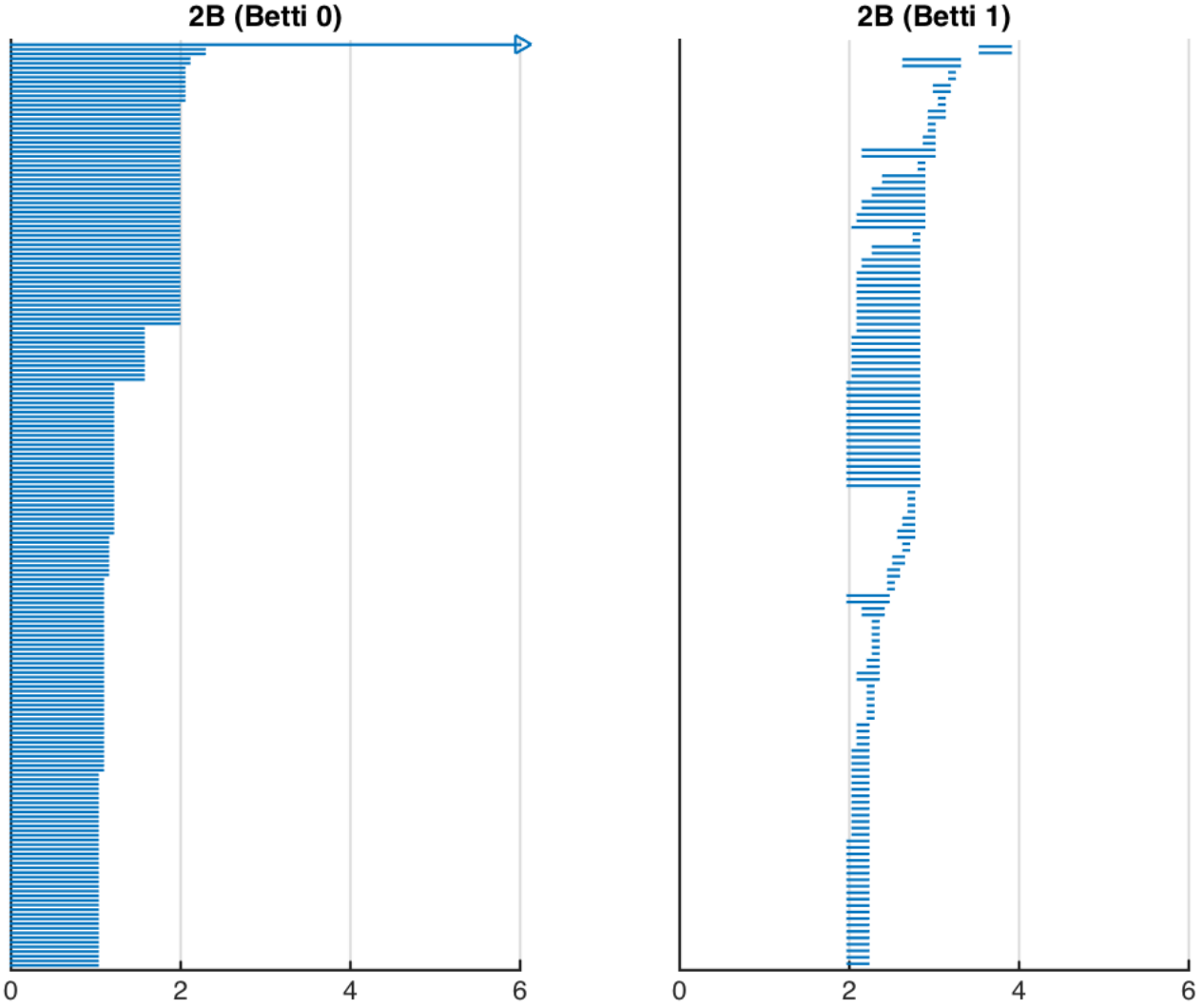}
\end{minipage}
\qquad
\begin{minipage}[b]{0.41\textwidth}\centering
\includegraphics[width=1\textwidth]{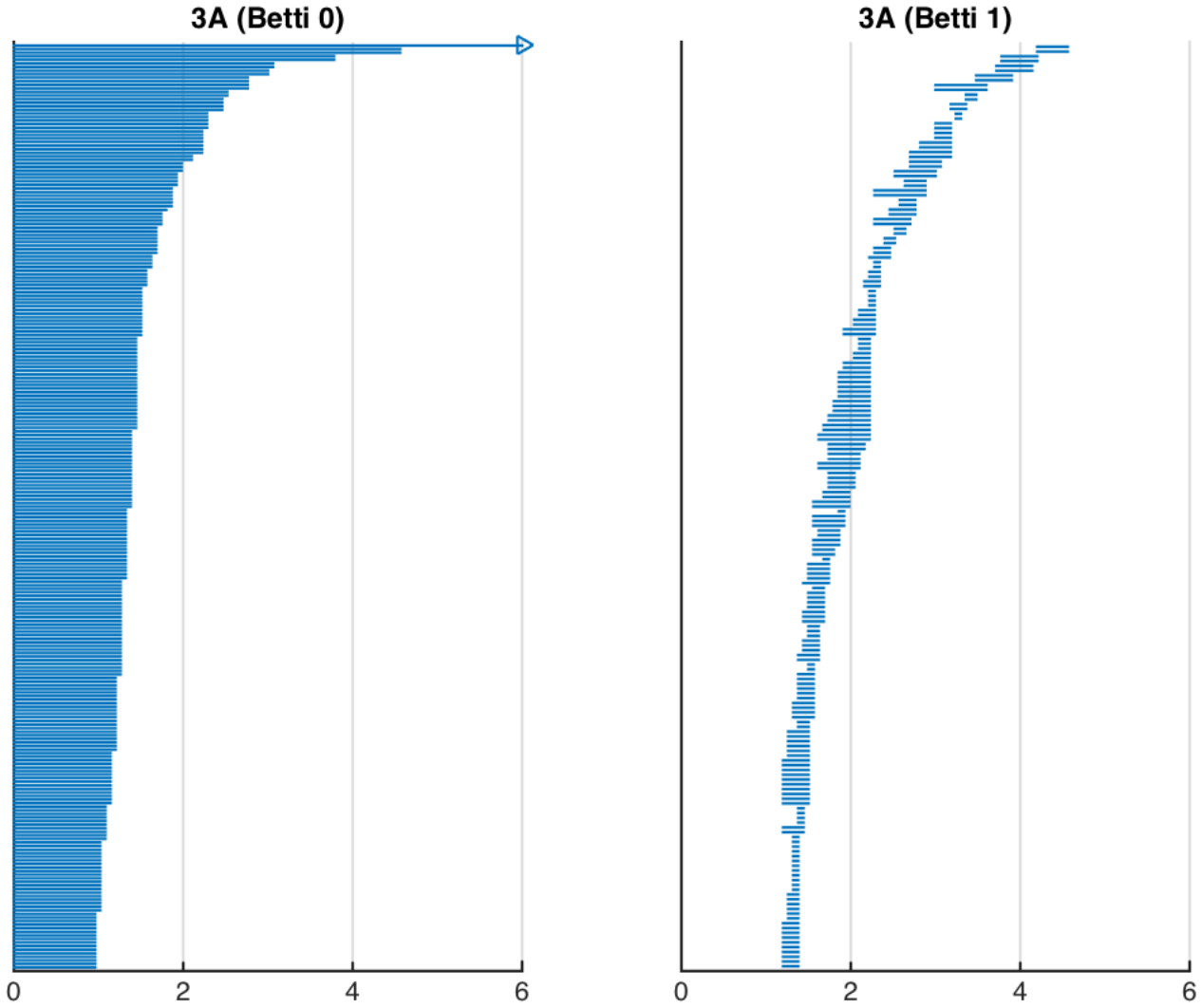}
\end{minipage}

\vspace{3mm}

\begin{minipage}[b]{0.41\textwidth}\centering
\includegraphics[width=1\textwidth]{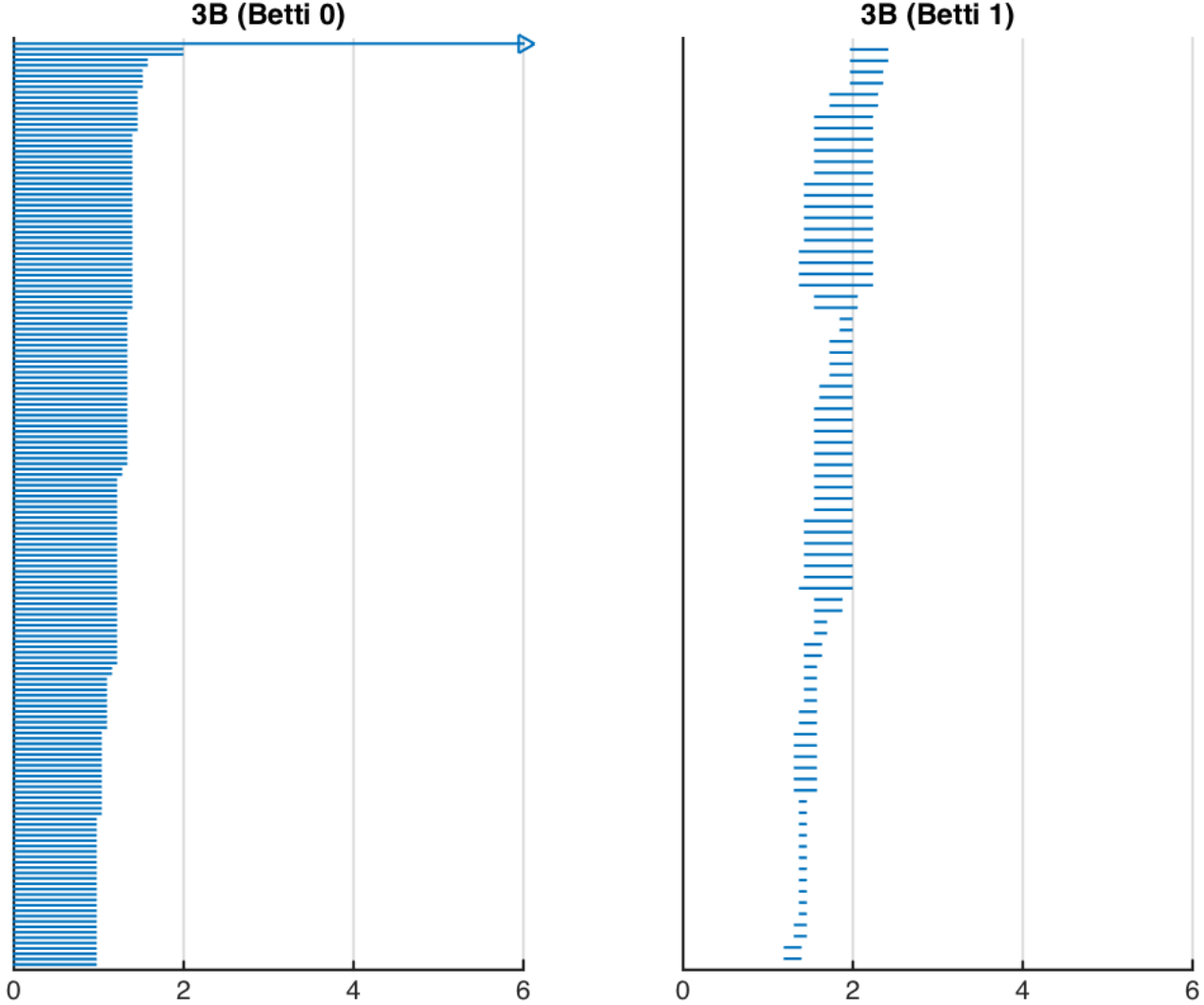}
\end{minipage}
\qquad
\begin{minipage}[b]{0.41\textwidth}\centering
\includegraphics[width=1\textwidth]{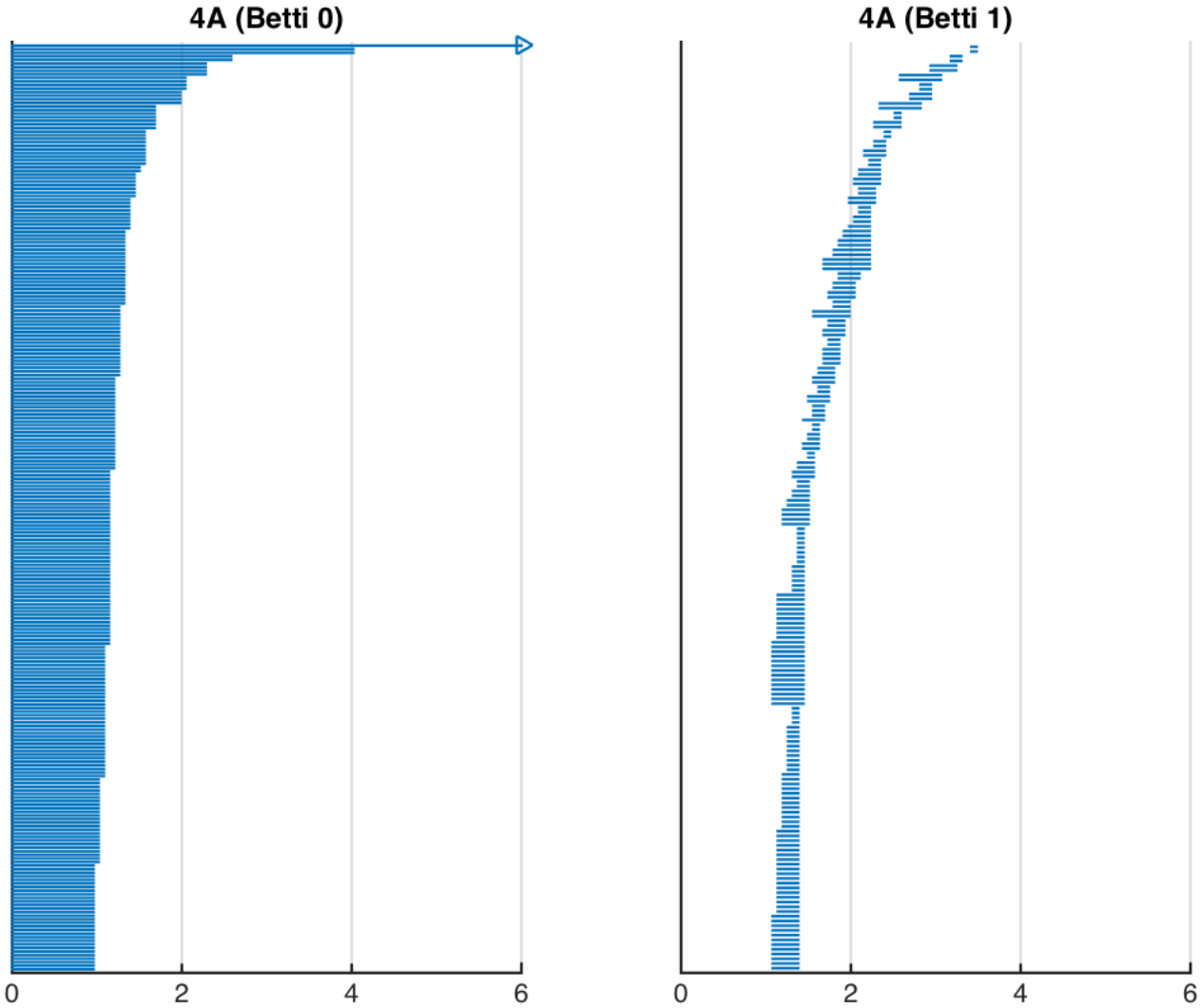}
\end{minipage}

\vspace{3mm}

\begin{minipage}[b]{0.41\textwidth}\centering
\includegraphics[width=1\textwidth]{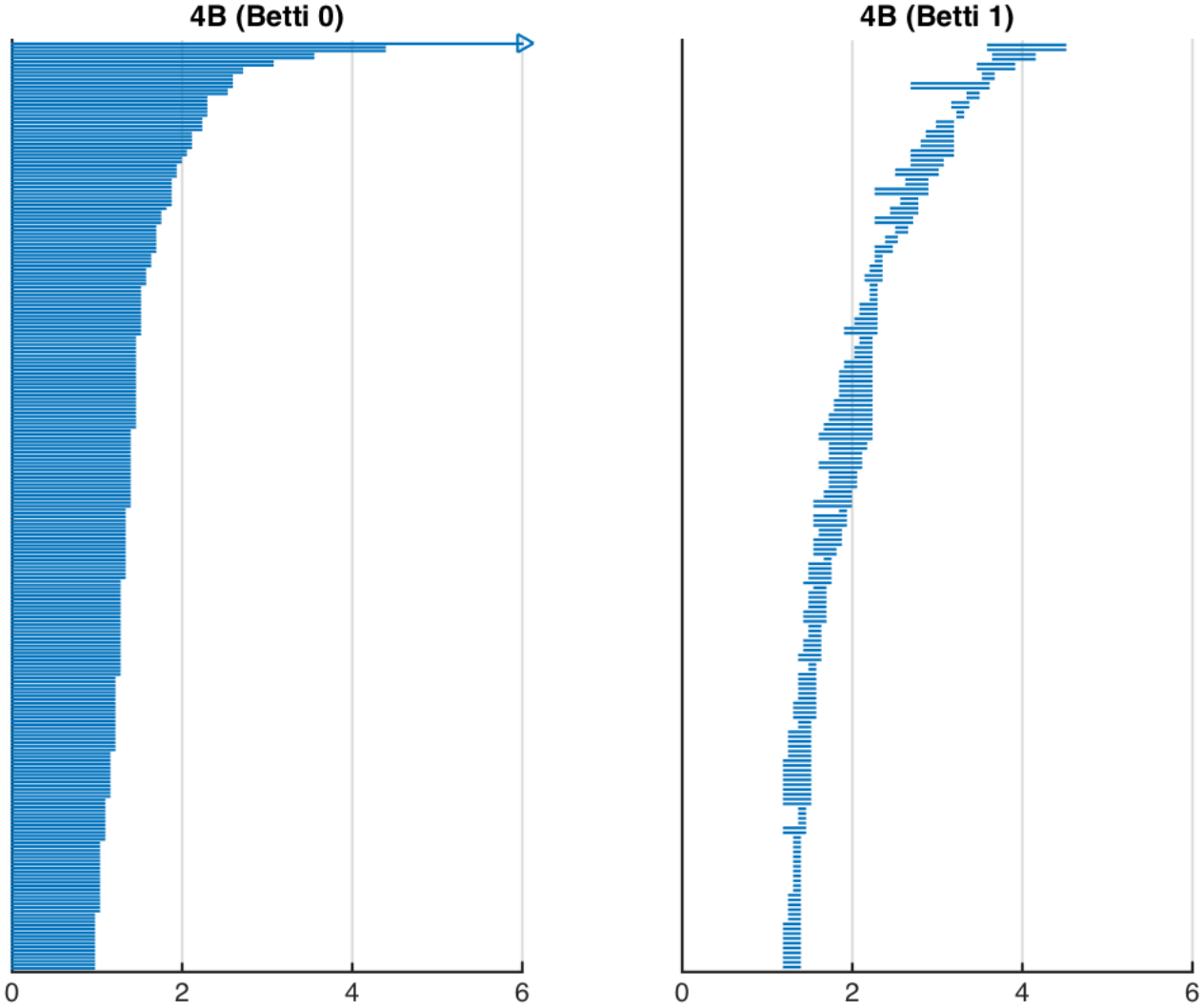}
\end{minipage}
\qquad
\begin{minipage}[b]{0.41\textwidth}\centering
\includegraphics[width=1\textwidth]{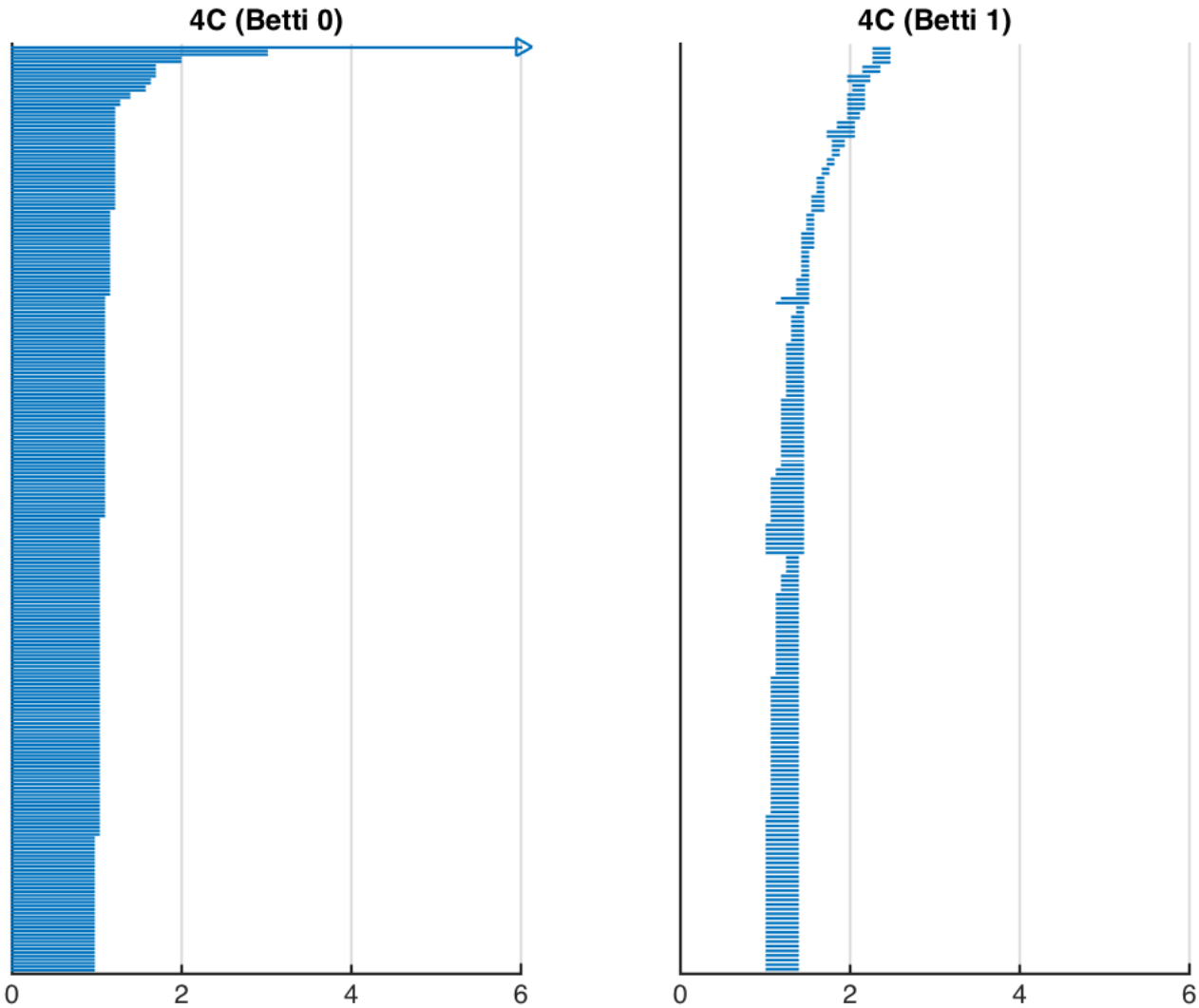}
\end{minipage}
\end{figure}

\begin{figure}[htbp]\centering
\begin{minipage}[b]{0.41\textwidth}\centering
\includegraphics[width=1\textwidth]{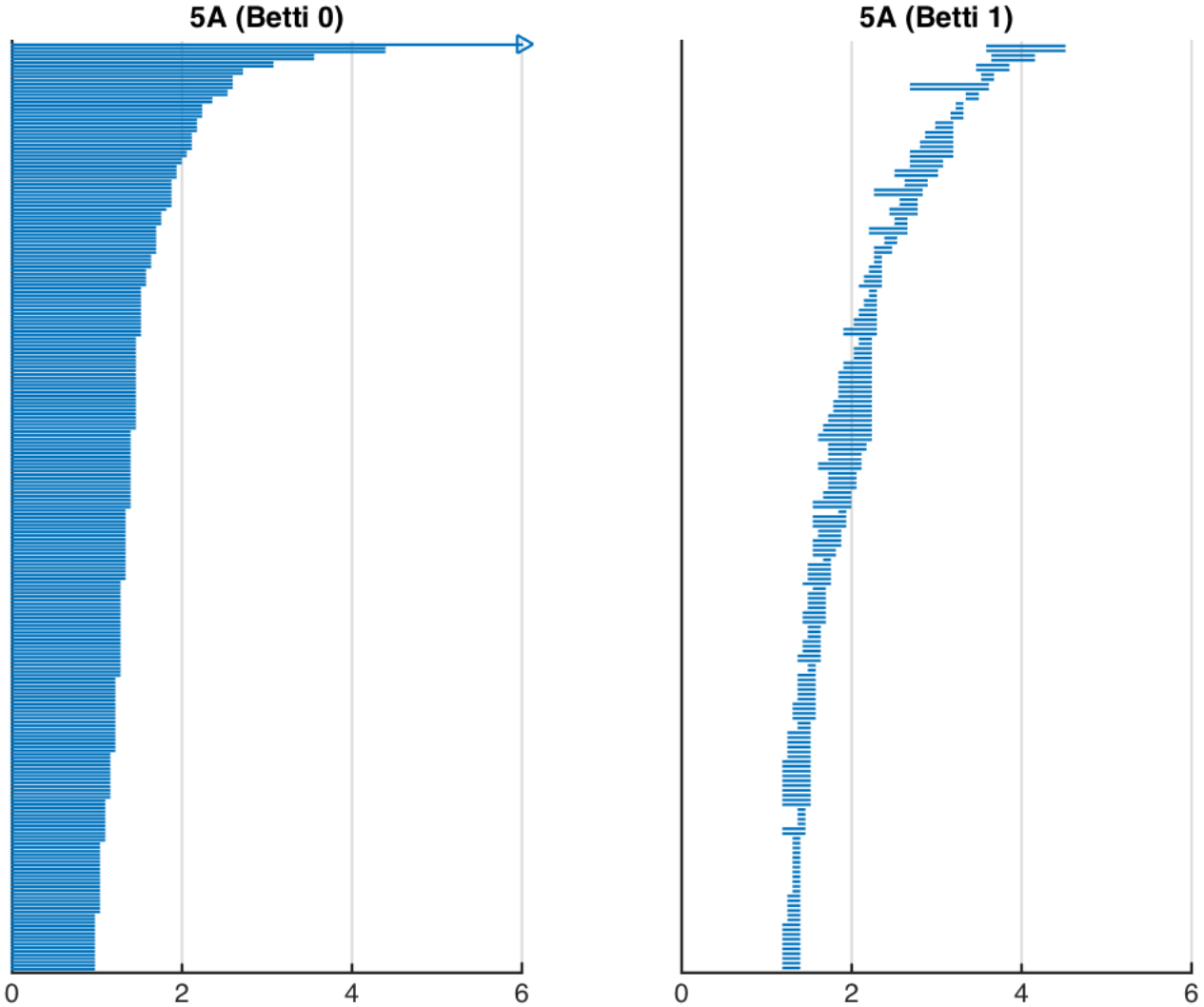}
\end{minipage}
\qquad
\begin{minipage}[b]{0.41\textwidth}\centering
\includegraphics[width=1\textwidth]{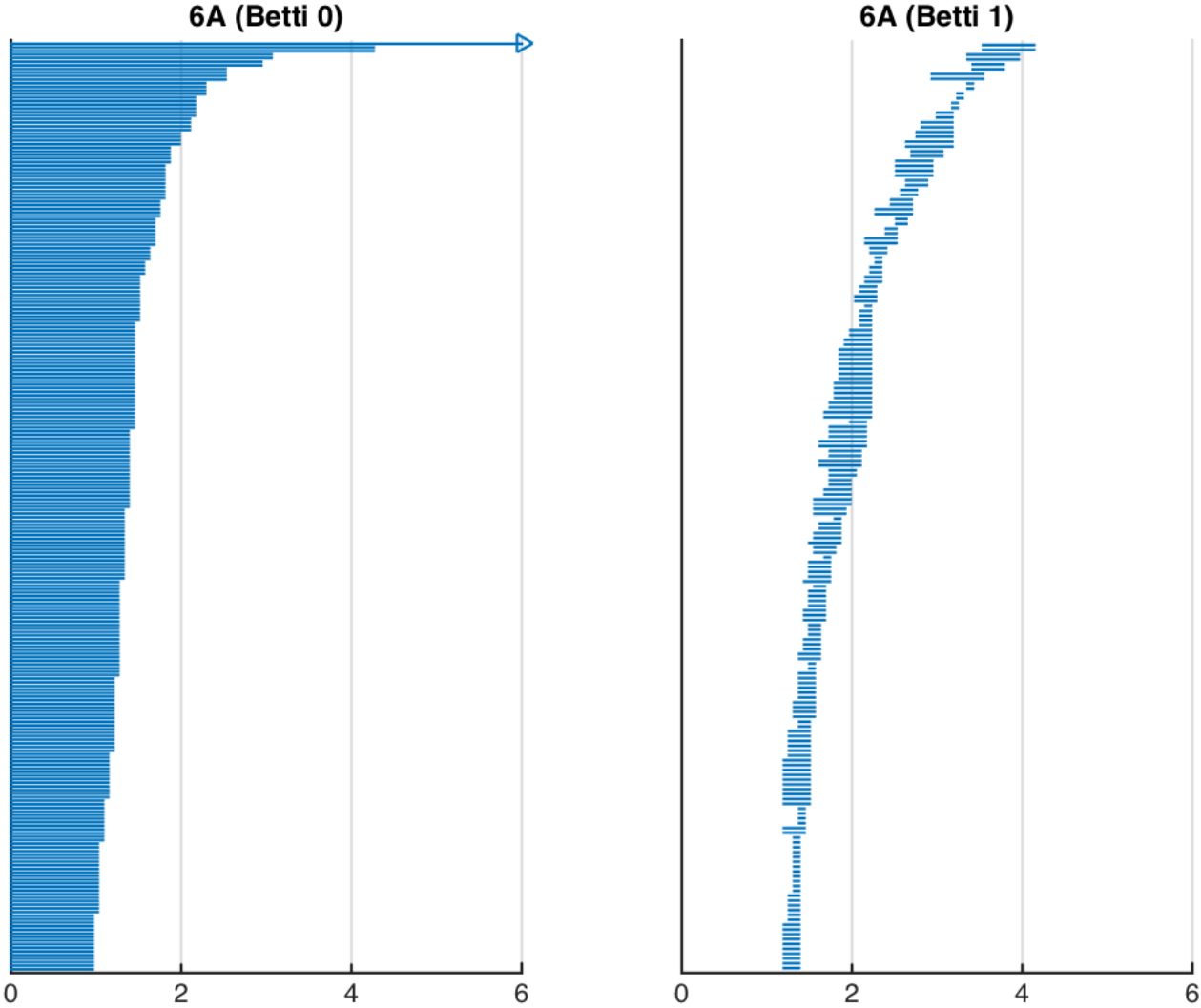}
\end{minipage}

\vspace{3mm}

\begin{minipage}[b]{0.41\textwidth}\centering
\includegraphics[width=1\textwidth]{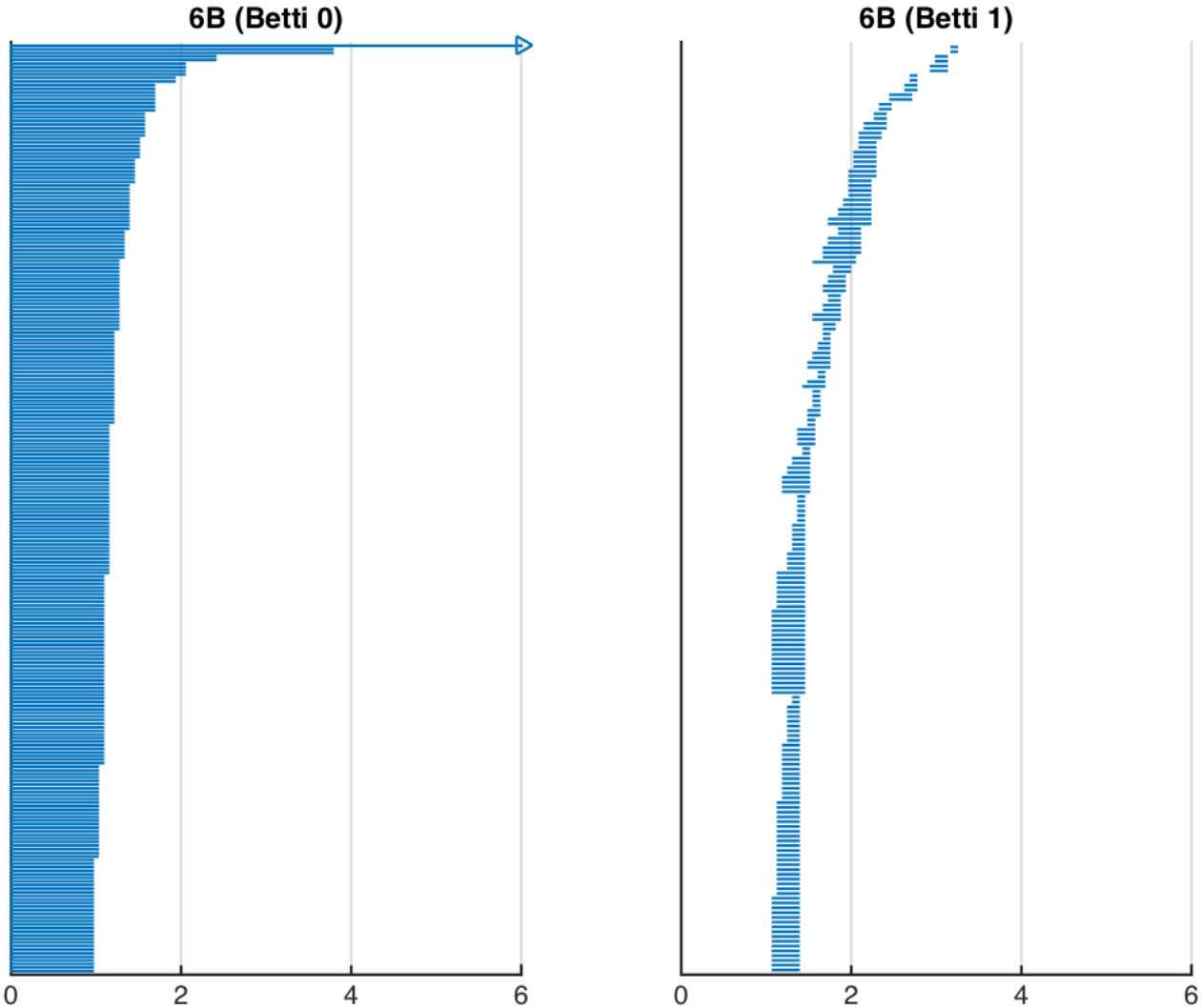}
\end{minipage}
\qquad
\begin{minipage}[b]{0.41\textwidth}\centering
\includegraphics[width=1\textwidth]{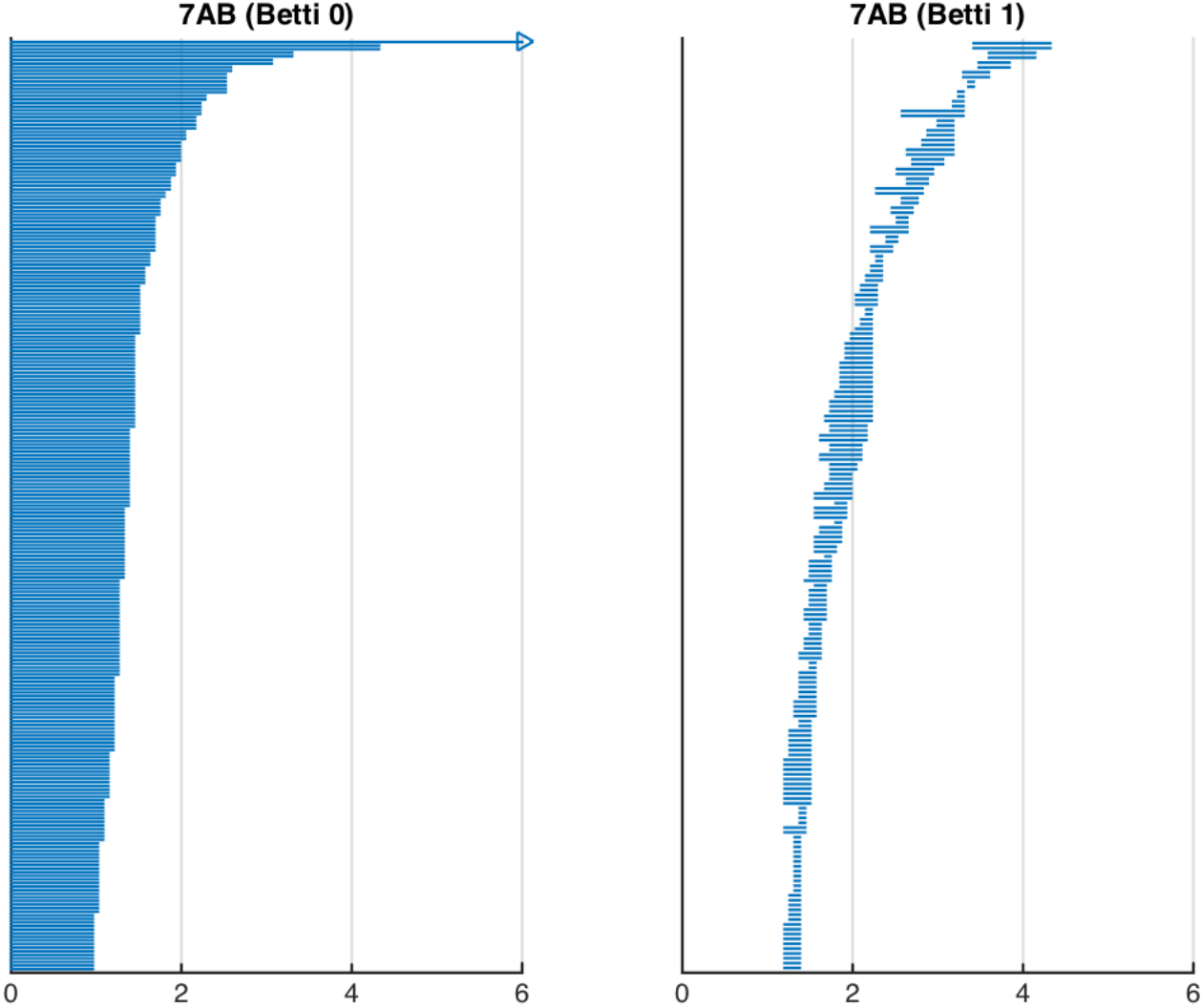}
\end{minipage}

\vspace{3mm}

\begin{minipage}[b]{0.41\textwidth}\centering
\includegraphics[width=1\textwidth]{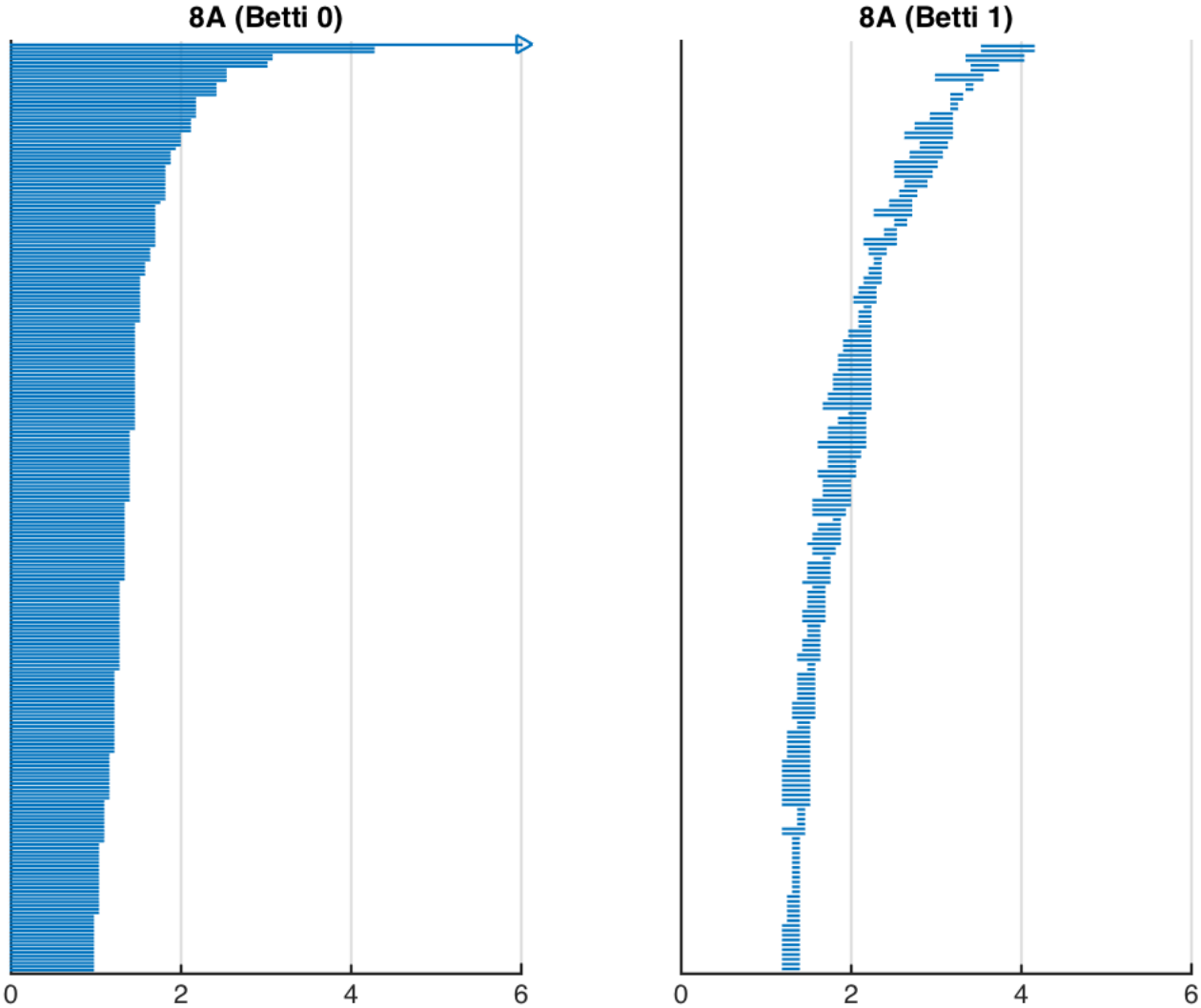}
\end{minipage}
\qquad
\begin{minipage}[b]{0.41\textwidth}\centering
\includegraphics[width=1\textwidth]{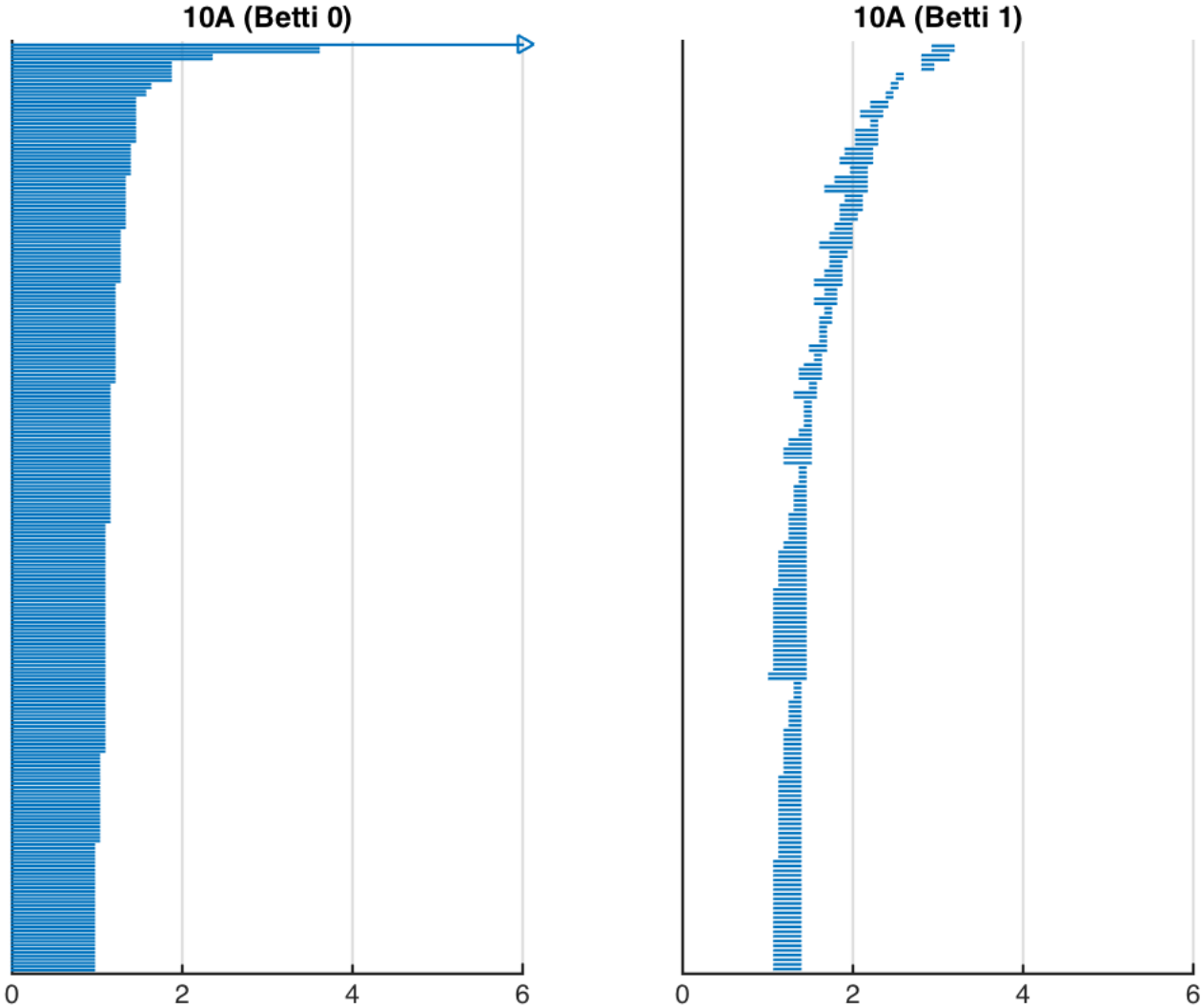}
\end{minipage}

\vspace{3mm}

\begin{minipage}[b]{0.41\textwidth}\centering
\includegraphics[width=1\textwidth]{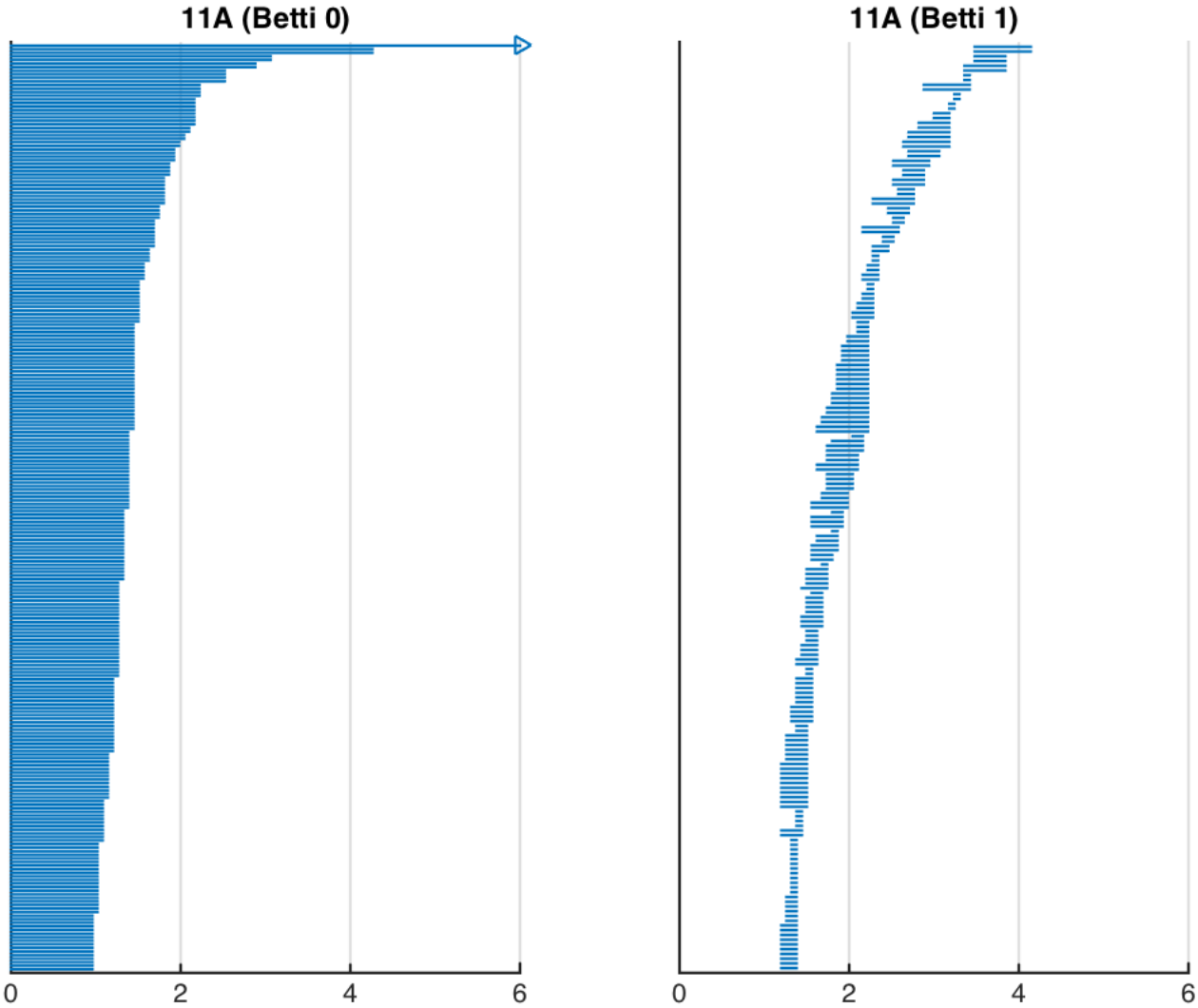}
\end{minipage}
\qquad
\begin{minipage}[b]{0.41\textwidth}\centering
\includegraphics[width=1\textwidth]{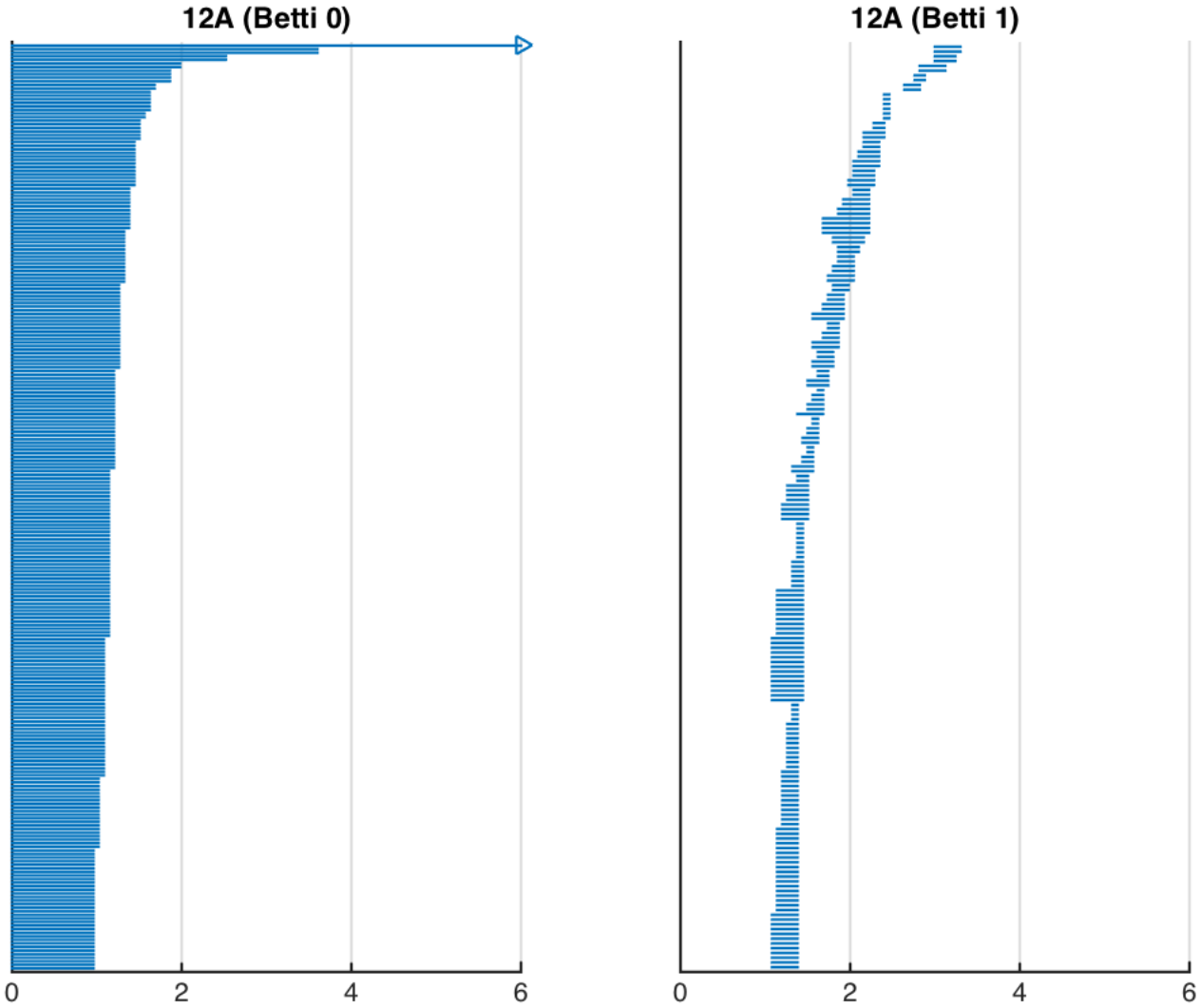}
\end{minipage}
\end{figure}

\begin{figure}[th!]\centering
\begin{minipage}[b]{0.41\textwidth}\centering
\includegraphics[width=1\textwidth]{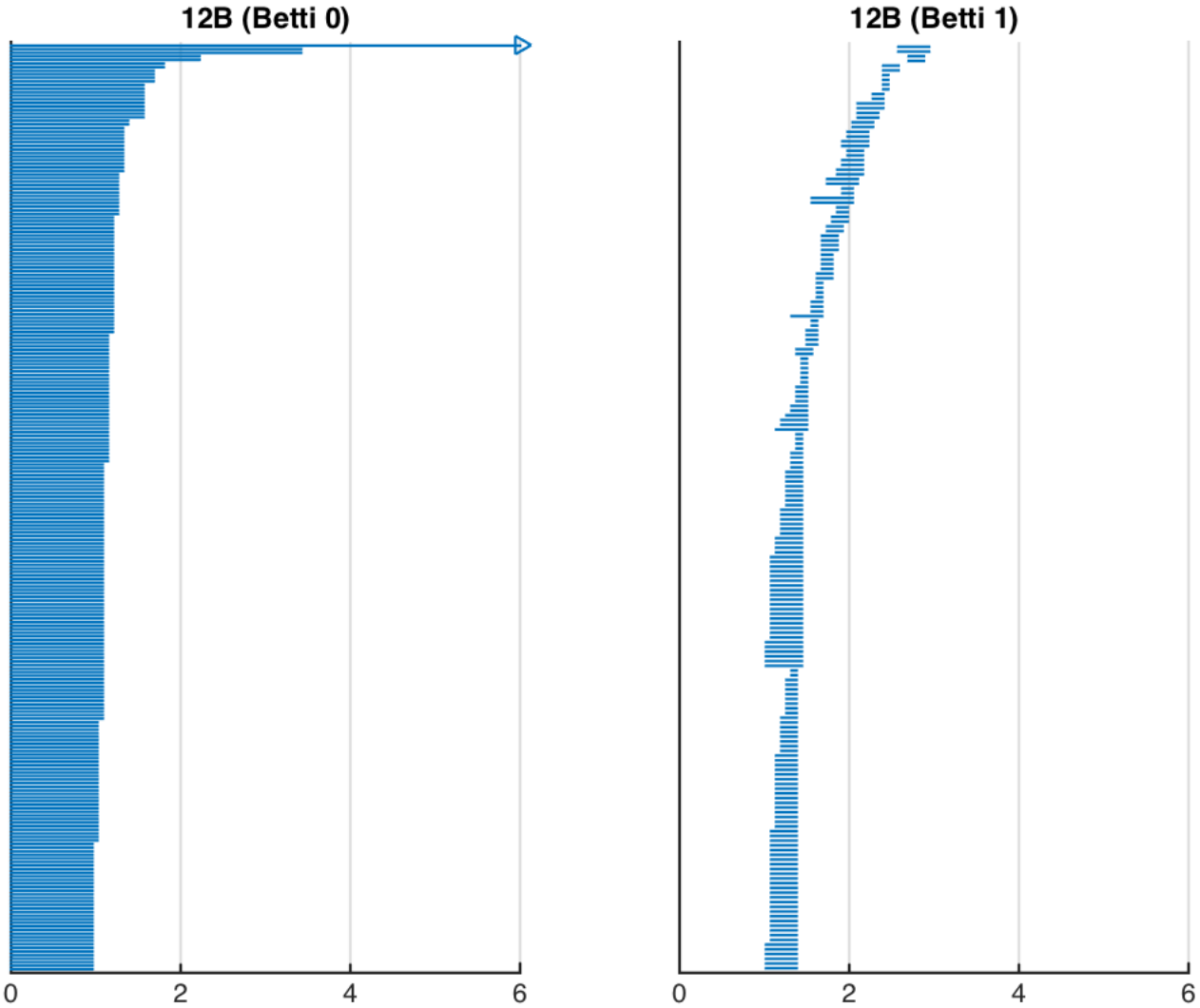}
\end{minipage}
\qquad
\begin{minipage}[b]{0.41\textwidth}\centering
\includegraphics[width=1\textwidth]{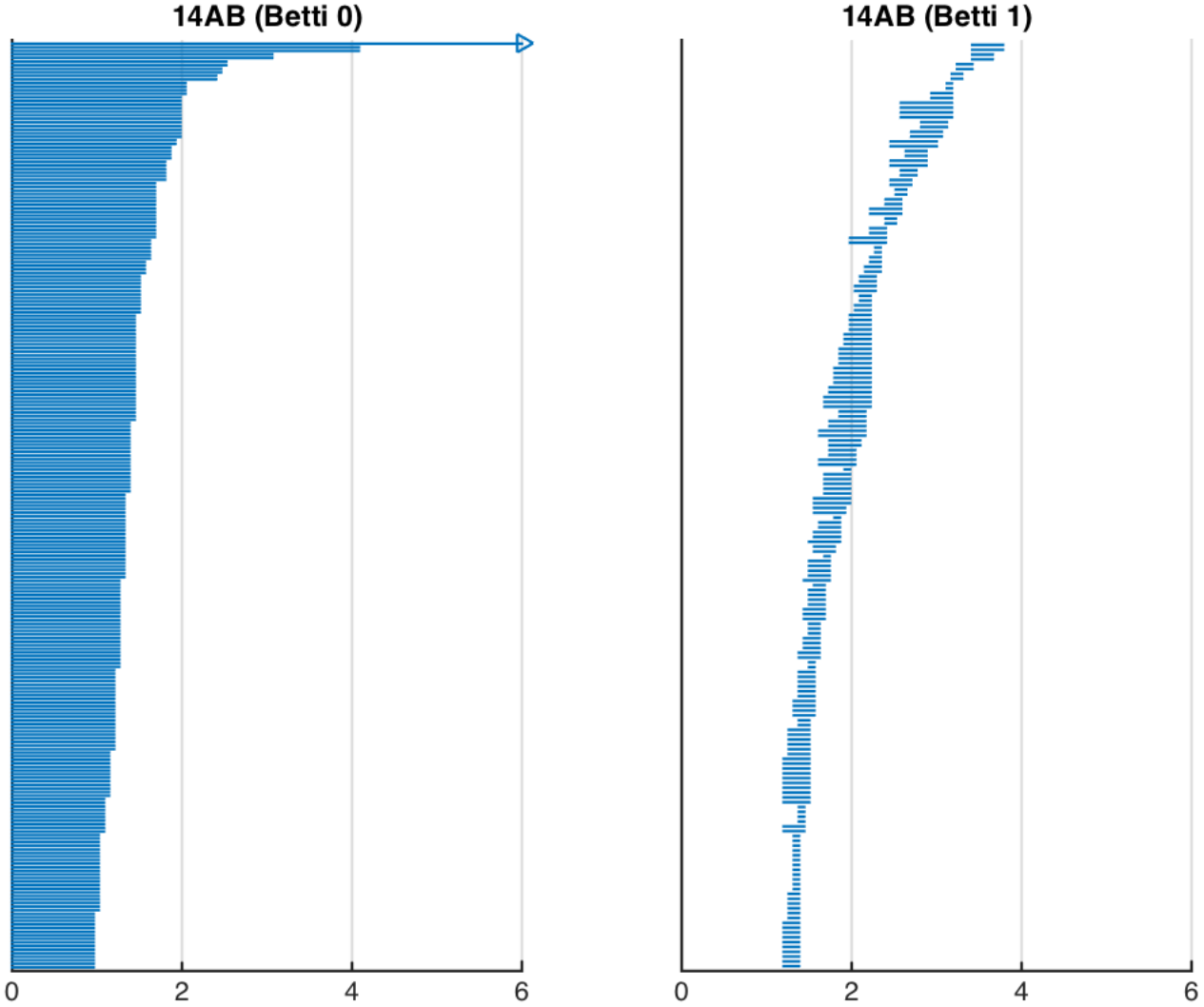}
\end{minipage}

\vspace{3mm}
\begin{minipage}[b]{0.41\textwidth}\centering
\includegraphics[width=1\textwidth]{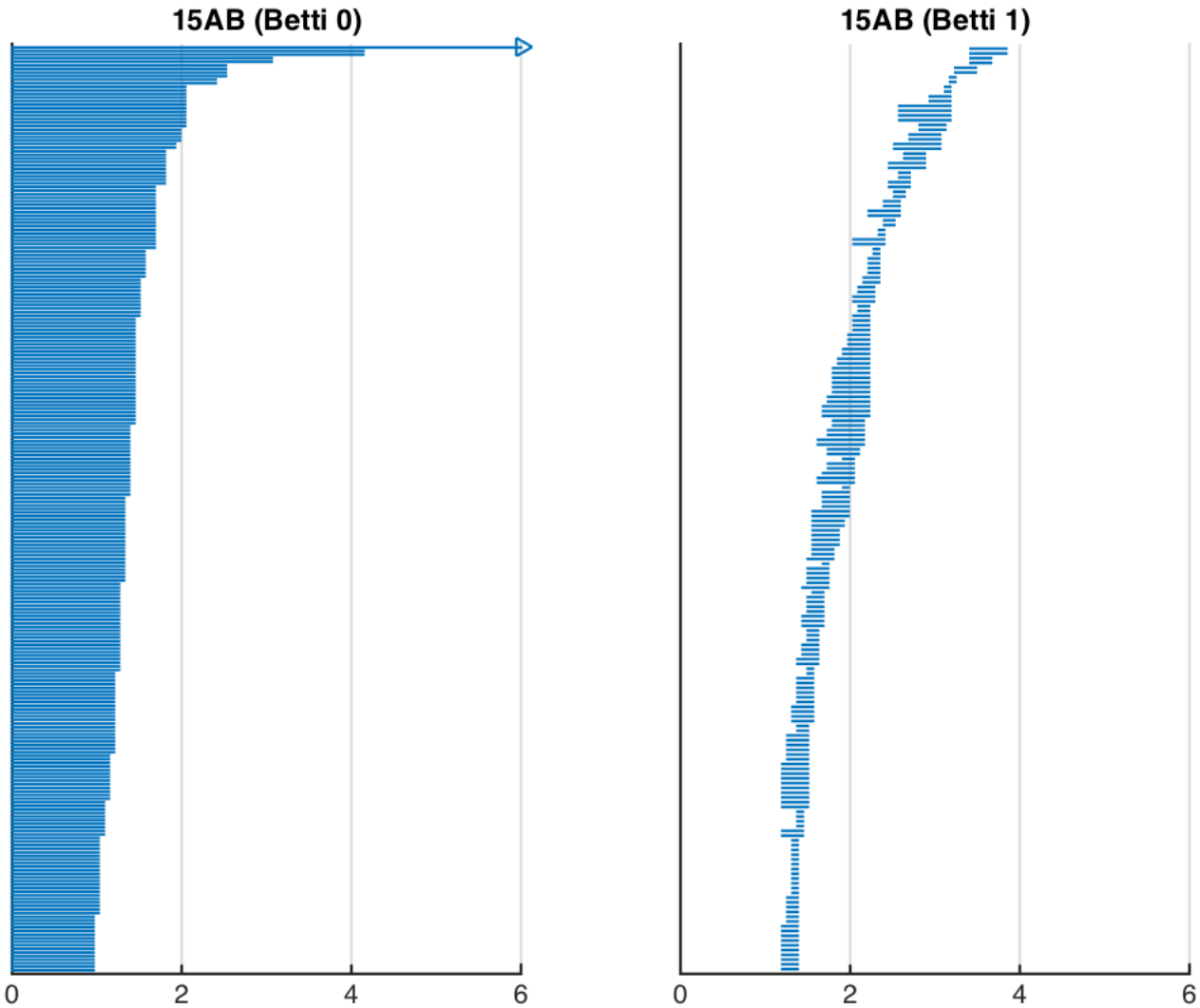}
\end{minipage}
\qquad
\begin{minipage}[b]{0.41\textwidth}\centering
\includegraphics[width=1\textwidth]{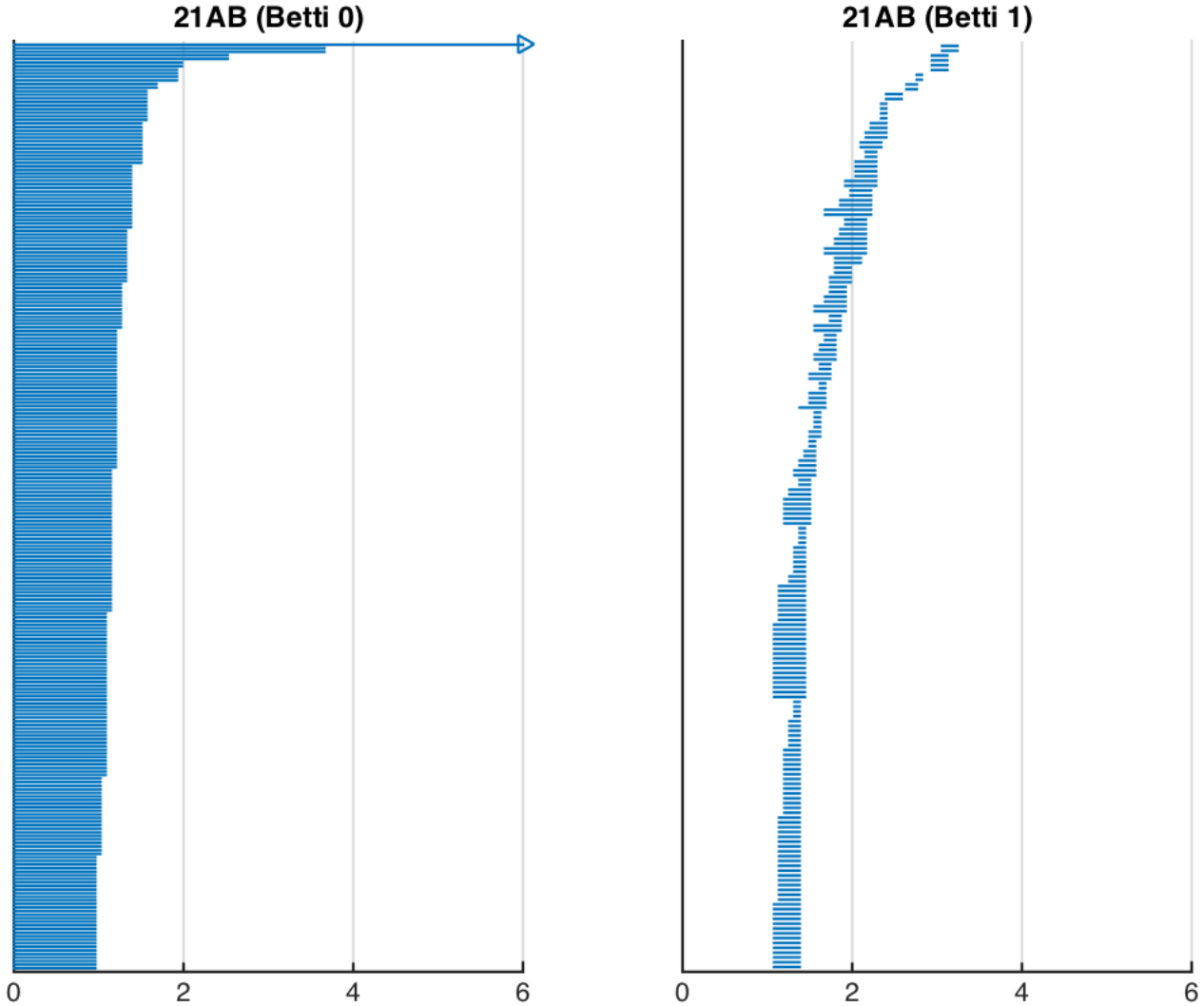}
\end{minipage}

\vspace{3mm}

\includegraphics[width=0.41\textwidth]{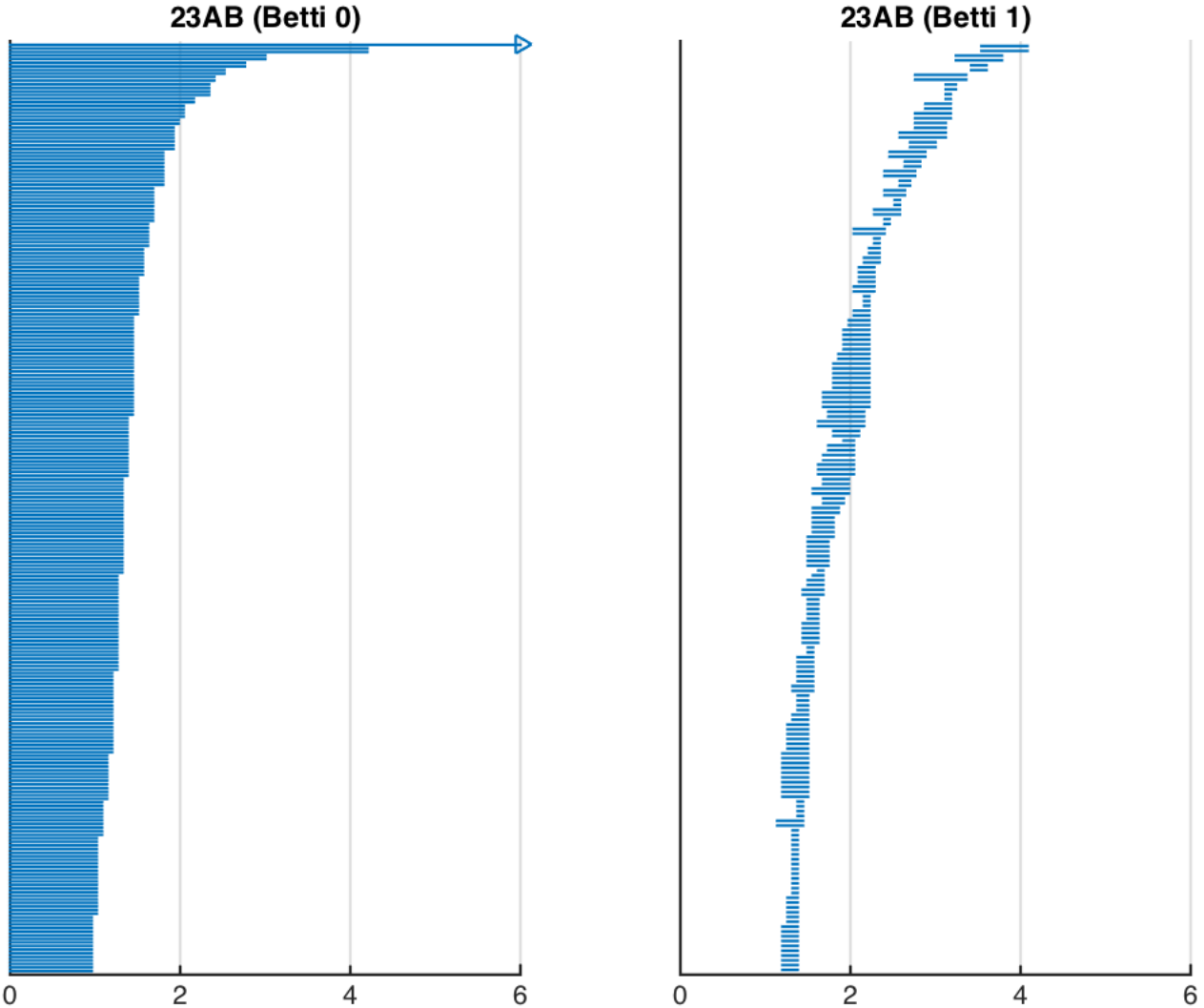}
\end{figure}

\subsection*{Acknowledgements}

I wish to thank the Applied Topology group at Stanford University for making \textsc{javaplex} available in~\cite{javaplex}. I also thank the theory division at CERN and IHES for hospitality and support during the preparation of this note. This work was partially supported by FCT/Portugal and IST-ID through UID/MAT/04459/2013, EXCL/MAT-GEO/0222/2012 and the program Investigador FCT IF2014, under contract IF/01426/2014/CP1214/CT0001. I am a member of INDAM-GNFM, I am supported by INFN via the Iniziativa Specifica GAST and by the FRA2018 project ``K-theoretic Enumerative Geometry in Mathematical Physics''.

\pdfbookmark[1]{References}{ref}
\LastPageEnding

\end{document}